\documentclass{aa}  
\usepackage{amsmath}
\usepackage{graphicx}
\usepackage{natbib}
\usepackage{multicol}
\usepackage{longtable}
\usepackage{graphicx}
\usepackage{natbib,twoopt}
\usepackage{supertabular,booktabs}
\usepackage{sidecap} 
\usepackage{enumitem}
\usepackage{txfonts}
\usepackage{epstopdf}
\usepackage{tabularx}
\usepackage{longtable}
\usepackage{ltablex}
\usepackage[normalem]{ulem}
\usepackage[pdftex]{color}
\newcommand{\Msun}{M$_{\odot}$}

\begin{document}

\title{Low-mass low-metallicity AGB  stars as an efficient i-process site explaining CEMP-rs stars \thanks{This paper is partly based on observations made with ESO/VLT at Paranal Observatory, under program 69.D$-$0063.}
}

\author{D. Karinkuzhi
          \inst{1,2,3}
          \and
          S. Van Eck\inst{2}
          \and
          S. Goriely\inst{2}
          \and
          L. Siess\inst{2}
                    \and
          A. Jorissen\inst{2}
          \and
          T. Merle \inst{2}
          \and
          A. Escorza\inst{2,4}
          \and
          T. Masseron\inst{5,6}
          }

\institute{Department of Physics, Indian Institute of Science, Bangalore 560012, India.
           \and
            Institut d'Astronomie et d'Astrophysique, Universit\'e Libre de Bruxelles, ULB, Campus Plaine C.P. 226, Boulevard du Triomphe, B-1050 Bruxelles, Belgium. \email{dkarinku@ulb.ac.be, svaneck@astro.ulb.ac.be}
            \and
            Department of Physics, Jnana Bharathi Campus, Bangalore University, Bangalore 560056, India. 
              \and
              Institute of Astronomy, KU Leuven, Celestijnenlaan 200D, B-3001 Leuven, Belgium
              \and
              Instituto de Astrofísica de Canarias, E-38205 La Laguna, Tenerife, Spain
              \and
              Departamento de Astrofísica, Universidad de La Laguna, E-38206 La Laguna, Tenerife, Spain
             }
\date{Received X, 2020; accepted Y, 2020}
   
\abstract
   {Among carbon-enhanced metal-poor (CEMP) stars, some are found to be enriched in slow-neutron capture (s-process) elements
   (and are then tagged CEMP-s), some have overabundances in rapid-neutron capture (r-process) elements (tagged CEMP-r), and some are characterized by both s- and r-process enrichments (tagged CEMP-rs).
   The current distinction between CEMP-s and CEMP-rs is based on their [Ba/Fe] and [Eu/Fe] ratios, since barium and europium are predominantly produced by the s- and the r-process, respectively.
   The origin of the abundance differences between CEMP-s and CEMP-rs stars is presently unknown. It has been claimed that the i-process, whose site still remains to be identified, could better reproduce CEMP-rs abundances than the s-process.
   }
   { We  propose a more robust classification method for CEMP-s and  CEMP-rs stars using additional heavy elements other than Ba and Eu. Once a secure classification is available, it should then be possible to assess whether the i-process or a variant of the s-process better fits the peculiar abundance patterns of CEMP-rs stars.
   }
   {We analyse high-resolution spectra of 24 CEMP stars and one r-process enriched star without carbon-enrichment, observed mainly with the high-resolution HERMES spectrograph mounted on the Mercator telescope (La Palma) and also with the UVES spectrograph on VLT (ESO Chile) and HIRES spectrograph on KECK (Hawaii).
   Stellar parameters and abundances are derived using MARCS model atmospheres. Elemental abundances are computed through spectral synthesis using the TURBOSPECTRUM radiative transfer code.
   Stars are re-classified as CEMP-s or -rs according to a new classification scheme using eight heavy element abundances.
   }
  {Within our sample of 25 objects, the literature classification is globally confirmed, except for HE 1429$-$0551 and HE 2144$-$1832, previously classified as CEMP-rs and now as CEMP-s stars.
  The abundance profiles of CEMP-s and CEMP-rs stars are compared in detail, and no clear separation is found between the two groups; it seems instead that there is an abundance continuum between the two stellar classes.
 There is an even larger binarity rate among CEMP-rs stars than among CEMP-s stars, indicating that CEMP-rs stars are extrinsic stars as well.
 The second peak s-process elements (Ba, La, Ce) are slightly enhanced in CEMP-rs stars with respect to first-peak s-process elements (Sr, Y, Zr), when compared to CEMP-s stars.
 Models of radiative s-process nucleosynthesis during the interpulse phases reproduce well the abundance profiles of CEMP-s stars, whereas those of CEMP-rs stars are  explained well by low-metallicity 1~\Msun\ models experiencing proton ingestion. 
 The global fitting of our i-process models to CEMP-rs stars is as good as the one of our s-process models to CEMP-s stars.
 Stellar evolutionary tracks of an enhanced carbon composition (consistent with our abundance determinations) are necessary to explain the position of CEMP-s and CEMP-rs stars in the Hertzsprung--Russell (HR) diagram using Gaia DR2 parallaxes; they are found to lie mostly on the red giant branch (RGB).
 }
 {CEMP-rs stars present most of the characteristics of extrinsic stars such as CEMP-s, CH, barium, and extrinsic S stars; they can be explained as being polluted by a low-mass, low-metallicity thermally-pulsing asymptotic
giant branch (TP-AGB)  companion experiencing i-process nucleosynthesis after proton ingestion during its first convective thermal pulses. As such, they could be renamed   CEMP-sr stars,  since  they  represent  a  particular manifestation of the s-process at low-metallicities. For these objects a call for an exotic i-process site may not necessarily be required anymore.}
\maketitle
\section{Introduction}
\label{Sect:Intro}

Low- and intermediate-mass stars (LIMS) dominate the stellar populations in galaxies. They are the siege of a rich nucleosynthesis which produces most of the elements heavier than iron, and thus play a crucial role in the chemical evolution of galaxies. The dominant nucleosynthetic processes
are the slow neutron-capture process (s-process) and the rapid neutron-capture process (r-process).
Detailed spectroscopic analyses of stars enhanced in heavy elements constitute an important method for investigating the physical and chemical conditions required for these processes to operate. While the nuclear mechanisms responsible for the s- and r-processes are  clearly identified, the origin of the peculiar abundance
patterns in the so-called rs-stars  showing enhancements of both s-process and r-process elements is still an open question \citep[][and references therein]{Barbuy1997,Hill2000,Beer2005,Jonsell2006,Masseron2010,Gull2018}.

Stars enriched in both r- and s-elements are  also carbon-enriched and are mostly found among metal-poor stars; they are thus tagged CEMP-rs (CEMP stands for carbon-enhanced metal-poor) stars \citep[e.g.][]{Beer2005,Masseron2010} but also among CH stars. \cite{Keenan1942}
coined the term `CH stars' to refer to high-velocity carbon stars that exhibit very strong G bands due to the CH molecule and otherwise weak metallic lines.
Both CEMP-s and CEMP-rs stars seem to belong to binary systems  \citep{Lucatello-2005,Starkenburg-2014,Hansen2016a}.
The current criterion used to distinguish CEMP-rs from CEMP-s stars is based on  [Ba/Eu] or [La/Eu] ratio \citep{Beer2005, Masseron2010}.
However, this classification could probably be  more robust,
and it is one of the goals of this paper to devise a better classification criterion.

 The scenarios proposed to explain CEMP-rs overabundances \citep{Jonsell2006, Masseron2010,Hampel2016} involve 
a primordial origin (pollution of the birth cloud by an r-process source);  pollution
of the binary by a third massive star (triple system);  pollution by the primary
exploding as a type 1.5 supernova or accretion-induced collapse; or  pollution by the intermediate-process or i-process, i.e. a
 process leading to neutron densities of the order of $N_n \sim 10^{14-15}$ cm$^{-3}$,
intermediate between those of the s-process ($N_n \sim 10^{8}$ cm$^{-3}$)
and those required by the r-process ($N_n \gg 10^{20}$ cm$^{-3}$).
 This process could also explain why low-metallicity stars,  including post-asymptotic
giant branch (post-AGB) stars,
 show lower abundance in lead and other heavy-s nuclei with respect to what is expected from standard s-process model predictions.  More precisely, below [Fe/H]$\sim -0.7$,
a very wide range of neutron irradiations is needed to explain the spread in heavy-s
abundances measured \citep{Lugaro2015,Desmedt2016}.
These authors suggested that the i-process could meet this challenge.
While one-zone parametric simulations of the i-process tend to reproduce convincingly the abundances
in CEMP-rs stars \citep{Hampel2016},
its astrophysical site has not yet been identified. 
 A number of suggestions have been proposed, among which the  proton ingestion in C-rich layers heated to relatively high temperatures, such as during core He flash in very low-metallicity low-mass stars 
 \citep[e.g.][]{Campbell10}, during the thermal pulse phase of intermediate-mass or massive AGB (super-AGB) stars of low metallicity \citep[e.g.][]{Goriely2005,Jones16}, during the post-AGB phase (`final thermal pulse') \citep[e.g.][]{Herwig11}, during rapid accretion of H-rich material on WDs \citep[e.g.][]{Denissenkov-2017,Denissenkov18}, or during shell He burning in massive very low-metallicity Pop II or Pop III stars \citep[e.g.][]{Clarkson18}.
Another possibility for the i-process operation would be the ingestion of protons
 in the He-burning convective region of low-metallicity low-mass AGB stars, as first described by \citet{Cowan1977} and \citet{Iwamoto04}. 
 This specific astrophysical site is the one we  focus on in the present study.  We note, however, that our recent identification and
analysis \citep{Karinkuzhi2018}  of a star enriched in s- and r-elements at
comparatively high metallicity ([Fe/H]$\sim~-0.7$) puts a strong constraint on the occurrence of
the i-process, which may not be limited to very low-metallicity environments.

The paper is structured as follows. The sample of CEMP-s and CEMP-rs stars and their binarity are described in Sect.~\ref{Sect:observations}, while the stellar parameter and abundance determinations are derived in Sects.~\ref{Sect:parameters} and \ref{Sect:abundances}. The new methodology to classify CEMP-s and -rs stars is presented in Sect.~\ref{Sect:assignment}. The abundances of the programme  stars are compared to literature determinations in Sect.~\ref{Sect:indiv-stars} and are  to nucleosynthesis computations in Sect.~\ref{Sect:nucleosynthesis}. 
Section \ref{Sect: abund-analysis} focuses on the discussion of specific abundance ratios concerning CNO and heavy elements. Finally, the stars are located in the Hertzsprung--Russell (HR) diagram in Sect.~\ref{Sect: HR} using Gaia DR2 parallaxes, and conclusions are drawn in Sect.~\ref{Sect: conclusion}.

\section{Stellar sample and observations}
\label{Sect:observations}
We analysed a sample of 25 metal-poor stars, 24 of which are carbon-enriched and one an r-process-enriched star without carbon enrichment. They were selected  from the catalogue of carbon stars from the Hamburg-ESO survey by \citet{Christlieb2001} and HK survey by \citet{Beers1992}, as well as from CEMP and CH stars in the literature, as listed in Table~\ref{Tab:program_stars}, which includes 15 CEMP-s and -rs stars already analysed in the literature. Therefore, we re-analysed these 15 stars and we present the first complete abundance pattern for ten new objects. Furthermore, many of the stars listed in Table~\ref{Tab:program_stars},  already classified as CEMP-s, in fact had no measured abundance for either Eu or any other r-process elements; their classification is therefore purely based on s-process elements, and their possible rs nature would therefore remain undetected.
It was the aim of our analysis to derive these missing abundances so as to decide whether these stars are CEMP-s or -rs.
Column 8  of Table~\ref{Tab:program_stars}, `Orig. class', 
lists the classification using our newly derived abundances and the [La/Eu] criterion of \cite{Beer2005} as described in Sect~\ref{Sect:assignment}.

High-resolution spectra for most of the objects were acquired from the HERMES spectrograph \citep{Raskin2011} mounted on the 1.2m Mercator telescope at the Roque de los Muchachos Observatory, La Palma, Canary Islands, which covers the wavelength range 3800 -- 9000 \AA~ at a resolution $R\sim 86\,000$. These objects were observed as part of a long-term monitoring of radial velocities to detect the binary nature and derive orbital parameters of specific families of stars \citep{Gorlova2014}.   Spectra  observed on  different nights were co-added after correcting for the Doppler shifts to  maximize the signal-to-noise ratio (S/N). We also used a few spectra obtained with the UVES spectrograph mounted on the ESO {\it Very Large Telescope}  with a wavelength coverage from 3300 to 6700 \AA.  Three stars were taken from the HIRES/KECK data archive and have wavelength coverage from 3700 to 8000 \AA.

\begin{table*}\small
\caption{Programme stars. }
\label{Tab:program_stars}
\begin{tabular}{llccccccccccc}
\hline
\\
Name &  $T_{\rm eff}$&$\log g$           & $\xi$         &[Fe/H] &$d_\text{S}$&$d_\text{rms}$&  Orig. & New  & $\chi^2$& Spec&Bin& Ref.\\
     &    (K)       & (cgs)    &  (km s$^{-1}$)&    && & Class. & Class. & & & & \\
\hline\\
CS 22887$-$048  &   6500 $\pm$ 50 &  3.20 $\pm$ 0.15  & 1.00 $\pm$ 0.05& $-$2.10 $\pm$ 0.09&0.784&0.861   & s & s & 4.8 &U& $-$ & $-$\\
CS 22891$-$171  &   5215 $\pm$ 68 &  1.24 $\pm$ 0.09  &2.14 $\pm$ 0.14 &$-$2.50 $\pm$ 0.10&0.530&0.756    &rs& rs  &5.3 &U& $-$ & $-$ \\
CS 22942$-$019  &   5100 $\pm$ 98 &  2.19 $\pm$ 0.20  &1.73 $\pm$ 0.10 & $-$2.50 $\pm$ 0.09&1.084&1.192  & s & s  & 9.9& U & Y&d,g\\
CS 30322$-$023  &   4500 $\pm$ 100 &  1.00 $\pm$ 0.50  & 2.80 $\pm$ 0.10 &$-$3.35 $\pm$ 0.09  &0.948&1.041& s & s & 4.8& U& $-$ &$-$\\
HD 26         &   5169 $\pm$ 108 &  2.46 $\pm$ 0.18  & 1.46 $\pm$ 0.08 &$-$0.98 $\pm$  0.09 &1.344&1.368& s & s  & 6.6& H & Y & b \\
HD 5223       &   4650 $\pm$ 120 &  1.03 $\pm$ 0.30  &2.16 $\pm$ 0.14  &$-$2.00 $\pm$ 0.08 &0.471&0.654 & rs&  rs &2.2& H& Y & e\\
HD 55496      &   4642 $\pm$ 39 &  1.65 $\pm$ 0.14  &1.33 $\pm$ 0.08  &$-$2.10 $\pm$ 0.09&1.350&1.416  & s & s  &9.1 &H&Y?&g,h\\
HD 76396      &   4750 $\pm$ 100 &  2.00 $\pm$ 0.30  & 2.00 $\pm$ 0.10 &$-$2.27 $\pm$ 0.10 &0.578&0.668 & rs& rs  &1.8 &H&Y&   b\\
HD 145777     &   4443 $\pm$ 57  &  0.50 $\pm$ 0.10 &2.63 $\pm$ 0.10    &$-$2.32 $\pm$ 0.10&0.448&0.575  & rs & rs  &3.9&  H&Y&b,c   \\
HD 187861     &   5000 $\pm$ 100 &  1.50 $\pm$ 0.25 &2.00 $\pm$ 0.20  &$-$2.60 $\pm$ 0.10&0.037&0.506  & rs&  rs  &10.2 &U& $-$ & $-$\\
HD 196944     &   5168 $\pm$ 48 &  1.28 $\pm$ 0.16  &1.68 $\pm$  0.11 &$-$2.50 $\pm$ 0.09&0.438&0.547  & rs&  rs &1.3 &U, H& Y & a\\
HD 198269     &   4458 $\pm$ 15 &  0.83 $\pm$ 0.08  & 1.64 $\pm$ 0.09 &$-$2.10 $\pm$ 0.10&0.856&0.946  & s&  s &4.9 & H&Y&e \\
HD 201626     &   5084 $\pm$  77 &2.18 $\pm$  0.16 & 1.60 $\pm$  0.09&$-$1.75  $\pm$ 0.10&1.106&1.163 & s&  s &3.9 & H&Y&e \\
HD 206983     &   4200 $\pm$ 100 &  0.60 $\pm$ 0.20  & 1.50 $\pm$ 0.10  &$-$1.00 $\pm$ 0.10 &0.630&0.702 & s&  s &2.7 & H&Y?&g \\
HD 209621     &   4740 $\pm$ 55 &  1.75 $\pm$ 0.25  & 1.94 $\pm$ 0.13 &$-$2.00 $\pm$ 0.09&0.314&0.517 &rs& rs  &6.3 & H&Y&e \\
HD 221170$^1$     &   4577 $\pm$ 50 &  0.77 $\pm$ 0.20  & 1.84 $\pm$ 0.10 &$-$2.40 $\pm$ 0.10&0.101&0.208  &r & r & --& H&N&i\\
HD 224959     &   4969 $\pm$ 64 &  1.26 $\pm$ 0.29  &1.63 $\pm$ 0.14  &$-$2.36 $\pm$ 0.09&0.221&0.647  &rs& rs  &10.6 & U, H&Y&e \\
HE 0111$-$1346  &   4687 $\pm$ 102 &  1.26 $\pm$ 0.30  &1.77 $\pm$ 0.16  &$-$2.10 $\pm$ 0.09&0.673&0.825& s &  s &10.9 & H&Y&b,f\\
HE 0151$-$0341  &   4820 $\pm$ 112 &  1.15 $\pm$ 0.08  &1.72 $\pm$ 0.16  & $-$2.89 $\pm$ 0.08&0.303&0.607&rs &  rs &8.9 & HI&Y&f\\
HE 0319$-$0215  &   4738 $\pm$ 100 & 0.66 $\pm$ 0.40   & 2.16 $\pm$ 0.10&$-$2.90 $\pm$ 0.10&0.540&0.780&rs &  rs &9.0 & HI&Y&f\\
HE 0507$-$1653  &  5035 $\pm$ 53  &  2.39 $\pm$ 0.16  & 1.53 $\pm$ 0.14  &  $-$1.35 $\pm$  0.10&1.038&1.073  &s& s  &9.0 & H& Y&b,f\\
HE 1120$-$2122  &  4500 $\pm$ 100 &   0.50 $\pm$ 0.50  &1.50 $\pm$ 0.10&$-$2.00 $\pm$ 0.10&0.331&0.470& rs &  rs &7.3 & H& Y&b,c \\
HE 1429$-$0551  &   4832 $\pm$ 41 &  1.14 $\pm$ 0.20  & 2.01 $\pm$ 0.14& $-$2.70 $\pm$ 0.10&0.714&0.777& rs&  s &4.1 & H& N & b,c\\
HE 2144$-$1832  &  4250 $\pm$ 100  &0.50  $\pm$ 0.30  &2.00 $\pm$ 0.15&$-$1.85 $\pm$ 0.10 &0.753&0.798& rs &  s  &5.1 & H&Y?&b,c  \\
HE 2255$-$1724  &  4776 $\pm$ 51  & 1.64  $\pm$ 0.15  &1.84 $\pm$ 0.15&$-$2.32 $\pm$ 0.09 &0.744&0.875& s & s & 7.4 & HI& $-$ &$-$\\
\hline
\end{tabular}
\medskip\\
Notes. Columns 2--5 list the atmospheric parameters ($\xi$ is the microturbulence velocity).
The $d_\text{S}$ and $d_\text{rms}$ distances are described in Sect.~\ref{Sect:assignment}.
 Column 8, labelled `Orig. Class.', specifies the classification based on  [La/Eu], adopting the criterion suggested by  \citet{Beer2005}, whereas  Col. 9, labelled `New Class.', refers to the star assignment adopted in the present paper (see Sect.~\ref{Sect:assignment}). The $\chi^2$ indicator (Eq.~\ref{eq:chi2} in Sect.~\ref{Sect:nucleosynthesis}) quantifies the agreement between the AGB model predictions and the measured abundances of eight heavy  elements. Column 11, labelled `Spec', refers to the spectrograph used (U, H, and HI correspond to UVES, HERMES, and HIRES, respectively). A question mark in the   `Bin' column flags a possible spectroscopic binary. $^1$ HD 221170 is an r-process-enriched star not enriched in carbon.\\
References for the `Bin' column: (a) HERMES unpublished data; (b) \citet{Jorissen2016}; (c) Jorissen et al. (in preparation); (d) \citet{Preston2001}; (e) \citet{Mcclure1990}; (f) \citet{Hansen2016a}; (g) \citet{Jorissen2005}; (h) \citet{Pereira2019}; (i) \citet{Hansen2015}.\\

\end{table*}

Information about radial-velocity variability is provided in the  `Bin' column in Table~\ref{Tab:program_stars}.
For many stars information was already available in the literature, but for five objects additional radial velocities and/or new orbits will be published in a forthcoming paper.
Table~\ref{Tab:program_stars}  reveals  the  high   incidence of binarity among  CEMP-rs  objects;  according to our new classification scheme (see  Sect.~\ref{Sect:assignment}), 9 out of 11 CEMP-rs stars (82\%) have been detected as binaries, compared to 6~--~9 out of 13 CEMP-s stars (46~--~69\%). Thus, the binarity rate of CEMP-rs stars is reminiscent of the value found in extrinsic-star classes, like extrinsic S stars, barium stars, CH stars, and their lower metallicity counterparts, the CEMP-s stars.

\section{Derivation of the atmospheric parameters}
\label{Sect:parameters}

Our programme stars are low-metallicity carbon stars
with strong molecular bands throughout their spectra. It is therefore difficult to find unblended neutral and ionized Fe lines
in their spectrum. Adopting the usual excitation and ionization equilibrium techniques to estimate the atmospheric parameters is challenging.
As an initial estimate we derived the photometric temperatures using  the relations given by \citet{Alonso1996,Alonso1999} for dwarfs and giants. These calibrations use Johnson $UBVRI$ colours, and the TCS (Telescopio Carlos Sanchez) system for the infrared colours $J-H$ and $J-K$. 
We performed the necessary transformation to these photometric systems using the transformation relations from \cite{Ramirez2004} and \cite{Alonso1996,Alonso1999}.

\begin{figure*}
\includegraphics[clip,trim=2cm 1cm 1cm 3cm, width=0.95\linewidth]{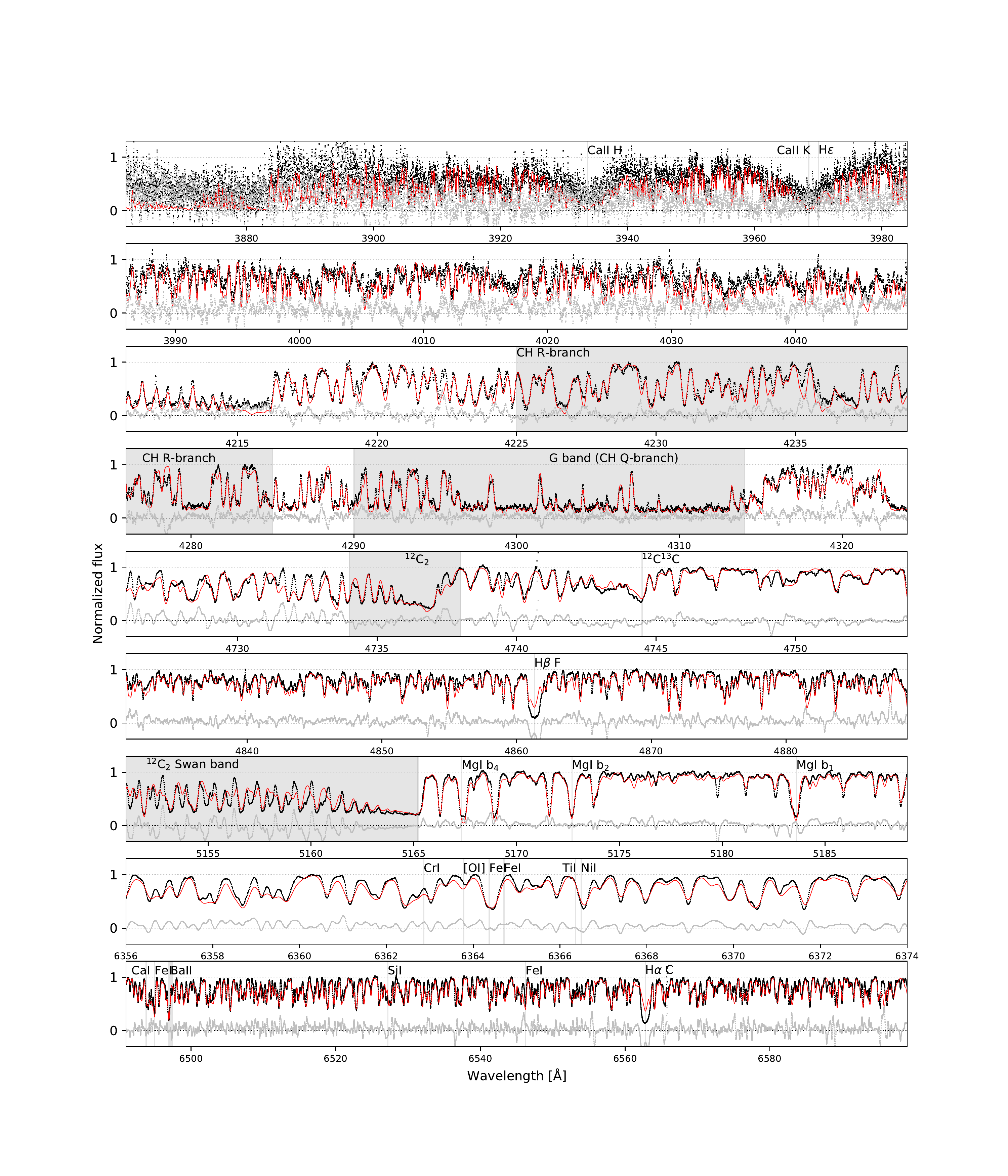}
\caption{Spectral fitting of HE 1120$-$2122 using KASTEEL. Different spectral regions used to fine-tune the atmospheric parameters are illustrated. The black dots represent the normalized observed flux, the red line corresponds to the synthetic fluxes, and the grey dots are the residuals. The shaded areas correspond to molecular bands, as labelled.}
\label{Fig:HE1120_kasteel}
\end{figure*}

With the help of the {\sc BACCHUS} pipeline \citep{Masseron2016}, which uses interpolated MARCS model atmospheres \citep{Gustafsson2008} and the 1D local
 thermodynamical equilibrium (LTE) spectrum-synthesis code {\sc TURBOSPECTRUM} \citep{Alvarez1998,Plez2012},
 we derived the atmospheric parameters for the programme stars. {\sc } The Brussels Automatic Code for Characterizing High accUracy Spectra (BACCHUS) derives the stellar parameters ($T_{\rm eff}$, [Fe/H], $\log g$, microturbulence velocity $\xi$, and rotational velocity).
 The code includes on-the-fly spectrum synthesis, local continuum normalization, estimation of local S/N, and automatic line masking.
 It computes
abundances using equivalent widths or spectral synthesis, allowing us to check for excitation and ionization equilibria, thereby constraining $T_{\rm eff}$ and $\log g$. The microturbulent velocity $\xi$ is calculated by ensuring consistency between Fe abundances derived from lines of various reduced equivalent widths. We manually selected clean Fe lines to avoid possible errors due to blends.
 We performed the spectral fitting of the C$_2$ band at 5165 \AA\ and of the CN  region around 6320 \AA\  using these parameters to get an initial estimate of the
 C and N abundances. We included these values in the {\sc BACCHUS} initial abundances so as to
 derive the equivalent widths of Fe lines by taking care of the possible blends caused by
 these elements, and we then derived the final parameters. In order to check the quality of the spectral fit with the adopted atmospheric  parameters, we used the KASTEEL code, which relies on a spectral fitting method of selected wavelength
ranges sensitive to the atmospheric parameters (see Fig.\ref{Fig:HE1120_kasteel}). It uses the same
radiative transfer code and line lists as {\sc BACCHUS}
to minimize systematics. Here we focus on
eight wavelength ranges, namely around the CN bands and Ca II H\&K lines
[3860 - 3985~\AA], the CH band [3985 - 4050~\AA], the CN blend at 4215~\AA\
and the Ca I line at 4226~\AA\ [4210 - 4240~\AA], the $^{12}$C$^{12}$C and
$^{12}$C$^{13}$C bands [4725 - 4755~\AA], the region around H$\beta$ [4830 - 4890~\AA], the C$_2$ Swan band and the \ion{Mg}{I}~b triplet [5150 - 5190~\AA],
the [\ion{O}{I}] line [6355 - 6375~\AA], and the region around H$\alpha$ including
the \ion{Ba}{II} line at 6496.9~\AA\ [6490 - 6600~\AA].
A few \ion{Fe}{I} and \ion{Fe}{II} lines in the spectral regions 5200 -- 5400~\AA\ and 5800 -- 5900~\AA, which are comparatively free from molecular blending, were visually tested to validate the atmospheric parameters, especially $\log g$ and  metallicity.

\section{Abundance determination}
\label{Sect:abundances}

Abundances were derived by comparing observed and synthetic spectra generated
with the TURBOSPECTRUM  radiative transfer code \citep{Alvarez1998} using MARCS model atmospheres.
We used carbon-enriched MARCS models computed by Masseron  whenever available.
We adopted the solar abundances from \citet{Asplund2009}. 
 The line lists assembled in the framework of the Gaia-ESO survey  were used \citep{Heiter2015,Heiter2020}.
 The individual atomic lines,  including hyperfine  (HF) splitting when available,  used for the abundance determination are listed in Table~\ref{Tab:linelist}.
We derived the abundances under the assumption of   LTE, but non-LTE corrections are applied whenever available. In the following we comment on individual elemental abundances.

\subsection{{\it C, N}, and {\it O}}

 Oxygen abundances were derived first
from the [\ion{O}{I}] line at 6300.303~\AA\ whenever it could be satisfactorily reproduced. We also used
the \ion{O}{I} resonance triplet at 7774~\AA. This feature yields 
higher abundances (by 0.2 to 0.3~dex) in all the objects compared to the abundance
derived from the [\ion{O}{I}] line. This difference is usually due to
 NLTE effects strongly affecting the resonance line as described by \citet{Asplund2005} and \citet{Amarsi2016}. Hence, we used the O
abundance derived from the [\ion{O}{I}] line, or if not available, from
the NLTE-corrected abundance from the \ion{O}{I} resonance triplet. In Table~\ref{Tab:abundances}, we present the oxygen abundances only for the objects for which we could use either the  6300.303~\AA\ or the 7774~\AA~line.
For a few cool stars for which we could detect neither the 6300.303~\AA\ [\ion{O}{I}]
line nor the triplet at 7774~\AA, we used the
$\alpha$-element abundances
scaled to the corresponding stellar metallicity.

\begin{figure}
\includegraphics[height=10.0cm,width=9.0cm]{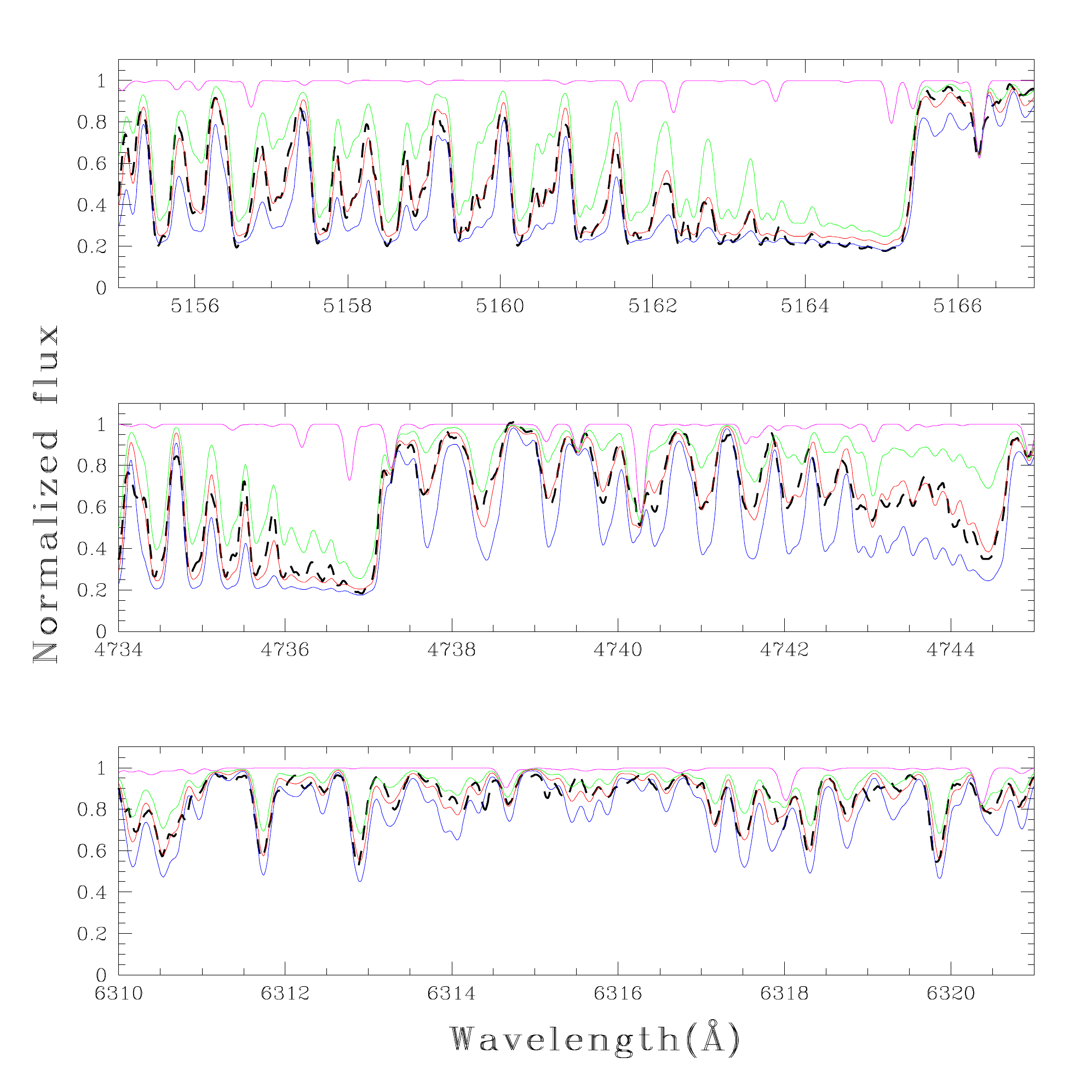}
\caption{Spectral fits of the C$_2$ and CN regions in HD 187861. Upper panel:   C$_2$ band around 5165 \AA\ where the red curve  represents the synthesis corresponding to the abundance $\log \epsilon_C = 8.3$, with $\pm0.3$~dex on either side represented by the blue and green curves. The black dashed line corresponds to the observed spectrum and the magenta line  to a spectral synthesis without carbon. The middle panel displays the spectral synthesis of the $^{12}$C$^{12}$C feature at 4737~\AA\ and the $^{12}$C$^{13}$C
feature at 4744~\AA\, with the adopted C abundance and derived isotopic ratio of 16. The spectral synthesis of the CN band around 6315 \AA\ is shown in the bottom panel with the adopted abundances of 8.3 and 7.85 for C and N, respectively. 
\label{Fig:HD187861_CN} }
\end{figure}

C and N abundances were derived for all programme stars.
The carbon abundance  was  derived mainly from the C$_2$ bands at 4737, 5165, and 5635~\AA, avoiding the band heads since they are often saturated and their depth is difficult to reproduce, and
also from the CH G band at 4310~\AA\ (Fig.~\ref{Fig:HD187861_CN}). We were able to get
 abundances consistent within 0.2 dex from all these bands.

The nitrogen
abundance for the programme stars observed with the UVES spectrograph
was derived from the CN bands at 3883, 4215, and 6323~\AA. 
For objects with spectra acquired with the HERMES spectrograph,
 we also used CN lines above 7500~\AA. 
 N abundances
 derived from these different bands are found to be consistent within 0.15 dex.

 The $^{12}$C/$^{13}$C ratio was derived using  $^{12}$CN features at 8003.553 and
 8003.910~\AA\, and $^{13}$CN features at 8004.554, 8004.728 and 8004.781~\AA.
We also used  the $^{13}$CN features at 8010.458 and 8016.429~\AA,
the $^{12}$C$^{12}$C feature at 4737~\AA,\ and the $^{12}$C$^{13}$C
feature at 4744~\AA\ as a consistency check. The $^{12}$C/$^{13}$C values from these
various bands are
found to agree within $\pm 5$.

\subsection{Light elements}
 The abundances of Na, Mg, Ca, Ti, Sc, Cr, Mn, Co, Cu, and Zn were derived (Table~\ref{Tab:light_abundances}) using the lines listed in Table~\ref{Tab:linelist}.

 \subsubsection{Na}
The Na abundance was derived using mainly two \ion{Na}{I} lines at 5682.633 and 5688.205 \AA. We also used the \ion{Na}{I} lines at 6154.226 and 6160.747 \AA\ whenever available. The lines  used to derive the Na abundance are weak and are known to be free from NLTE effects. 
\subsubsection{Mg, Ca, Sc, Ti}
 We could measure the Mg abundance only in a few stars using the Mg lines at 4571.096, 4702.991, 5528.405, and 5711.088~\AA.
Although  many lines of the elements Ca, Sc, and Ti could be identified, only those listed in Table~\ref{Tab:linelist} were used to derive abundances. We note that in all our programme stars these $\alpha$ elements are
enhanced with an average [$\alpha$/Fe] of 0.5 dex.

\subsection{Cr, Mn, Cu, Ni, Zn}
While Cr, Mn, Cu, and Ni
have either subsolar or solar abundances in all our programme stars, we found that Zn is slightly enriched. HD 55496
shows a rather high enrichment of Zn with [Zn/Fe] = 0.74. This is reminiscent of the depletion pattern observed in some RV Tauri stars and related objects
\citep[][and references therein]{Reyniers-2007}, i.e. of an anti-correlation between condensation temperatures in grains and elemental abundances.
Zinc is a relatively volatile element given its low condensation temperature \citep[$T_{\rm cond, 50\%} = 726$~K,][]{Lodders-2003}.
We  checked among all our stars that there is no obvious anti-correlation between condensation temperatures and elemental abundances. If there was any (as super-solar Zn abundances would tend to indicate), it has largely been  erased by a subsequent heavy-element enrichment.

\subsection{Light s-process elements: Sr, Y, and Zr}

Several clean (i.e. unblended) lines are available to derive the abundances of light s-elements. 

The Sr abundance was derived using the \ion{Sr}{I} lines at 4607.327 and 7070.070~\AA\ or the \ion{Sr}{II} lines at 4077.707 and 4215.519~\AA. 
These Sr lines (listed in Table~\ref{Tab:linelist}) are affected by NLTE effects; we now detail star by star the NLTE corrections that were adopted from \citet{Mashonkina2008}, \citet{Andrievsky2011}, and \citet{Bergemann2012}.

\noindent
{\bf CS 22887$-$048, CS 22942$-$019, HE 0507$-$1653, and HE 2144$-$1832}. We
used only the \ion{Sr}{II} line at 4077.707~\AA. NLTE corrections corresponding to the stellar parameters of these objects are small, however, ranging from 0.0 to $-$0.05~dex \citep{Bergemann2012}.\\

\noindent {\bf HD~209621, HD~224959, and HE~1120$-$2122}. The 
strontium abundance is derived using
the \ion{Sr}{I} line at 4607.327~\AA, known to show very large ($\approx$ 0.3 to 0.5 dex) NLTE corrections at low metallicities ($-3.9 \le  \mathrm{[Fe/H]} \le -1.2$), as listed in Table~3 of \citet{Bergemann2012}. These authors estimated a NLTE  correction of 0.47 dex for 
HD 122563 ($T_{\rm eff} = 4600$~K, $\log g = 1.6$, [Fe/H] = $-$2.5). We adopted this value 
 to correct the Sr abundance in these three stars. 

\noindent  {\bf HD~5223 and HD~145777}. Their Sr abundance was derived using the \ion{Sr}{II} line at 4215.52~\AA\ which is known to have a small NLTE correction ($\approx$ $-$0.05 dex) in metal-poor stars \citep{Mashonkina2008,Andrievsky2011}.

\noindent {\bf HD~26 and HD~55496}.  We used the \ion{Sr}{I} line at 7070.070 \AA\, for which
NLTE investigations are not available in the literature. 

The Y abundances for the programme stars were derived from the \ion{Y}{II} lines listed in Table~\ref{Tab:linelist}. No NLTE correction is available for these lines.
The Zr abundance was derived mainly using the \ion{Zr}{II}  lines at 5112.270 and 5350.350~\AA. We could not find any useful \ion{Zr}{I} lines; in particular, we could not use the lines at 7819.374 and 7849.365~\AA\ with laboratory $\log gf$ values as in \citet{Karinkuzhi2018}.

\subsection{Heavy s-process elements:  Ba, La, Ce, Pr, Nd}
Since most of the Ba lines are very strong in our programme stars, the Ba abundance is derived mainly from the spectral synthesis
of the weak \ion{Ba}{II} line at 4524.924~\AA.
For a few objects, we could also
use the line at 5853.673~\AA.
In the case of CS22887$-$048 we used the
\ion{Ba}{II} line at 4166.000~\AA.
Ba lines are strongly affected by HF splitting.
HF splitting data is not available for the 4524.924~\AA\  line, but it was
taken into account for the \ion{Ba}{II} line at 5853.673~\AA .
NLTE corrections for the Ba line at 5853.673 \AA\ is negligible for dwarfs and giants, as concluded by \citet{Mashonkina1999}, \citet{Mashonkina2000}, and \citet{Andrievsky2009} for the metallicity and temperature ranges of our objects. For CS 30322$-$023, HD 55496, HD 221170, and HE 2255$-$1724 we also used the \ion{Ba}{II} line at 5853.673~\AA\  to derive the Ba abundance. NLTE corrections corresponding to the parameter range of these four objects are   from \citet{Andrievsky2009} and range between 0.0 and 0.07. We present LTE and NLTE Ba abundances separately in  Table~\ref{Tab:abundances} for these four objects, where Ba$_{\rm LTE}$  indicates the Ba abundance derived using the \ion{Ba}{II} line at 4524.924 \AA\ and Ba$_{\rm NLTE}$ indicates that from 5853.673 \AA\ line after considering the NLTE correction.  

\begin{figure}
\includegraphics[width=9.0cm]{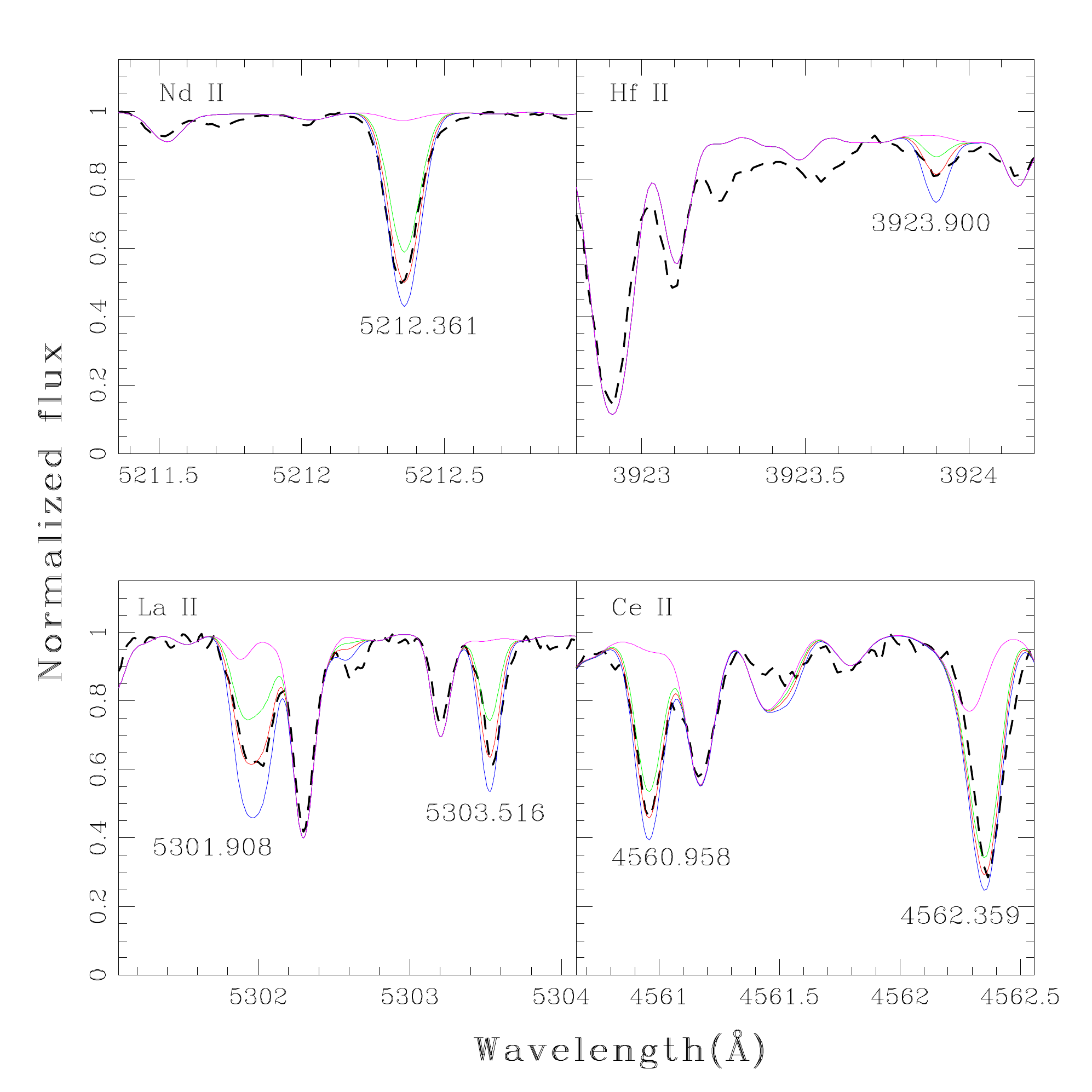}
\caption{Spectral fitting of the \ion{La}{II}  and  \ion{Ce}{II} lines is shown for HD~198269 in the lower panel. Spectral synthesis of \ion{Nd}{II} and \ion{Hf}{II} is shown for HD~224959 in the upper panel. Red curves correspond to spectral syntheses with the adopted abundances for these elements, as listed in Table~\ref{Tab:abundances}. The blue and green curves correspond to syntheses with abundances deviating by  $\pm$ 0.3~dex from the adopted abundance.  The black dashed line represents the observed spectrum. The magenta line corresponds to the synthesis with a null abundance for the corresponding element.
\label{Fig:HD224959_s} }
\end{figure}

The La abundance is derived mainly using the lines for which HF splitting is available. Although \ion{Nd}{II} lines are available throughout the spectrum, we used mainly \ion{Nd}{II} lines in the range 5200 -- 5400~\AA\, since they are
relatively free from molecular blends. The  \ion{Ce}{II} and \ion{Pr}{II} abundances were also  derived in all the programme stars.  Spectral fitting of a few lines is shown in Fig.~\ref{Fig:HD224959_s}.

\subsection{r-process-dominated elements: Sm, Eu, Dy, Gd, Er, Hf, Os, Ir}
The  Solar  System  enrichment  in  Sm,  Eu,  Dy,  Gd,  Er,  Os, and  Ir  is  believed  to   originate  from  the r-process nucleosynthesis by more than 80\% \citep{Goriely1999}.  The solar Hf is expected to be produced by the s- and r-process in approximately the same amount. Globally, these elements play a key role in tracing large neutron irradiations responsible for their production; they are thus often referred to as r-process elements.

Most of the \ion{Sm}{II} lines used to derive the Sm abundance are
blended with bands from carbon-bearing molecules.
These were carefully reproduced before deriving Sm abundances.

The Eu abundance was derived using four \ion{Eu}{II} lines with HF splitting available. As for the Sm lines, the Eu lines are blended with the CH and CN lines. The Eu abundance, when derived using two blue lines at 4129.720 and 4205.065~\AA,  is always lower by approximately 0.1 to 0.2~dex  
compared to the abundance derived from the two red lines. \citet{Mashonkina2008} calculated the NLTE corrections for the \ion{Eu}{II}  line at 4129.720~\AA\ in the two metal-poor stars HD~122563 ($T_{\rm eff}$ = 4600~K, $\log g = 1.50$, [Fe/H] = $-$2.53), and HD~84937
($T_{\rm eff}$ = 6365~K, $\log g = 4.00$, [Fe/H] = $-$2.15) and found corrections of respectively 0.16 and 0.12 dex. Since our programme star parameters are close to those of these two objects, the differences in Eu abundances when measured using the 4129.720~\AA\ line instead can be explained with these NLTE corrections. We could use the 4129.720~\AA\ Eu line in four objects, namely CS 22887$-$048, CS 22891$-$171, CS 22942$-$019, and HD 221170. We present LTE and NLTE Eu abundances separately in  Table~\ref{Tab:abundances} for these objects. Hence, Eu$_{\rm NLTE}$ denotes the abundance derived using the 4129.720~\AA\ line after applying the corresponding NLTE correction, while Eu$_{\rm LTE}$ denotes the average Eu abundance derived using all Eu lines available excluding the 4129.720~\AA\ line. For HD~196944 we used the two \ion{Eu}{II} lines at 3907.10 and 4205.065 \AA\ to derive the Eu abundance. \citet{Mashonkina2014} listed the NLTE corrections of 0.28 and 0.15 dex respectively for these two lines, corresponding to the parameters $T_{\rm eff}$ = 4750~K, $\log g = 1.00$, [Fe/H] = $-$3.00, which are closest to those of HD~196944. The Eu$_{\rm NLTE}$ abundance presented in Table~\ref{Tab:abundances} for HD~196944 is the average of the abundances derived from these two lines after applying NLTE corrections. For HE~0151$-$0341 the Eu abundance is derived using \ion{Eu}{II} line at 4205.065 \AA. We applied NLTE correction of 0.15 dex corresponding to $T_{\rm eff}$ = 4750~K, $\log g = 1.00$, [Fe/H] = $-$3.00 from \citet{Mashonkina2014}. For the 6645.130~\AA\  \ion{Eu}{II} line these corrections are very small ($\approx$ 0.06 to 0.08 dex in main-sequence and turn-off stars with metallicities higher than $-3.0$ \citep{Mashonkina2000,Mashonkina2014}. NLTE correction is not available for \ion{Eu}{II} line at 6437.640 \AA. The final Eu abundance given in Table~\ref{Tab:abundances} for all other targets is the average of the abundances from all the available Eu lines.

\begin{figure}
\includegraphics[width=9.0cm]{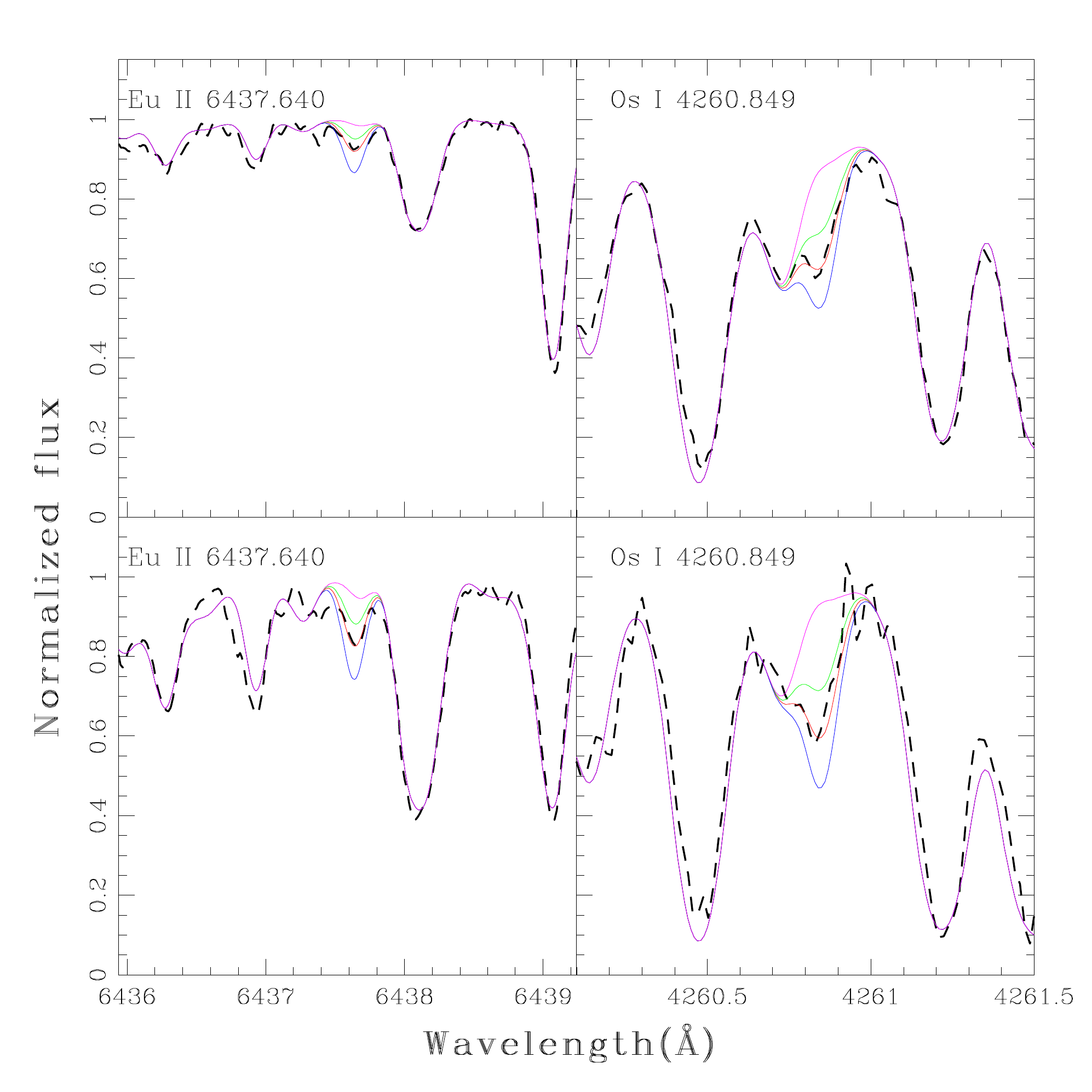}
\caption{Spectral fits of the \ion{Eu}{II} line at 6437.640~\AA\ and of the Os  line at 4260.849~\AA\ are shown for HD~5223 in the lower panel and for HD~198269 in the upper panel. The red curve corresponds to syntheses with the adopted abundances ($-$0.3~dex for Eu and 0.4~dex for Os in HD~5223, and $-$0.91~dex and $-$0.15~dex in HD~198269). The blue, green, magenta, and black curves are as in Fig.~\ref{Fig:HD224959_s}.
\label{Fig:HD5223_r}}
\end{figure}

Abundances of Gd and Os  were measured using a  single \ion{Gd}{II} line at 4251.731 \AA\ and a \ion{Os}{I} line at 4260.849 \AA, respectively. We present the spectral fitting of the \ion{Os}{I} line for HD~5223 and HD~198269 in Fig.~\ref{Fig:HD5223_r}. Since most of the good Er  lines are located at the blue end of the spectrum, we could derive Er abundances in only a few stars. We used the \ion{Er}{II} line at 3906.311~\AA\ for most of the stars. The Ir abundance was measured only in HE~1120$-$2122, using the \ion{Ir}{I} line at 3992.121 \AA. 

\subsection{The third peak s-process element Pb}
The Pb abundance was derived using the \ion{Pb}{I} line at 4057.807~\AA, taking into account HF splitting.   NLTE corrections for this Pb line are listed in \citet{Mashonkina2012} for a range of atmospheric parameters. The NLTE corrections for our objects vary from 0.3 to 0.5~dex, and the adopted  correction corresponds to the model \citep[from Table 1 of][]{Mashonkina2012} closest to the atmospheric parameters of the considered star. As for Eu, we list both LTE and NLTE Pb abundances in Table~\ref{Tab:abundances}.

\subsection{Uncertainties on the abundances}

We  estimated the uncertainties on all the elemental abundances $\log \epsilon$ following Eq.~2 from \citet{Johnson2002}:
\begin{equation}
\begin{split}
\label{Eq:Johnson}
\sigma^{2}_{\rm tot}=\sigma^{2}_{\rm ran}
\;+\; \left(\frac{\partial \log\epsilon}{\partial T}\right)^{2}\sigma^{2}_{T}  \;+\; \left(\frac{\partial \log \epsilon}{\partial \log g}\right)^{2}\;\sigma^{2}_{\log g} \\
\;+\; \left(\frac{\partial \log \epsilon}{\partial \xi }\right)^{2}\;\sigma^{2}_{\xi}\;+\;
\left(\frac{\partial \log\epsilon}{\partial  \mathrm{[Fe/H]}}\right)^{2}\sigma^{2}_{\mathrm{[Fe/H]}} \;+\; \\ 2\bigg [\left(\frac{\partial \log\epsilon}{\partial T}\right) \left(\frac{\partial \log \epsilon}{\partial \log g}\right) \sigma_{T \log g}
\;+\; \left(\frac{\partial \log\epsilon}{\partial \xi}\right) \left(\frac{\partial \log \epsilon}{\partial \log g}\right) \sigma_{\log g \xi} \\
\;+\;\left(\frac{\partial \log\epsilon}{\partial \xi}\right) \left(\frac{\partial \log \epsilon}{\partial T}\right) \sigma_{ \xi T}\Bigg].
\end{split}
\end{equation}
In Eq.~\ref{Eq:Johnson} the values $\sigma_{T}$, $\sigma_{\log g}$, and $\sigma_{\xi}$  are
the typical uncertainties on the atmospheric parameters, which are estimated as $\sigma_{T}$ = 80~K, $\sigma_{\log g}$ = 0.3~dex, $\sigma_{\xi}$ = 0.1~km/s, and $\sigma_{\mathrm{[Fe/H]}}$ = 0.15~dex.
 The partial derivatives appearing in Eq.~\ref{Eq:Johnson} were evaluated in the specific cases of HD~196944 and HD~198269, varying the
atmospheric parameters $T_{\rm eff}$, $\log g$, microturbulence $\xi$, and [Fe/H] by $100$~K, 0.5, 0.5~km/s, and 0.5 dex, respectively.
The resulting changes in the abundances are presented in Table~\ref{Tab:uncertainties}.
The covariances $\sigma_{T \log g}$, $\sigma_{\log g \xi}$, and $\sigma_{\xi T}$ were derived using the same method as given in \citet{Johnson2002}. In order to calculate  $\sigma_{T \log g}$, we varied the temperature while fixing metallicity and microturbulence, and determined the $\log g$ value required to ensure the ionization balance. Then using Eq.~3 in \citet{Johnson2002}, we derived the covariance $\sigma_{T \log g}$ and found a value of $-$0.5. In a similar way, we found $\sigma_{\log g \xi}$ = 0.02 and $\sigma_{\xi T}$ = 3.

Finally, the random error $\sigma_{\rm ran}$ is the line-to-line scatter. For most of the elements, we could use more than four lines to derive the abundances. In those cases we   adopted $\sigma_\mathrm{ran} = \sigma_{l}/N^{1/2}$, where $\sigma_{l}$ is the standard deviation of the abundances derived from all the $N$ lines of the considered element. It involves uncertainties caused by factors like line blending, continuum normalization, and oscillator strength.
However, for elements like  Zr, Eu, Dy, Er, and Hf, only two or three lines could be used, so that the above method had to be adapted. For instance, in the case of Zr (but we used the same procedure for Eu, Dy, Er, and Hf), we first calculate  $\sigma_\mathrm{Zr,avg}$, which is the average standard deviation on Zr abundances on all the stars in our sample. Then we define $\sigma_\mathrm{Zr, ran}$ as $\sigma_\mathrm{Zr,avg}$/$N^{1/2}$, where $N$ is the number of considered Zr lines in a star with $N < 4$. For elements Sr, Ba, Gd, Os, and Pb we used only one line to measure the abundance in all the
programme stars. Hence, we   adopted  $\sigma_{\rm ran} = 0.1$~dex for these elements, which is the minimum abundance difference yielding a clear difference in the synthetic spectra.
Finally, the error on  [X/Fe] was derived from \\
\begin{equation}
\sigma^{2}_{\rm [X/Fe]} =\sigma^{2}_{\rm X} + \sigma^{2}_{\rm Fe} - 2\;\sigma_{\rm X,Fe},
\end{equation}
where $\sigma_{\rm X,Fe}$ was calculated using Eq.~6 from \citet{Johnson2002}.

\begin{table}
\caption{Sensitivity of the abundances ($\Delta \log \epsilon_{X}$) with variations in the atmospheric parameters (considering the atmospheric parameters of HD~198269).}
\label{Tab:uncertainties}
\begin{tabular}{crrrr}
\hline
\\
       & \multicolumn{4}{c}{$\Delta \log \epsilon_{X}$} \\
        \cline{2-5}
 Element&   $\Delta T_{\rm eff}$ & $\Delta \log g$ & $\Delta \xi_t$ &$\Delta$ [Fe/H]   \\
&   ($+$100 K) & ($+$0.5) & ($+$0.5  & ($+$0.5 \\
   &        &         & km~s$^{-1}$)& dex) \\
\hline\\
C  &  0.04  &  0.06   &    $-$0.03  &   $-$0.05 \\
N  &  0.21  &  0.19   &     0.55  &    0.14 \\
O  &  0.00  &  0.20   &    $-$0.10  &   $-$0.10 \\
Na &  0.08  &  0.05   &     0.13  &    0.23 \\
Mg &  0.10  &  0.00   &     0.00  &    0.00 \\
Fe &  0.12  &  0.07   &    $-$0.12  &    0.04 \\
Sr &  0.05  & $-$0.05   &    $-$0.20  &    0.00  \\
Y  &  0.01  &  0.16   &     0.04  &   $-$0.08 \\
Zr &  0.00  &  0.07   &    $-$0.10  &   $-$0.03 \\
Ba &  0.00  &  0.10   &     0.05  &    0.00 \\
La &  0.04  &  0.17   &    $-$0.08  &   $-$0.09 \\
Ce &  0.00  &  0.14   &    $-$0.12  &   $-$0.07\\
Pr &  0.03  &  0.14   &    $-$0.06  &   $-$0.10 \\
Nd &  0.01  &  0.10   &    $-$0.17  &   $-$0.13  \\
Sm & $-$0.01  &  0.08   &   $-$0.19  &   $-$0.13 \\
Eu & $-$0.03  &  0.13   &  $-$0.05  &   $-$0.18 \\
Gd &  0.00  &  0.20   &    $-$0.15  &   $-$0.30\\
Dy &  0.00  &  0.10   &     0.03  &    0.00 \\
Er &  0.00  &  0.10   &    $-$0.25  &   $-$0.05 \\
Hf &  0.05  &  0.15   &    $-$0.15  &   $-$0.18 \\
Os &  0.05  &  0.05   &     0.05  &    0.05 \\
Pb &  0.30  &  0.00   &     0.20  &    0.30\\
\hline
\end{tabular}
\medskip\\
Notes. The lines of a few elements could not be detected in HD~198269; we thus used HD~196944 (slightly higher in temperature, but with otherwise similar parameters) to estimate the abundance sensitivity to parameter changes. 
\end{table}


\section{Re-assignment of the programme stars}
\label{Sect:assignment}

CEMP stars are defined as low-metallicity stars (typically [Fe/H]~$\le -1$)  with typically  [C/Fe]~$>+1$ \citep{Beer2005}. However, different carbon abundance thresholds 
are used in the literature
to identify stars as CEMPs. 
\citet{aoki2007} use  the following criteria: [C/Fe] $> 0.7$ for stars with $\log (L/L_{\odot}) \leq 2.3$ and [C/Fe] $\ge 3 - \log(L/L_{\odot}$) for stars with $\log (L/L_{\odot}) > 2.3$. 
\citet{Masseron2010} use  instead a threshold of [C/Fe]~$>0.9$ to define CEMP stars. 
However,   a classification based on the  C abundance alone might be misleading when considering low-metallicitiy nucleosynthesis. Some stars on the red giant branch (RGB) experience CN processing which will reduce the surface C abundance. 
The [C/Fe] limits set by \citet{aoki2007} allow more evolved stars to be considered as participating in similar nucleosynthesis processes. 
In the same vein, among our 25 objects, 3   have [C/Fe] $<$ 0.5. HD 206983 and HD 55496 have low C abundances ([C/Fe] = 0.42 and 0.07 respectively), but definite s-process enrichments ([s/Fe]= 0.59 and 0.87, respectively). 
HD~221170 is an r-process-enriched star without carbon enrichment. 
This confirms that a classification involving enrichments in s- and r-process elements, as well as carbon enrichment (even modest levels) is needed to properly classify the full family of metal-poor objects.
Currently the abundances of two additional chemical elements are used, namely barium (proxy for s-process) and europium (proxy for r-process).
Here again, the definition of CEMP subgroups vary among authors \citep{Beer2005,Jonsell2006,Masseron2010,Hollek2015,Abate16}.
\citet{Beer2005} propose   classifying as CEMP-rs the stars with 0~<~[Ba/Eu]~<~0.5, with  no restrictions  on [Ba/Fe] or  [Eu/Fe].
\citet{Jonsell2006} 
modified the \citet{Beer2005} classification for CEMP-rs stars by imposing [Ba/Fe] $>$ 1.0 and [Eu/Fe] $>$ 1.0.
\citet{Masseron2010} followed \citet{Jonsell2006} for CEMP-s and -rs stars, but adopted the definition of CEMP-rI and rII stars from \citet{Beer2005}.

With the derived abundances becoming more accurate, stars with definite (albeit low) levels of (s- or r-) heavy element enrichments are detected. A limit on [Ba/Fe] or [Eu/Fe] might turn rapidly obsolete, while [Ba/Eu] is more robust.
This is why we  adopted here the original classification criteria proposed by \citet{Beer2005}, but replace barium by lanthanum, since La abundances are often considered to be more reliable than Ba abundances because they are often obtained from more numerous, less saturated lines. To summarize, we applied the following classification criteria in  Col. 8 of Table~\ref{Tab:program_stars}:

\begin{itemize}
\item CEMP-s: [La/Eu] $> 0.5$;
\item CEMP-rs: $0.0 <$ [La/Eu] $< 0.5$;
\item CEMP-rI: [La/Eu] $< 0$, $0 < $ [Eu/Fe] $< +1$;
\item CEMP-rII: [La/Eu] $< 0$, [Eu/Fe] $> +1$.
\end{itemize}

However, 
these classifications  are not without problems.
HD~196944 and HD~187861 have [La/Eu] very close to zero  and are classified as CEMP-rs (instead of CEMP-r). CS 22891$-$171, with [La/Eu]=0.55, has been classified as CEMP-rs (instead of CEMP-s), as in \citet{Masseron2010}. HD 206983 and HD~55496 have been classified as CEMP-s, as in \citet{Pereira-2011,Pereira2019}. 

\begin{figure}
\includegraphics[width=9.5cm]{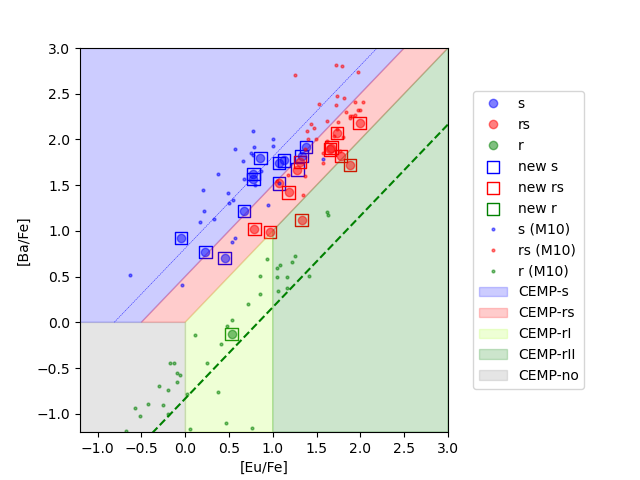}
\caption{
[Ba/Fe]  as a function of [Eu/Fe].  CEMP-s stars, CEMP-rs stars, and an r-process enriched star studied in this paper  are  represented  by blue,  red,  and  green filled circles when adopting the original classification (Table~\ref{Tab:program_stars}, Orig. Class.) and by large squares with the same colour-coding when adopting our new classification (Col. 9 of Table~\ref{Tab:program_stars} and Sect.~\ref{Sect:assignment}).  The   small dots represent objects compiled from the literature in \citet{Masseron2010}. The dashed green line  corresponds  to abundance-ratio scaling with a pure solar r-process  \citep{Goriely1999},  whereas the continuous blue line corresponds to s-process  nucleosynthesis abundance ratio scaling with the predictions for a 1.5  M$_\odot$ star of [Fe/H]~$=-1$ (5th pulse). 
\label{Fig:BaFeEuFe} }
\end{figure}

\begin{figure}
\includegraphics[width=9.5cm]{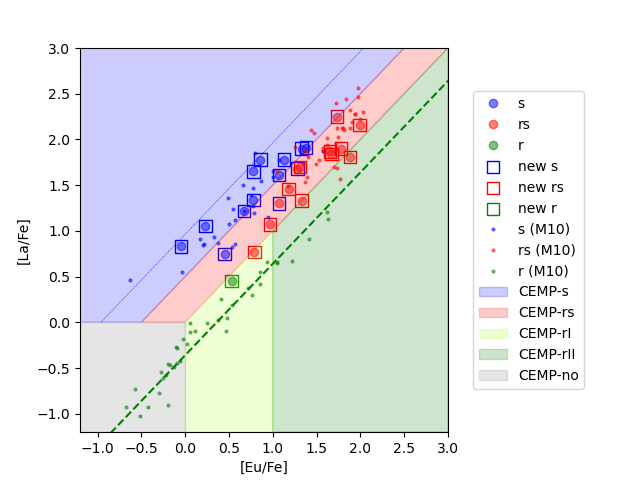}
\caption{Same as in Fig.~\ref{Fig:BaFeEuFe}, but for [La/Fe] as a function of [Eu/Fe].
\label{Fig:LaFeEuFe}}
\end{figure}

Finally, Figs.~\ref{Fig:BaFeEuFe} and \ref{Fig:LaFeEuFe} present the usual s-process diagnostics ([Ba/Fe] and [La/Fe]) as a function of an r-process diagnostic ([Eu/Fe]), classically allowing us to distinguish  CEMP-s, CEMP-rs, and CEMP-r stars. The plot includes stars from this paper and from the literature. Small circles represent 
our original assignment
(Table~\ref{Tab:program_stars}, Col. 8.). 
This classification is not as robust as one could wish. While r-stars occupy a distinct region, the exact limit between CEMP-s and CEMP-rs is difficult to define without a certain degree of arbitrariness.
Moreover, this classification
relies on only two species whose elemental abundances are derived from a small number of lines (\ion{Ba}{II} 4554, 5853, and/or 6141~\AA;\  \ion{Eu}{II} 4129, 4205, and/or 6645~\AA).
It would be desirable to use instead information provided by more chemical elements, as developed below.

\begin{figure}
\includegraphics[width=9.5cm]{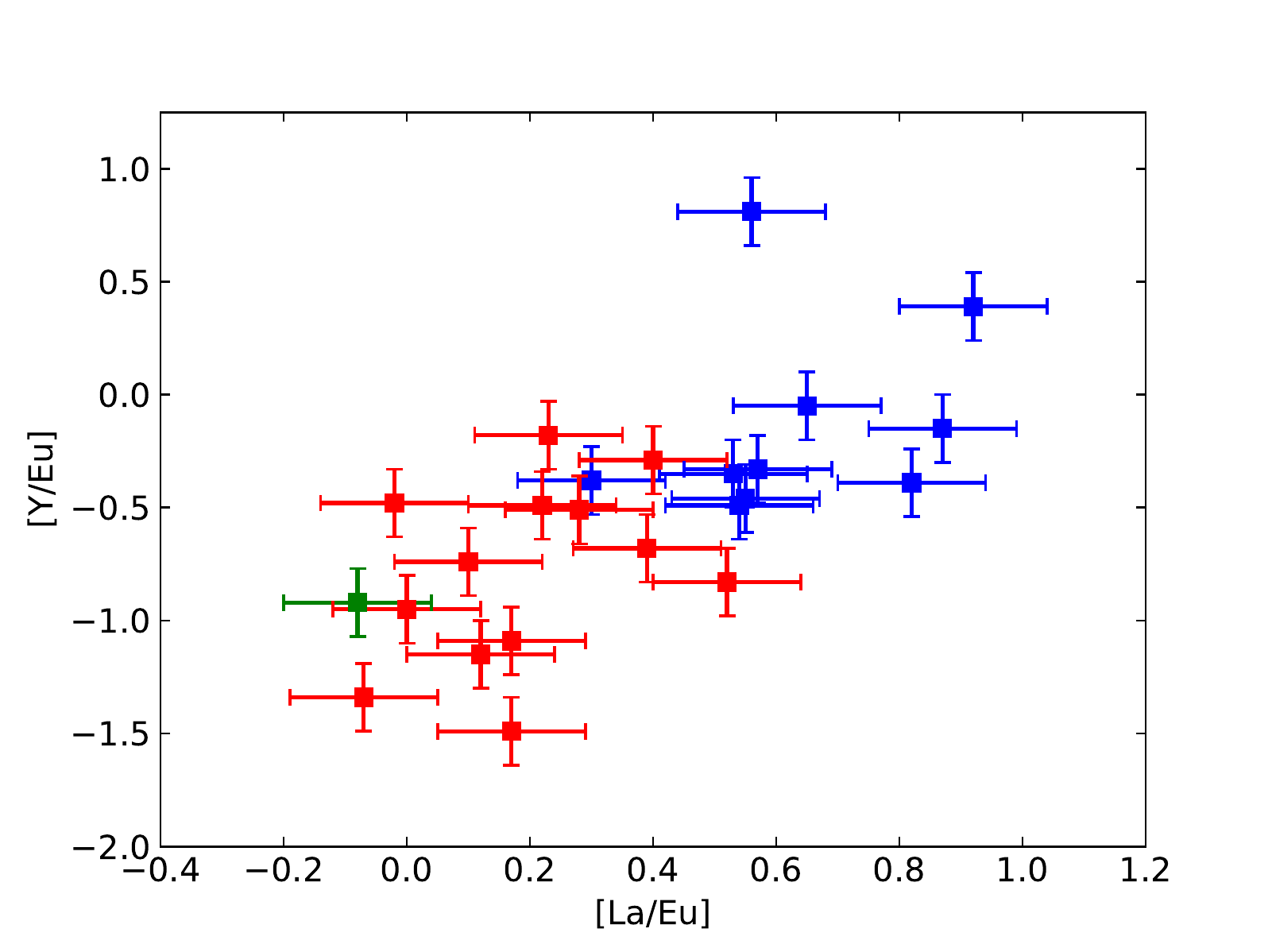}
\caption{[Y/Eu] as a function of [La/Eu] for CEMP-s stars (blue squares), CEMP-rs stars (red squares), and an r-process-enriched star (green square), according to Col. 8 of Table~\ref{Tab:program_stars}.
\label{Fig:CEMP_YEu} }
\end{figure}

\begin{figure}
\includegraphics[width=9.5cm]{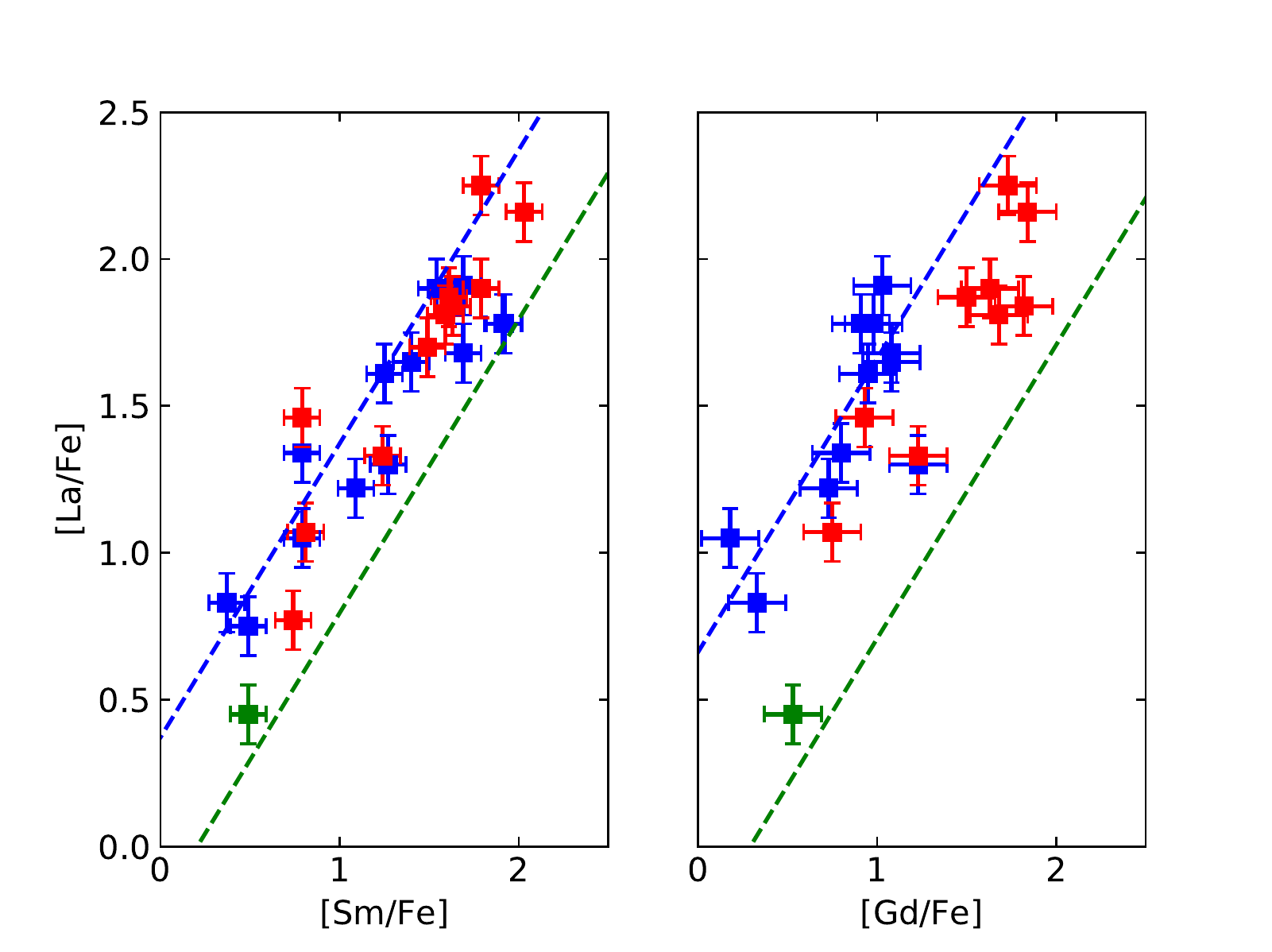}
\caption{Same as Fig.~\ref{Fig:LaFeEuFe}, but for [La/Fe] vs [Sm/Fe] (left panel) and [La/Fe] vs [Gd/Fe] (right panel). Symbols are as in Fig.~\ref{Fig:CEMP_YEu} but the stars are now classified according to the new classification (Col. 9 of Table~\ref{Tab:program_stars}). 
\label{Fig:LaGdSm} }
\end{figure}

Figure~\ref{Fig:CEMP_YEu} presents the [La/Eu] versus [Y/Eu] abundance ratio for CEMP-s and CEMP-rs objects  (according to Col. 8  of Table~\ref{Tab:program_stars}). A clear correlation is visible, showing that [Y/Eu] is almost as good as [La/Eu] (and as [Ba/Eu]) for  distinguishing CEMP-rs from CEMP-s stars.

We also tried to find whether an alternative to the r-process element europium could improve the separation between classes of stars. Figure~\ref{Fig:LaGdSm}, displaying [La/Fe] versus  [Sm/Fe] and [Gd/Fe] (Sm and Gd being two elements mostly produced by the r-process), is thus comparable to Figs.~\ref{Fig:BaFeEuFe} and \ref{Fig:LaFeEuFe}. The difference between CEMP-s and CEMP-rs is slightly less clear
than in the [La/Fe] versus [Eu/Fe] plane, but the same trends are visible. It thus suggests that a more robust classification could be obtained using the information from as many elemental abundances as possible.

Here we propose a new classification procedure, based on a distance from   the solar r-process abundance profile.
The r-process abundance profile is adopted as a reference because
the observation of some universality pattern has been invoked in many low-metallicity r-process-rich stars \citep{Roederer10}.
For example, the s-process is known to be dependent on metallicity, on the environment (radiative or convective), and on the details of the partial mixing of protons into the C-rich radiative intershell region at the time of the third dredge-up.  
It is thus impossible to define a universal s-process abundance profile.
On the contrary, although variations in the r-process distribution are observed \citep{Goriely97b,sneden-2010,Roederer10,Roederer11}, it appears less sensitive to its operation conditions, and is thus adopted here as a reference baseline.

We compute a signed distance
\begin{equation}
d_{\rm S} = \frac{1}{N}\sum_{x_i} (\log_{10} \epsilon_{x_i,\ast} - \log_{10} \epsilon_{x_i,\rm{norm(r,\ast)}})
\label{Eq:dist-signed}
\end{equation}
and an rms distance
\begin{equation}
d_{\rm rms} = \left( \frac{1}{N}\sum_{x_i} (\log_{10} \epsilon_{x_i,\ast} - \log_{10} \epsilon_{x_i,\rm{norm(r,\ast)}})^2 \right)^{1/2}
\label{Eq:dist-RMS}
,\end{equation}
where $\{x_1... x_N\}$ is the list of the $N$ considered heavy elements, and
we use the usual notation $\log_{10} \epsilon_{x_i} = \log_{10} (n_{x_i}/n_{\rm H}) + 12$, with $n_{x_i}$ the number density of element $x_i$. We denote
 $\log_{10} \epsilon_{x_i,\ast}$  the abundance of element $x_i$ as measured in the stars of the present paper (Table~\ref{Tab:abundances}),
and
$\log_{10} \epsilon_{x_i,\rm{norm(r,\ast)}}$  the standard r-process abundance profile $\log \epsilon_{x_i, r}$  normalized to the star abundance profile with respect to europium:
\begin{equation}
\log \epsilon_{x_i,\rm{norm(r,\ast)}} = \log \epsilon_{x_i, r} + (\log\epsilon_{Eu,\ast} - \log\epsilon_{Eu,r}).
\end{equation}
The adopted r-process abundances $\log \epsilon_{x_i,r}$ are listed in Table~\ref{Tab:r-process}.
We adopted europium as a normalizing element because it is mainly r-process; it is also easily measurable in most stars.

Here we consider the element set $x_i = \{\rm Y, Zr, Ba, La, Ce, Nd, Sm\}$ because it is the largest intersection of available heavy-element abundances derived in the present paper. The europium abundance is also used, but since it is the normalizing element, its corresponding distance to the r-process is null by definition.
Actually, the europium abundance could not be determined for HD~55496 (it was previously identified as a CEMP-s star; see Table~\ref{Tab:program_stars}). To determine its new classification, we assign the average [Eu/Sm] ratio of the stars previously classified as s-enriched  in our sample ([Eu/Sm]$_{\rm av, s}=-0.42$), leading to $\log\epsilon_{\rm Eu} = -1.63$~dex for HD~55496.
We note that even if we had assigned the average [Eu/Sm] of CEMP-rs stars ([Eu/Sm]$_{\rm av, rs}=0.01$), HD~55496 would still be classified as CEMP-s by our new procedure, with $d_{\rm S}=0.9$ (and $d_{\rm rms}=1.0$) well above the $d_{\rm S}= 0.6$ threshold.

\begin{figure}
\includegraphics[width=9.5cm]{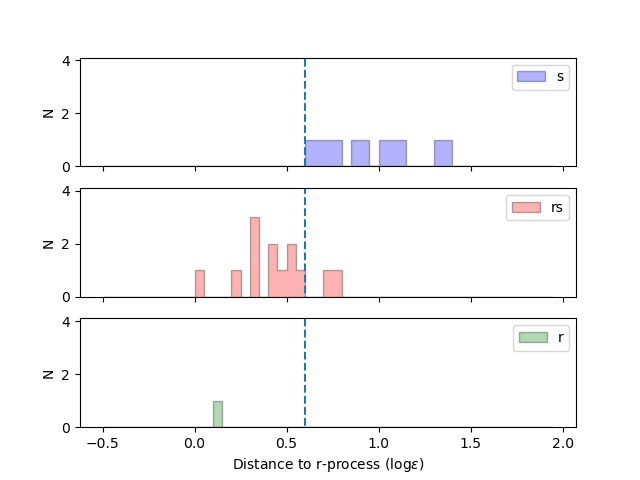}
\caption{Histograms of the signed distance $d_{\rm S}$ (Eq.~\ref{Eq:dist-signed}) for stars classified as CEMP-s stars,  CEMP-rs stars, or r star in Col. 8 of  Table~\ref{Tab:program_stars}.
The vertical dashed line at $d_{\rm S}= 0.6$ sets the separation between CEMP-s and CEMP-rs stars using our new classification (Col. 9 of Table~\ref{Tab:program_stars}).
\label{Fig:histo-signed-dist} }
\end{figure}

\begin{figure}
\includegraphics[width=9.5cm]{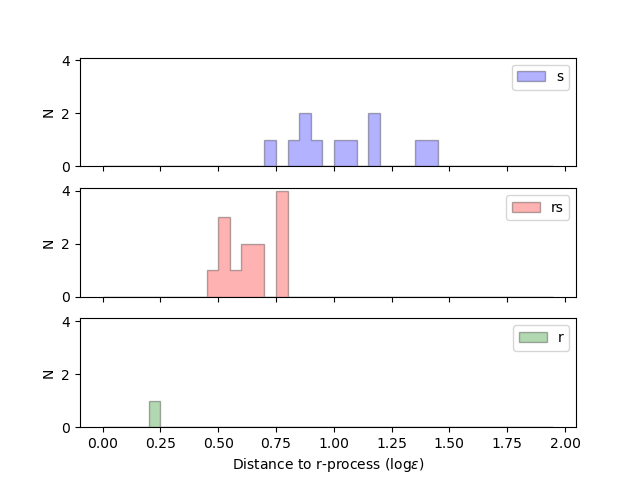}
\caption{Histograms of the rms distance
(Eq.~\ref{Eq:dist-RMS}) separately for stars classified as CEMP-s, CEMP-rs or r-star in Col. 8 of   Table~\ref{Tab:program_stars}.
Since the re-assignment procedure is based on a threshold value using the signed distance ($d_{\rm S}=0.6$,  Eq.~\ref{Eq:dist-signed} and Fig.~\ref{Fig:histo-signed-dist}), we cannot plot 
a similar threshold in the present $d_{\rm rms}$ histogram. }
\label{Fig:histo-RMS-dist} 
\end{figure}

We can interpret the two distances
$d_{\rm S}$ and $d_{\rm rms}$
as average abundance distances (in dex) between the abundance profile of a given star and the standard-r abundance profile.
Histograms of these two distances are presented in Figs.~\ref{Fig:histo-signed-dist} and ~\ref{Fig:histo-RMS-dist}.
The three groups of stars are nicely separated when using either $d_{\rm S}$ or $d_{\rm rms}$ distances.
As expected, the r stars have   the smallest distances, CEMP-rs stars are characterized by intermediate distances ($0.5 \le d_{\rm rms} < 0.8$), and  CEMP-s stars have the largest distances from the r-process ($0.7 \le d_{\rm rms} \le1.4 $).

The advantage of this new classification is that it uses eight abundances instead of two to assign a star to a given class.
Because the rms distance considers the observed minus computed (O$-$C) absolute values, and as such erases any information contained in the sign of this difference, we decided to adopt in the following the signed distance,
and $ d_{\rm S}=0.6$ as the limit between CEMP-s and CEMP-rs stars (represented by the dashed line in Fig.~\ref{Fig:histo-signed-dist}).
This new assignment is listed as  `New Class.' in Col. 9 of Table~\ref{Tab:program_stars}, and is hereafter adopted in all figures  in this paper.

Our new classification confirms the previous assignments except for two borderline objects, HE 1429$-$0551 and HE 2144$-$1832, previously classified as CEMP-rs stars, and now as CEMP-s objects. Moreover, CS 22887$-$048 was classified as a CEMP-rs star by \citet{Masseron2010}, but both the new [La/Eu] derived in the present paper and our new classification scheme agree to tag this object as a CEMP-s star.

In Figs.~\ref{Fig:BaFeEuFe} and \ref{Fig:LaFeEuFe}, the overlap between our sample of CEMP-s stars and those collected from the literature is very good.
We note, however, that our CEMP-s stars extend to higher s-process enrichments, especially CS 22887$-$048 and HE 0111$-$1346 with [La/Fe]$\sim$2. The [La/Fe] of CEMP-rs stars encompass those  of  CEMP-s  stars,  but  they  are  located  at  higher [Eu/Fe] values for  similar [La/Eu] enrichments.  In the literature CEMP-rs stars are restricted to very high [La/Fe] and [Eu/Fe] enrichments (higher than 1.5 and 1 dex, respectively), our sample of CEMP-rs stars extends to lower [La/Fe] and [Eu/Fe] ratios (Fig.~\ref{Fig:LaFeEuFe}).
One reason for this is that the Eu abundance was not systematically determined, and as a consequence
these stars were in most cases included in the CEMP-s group based on the enrichment of s-process elements; it was not realized  that this pattern
was merely a consequence of a strong r-process enrichment.  The case of the well-known CH star HD 5223 is a good example of this difficulty. Since the abundance of europium or of any other r-process element was not available for this object, it was previously considered  a CEMP-s star. In the present paper we derive [Ba/Eu] = 0.24 and [La/Eu] = 0.28 for this object, making it a definite CEMP-rs object, as confirmed by its $d_{\rm S}$ value of 0.471.
It also seems that europium  was  determined  only  in  the  most  enriched CEMP-s stars,  while in our study we systematically undertook its determination even in objects with low enrichment levels.

For  CEMP-r stars  the $d_{\rm rms}$ index allows a clearer separation than the $ d_{\rm S}$ index, which was  favoured  because it contains information on the sign of the O$-$C, as explained above. We note, however, that some CEMP-rs objects have abundance profiles resembling those of CEMP-r stars. For example, the CEMP-rs object HD~187861 has  $d_{\rm S}=0.037$, while showing very large overabundances of both r- and s-process elements. With [La/Eu]$= -0.05$, it is clearly a borderline case between CEMP-rs and CEMP-r stars, as discussed at the beginning of this section.

Another  striking  result  is  the abundance continuity  between  CEMP-s  and  -rs  stars. 
Drawing  a  line  between CEMP-s and CEMP-rs stars turns out to be somewhat artificial, which explains that several objects are actually borderline cases that could easily be assigned to either category.
Therefore, we note that the new limit drawn between CEMP-rs and CEMP-s classes at $d_{\rm S}=0.6$ is subject to a certain degree of arbitrariness, especially since we do not know yet whether these two classes originate from different physical mechanisms or represent two extreme manifestations of the same physical process. To answer this question we need to separate the two groups and to investigate their respective properties, as we do in Sect.~\ref{Sect: abund-analysis} after discussing individual objects in Sect.~\ref{Sect:indiv-stars} and the nucleosynthetic models in Sect.~\ref{Sect:nucleosynthesis}.

\section{Discussion on individual stars and comparison  with previous abundance studies}
\label{Sect:indiv-stars}

In this section we compare the results from the present abundance analysis (listed in Table~\ref{Tab:abundances}) with previous abundance studies.

\noindent{\bf CS 22887$-$048.} \citet{tsangarides2005} studied this object and estimated the atmospheric
parameters $T_{\rm eff}$,  $\log g$, and [Fe/H] as 6500~K, 3.35, and $-$1.70, respectively,
in close agreement with our values with the exception of metallicity for which we derived [Fe/H] = $-$2.10.
\citet{Masseron2010} classified this object as a CEMP-rs star based on their classification scheme. However, the [Ba/Eu] (their paper: 0.51; this paper: 0.42) and [La/Eu] ratios (their paper: 0.24; this paper: 0.50) point to a CEMP-s object. This illustrates the difficulty in distinguishing CEMP-s from CEMP-rs stars using only two or three individual abundances. It justifies our new classification scheme (Sect.~\ref{Sect:assignment}), which clearly assigns it to the CEMP-s class.

\noindent{\bf CS 22891$-$171.} This object was analysed by \citet{Masseron2010} who derived atmospheric parameters $T_{\rm eff}=5100$~K,  $\log g = 1.6$, and [Fe/H] = $-$2.25, close to our values of respectively 5215~K, 1.24, and $-$2.50. This is yet another object that is a borderline case between the CEMP-s and rs class. Our new classification assigns this star to the CEMP-rs class.
In addition to Eu, we report in Table~\ref{Tab:abundances} the first abundances for other r-process elements like Gd, Dy, Er, and Hf that strongly support the rs nature
of this object.\\
{\bf CS 22942$-$019.} This object was studied by both \citet{aoki2002} and \citet{Masseron2010}, who estimated the parameters ($T_{\rm eff}$,  $\log g$, [Fe/H]) as (5000 K, 2.4, $-$2.64) and  (5100 K, 2.5, $-$2.43), respectively.
In the present work we obtain (5100 K, 2.19, $-$2.50), closely matching the parameters from previous studies. This object was classified as a CEMP-s star in both papers. We confirm this classification both with our derived [La/Eu] ratio and with our new classification scheme. 
\\
{\bf CS 30322$-$023.} Various atmospheric parameters have been proposed for this object. \citet{Masseron2006}
found $\log g = -0.3$, implying that this object could be a thermally pulsing (TP) AGB star (but see the update about this issue in Sect.~\ref{sect:cno}). Moreover, its low C and high N abundances suggest that it may have experienced hot bottom burning (HBB).
However \citet{aoki2007} found a satisfactory spectral fit with $T_{\rm eff} = 4300$~K and $\log g = 1.0$. Our estimated values of 4500~K and $\log g = 1.0$ are closer to those of \citet{aoki2007} than to those of \citet{Masseron2006}.
The derived abundances suggest a pure s-process pattern in this object, with greater overabundances  than those  derived by \citet{Masseron2006}, a consequence of the difference in their adopted atmospheric parameters. We assign this object to the CEMP-s class based on our new classification criteria.\\
{\bf HD 26.} This is a well-known, prototypical CH star with a clear s-process enrichment pattern. It is also a binary with one of the longest orbital periods among the families of extrinsic stars ($\ge$ 54 yr, could be up to 100 yr) \citep[][]{Jorissen2016, Jorissen2019}, but the orbital parameters  still have to be confirmed.  Our  parameters and abundances are in good agreement with those from previous studies \citep{Vaneck2003,Vanture2003,Goswami2016}. Here we add the abundances of Gd, Hf, and Os that were not reported earlier. Carbon is enriched only at a moderate level with [C/Fe] = 0.65. Based on the derived [La/Eu] ratio and on our new classification, we confirm the CEMP-s nature of this object. \\
{\bf HD 5223.} \citet{Goswami2006} studied this object and estimated the atmospheric parameters as $T_{\rm eff} = 4500$~K, $\log g
= 1.0$, and [Fe/H]$ =  -2.06$. They derived the abundances of Ba, La, and Sm and classified this object as a CEMP-s star.
In the present study we  derived the abundances of the r-process elements Eu, Gd, Hf, and Os and 
assigned this object to the CEMP-rs group based on the derived [La/Eu] value. We confirm this assignment with our new classification scheme. 
\\
{\bf HD 55496.} \citet{Karinkuzhi2015} derived $T_{\rm eff} = 4850$~K, $\log g = 2.05$, and [Fe/H]$= -1.49$ for this star, as well as
abundances of Ba, Ce, and Pr. In the present analysis we estimated the atmospheric parameters to be 4642~K, 1.65, and $-$2.10, respectively.
Abundances of many s-process elements have been obtained for this object, along with the r-process elements Sm, Gd, Dy, and Hf. We also found a Pb overabundance of 1.92~dex (NLTE corrected). Light s-process elements are more enriched than heavy s-process elements in this star, and Fig.~\ref{Fig:CEMP_boxplot}   shows that this tendency seems to be a distinctive feature of CEMP-s stars contrasting with CEMP-rs objects.  
Because of its very low [C/Fe] (= 0.07) and high [N/Fe] (= 2.37) ratios, HD~55496 actually belongs to the nitrogen-enhanced metal-poor (NEMP) class rather than to the CEMP class.
Recently, \citet{Pereira2019} noted its high [Na/Fe] and [Al/Fe] abundance ratios,  similar to those of second-generation globular cluster stars.
They suggested that this object, probably polluted by an intermediate-mass AGB star that had activated the $^{22}$Ne neutron source, evaporated from a globular cluster or from a disrupted dwarf galaxy.
We note that there is currently no proof that HD 55496 is a binary (Table~\ref{Tab:program_stars}). Since the Eu abundance has not been measured in this object, the original classification based on the [La/Eu] ratio is not possible. However, with our new classification, we assign this object to the CEMP-s group. 
 \\
{\bf HD 76396.} This paper presents the first abundance analysis of this object.
It shows that HD~76396 is a very metal-poor star ([Fe/H] $= -2.27$) with large overabundances for both s-process and r-process elements. It is definitely a binary with a period of ~40 years (Jorissen et al., in prep.),  belonging to the CEMP-rs group, as derived from both the original and new classifications.
\\
{\bf HD 145777.}  We present the first abundance analysis of this object, with atmospheric parameters $T_{\rm eff} = 4444$~K, $\log g = 0.50$, and [Fe/H]$ = -2.32$. The star shows a clear enrichment in s-process elements (with [s/Fe]~$\approx 1.0$) and in r-process elements, albeit to a lower level. It is assigned to the CEMP-rs category based on the derived 
[La/Eu] ratio and on  our new classification. Its binarity is unclear.
\\
{\bf HD 187861.} This object was first studied by \citet{Vanture1992C}, and then re-analysed by  \citet{Vaneck2003} (as one of their Pb-rich stars) and by \citet{Masseron2010}. The range of effective temperatures obtained by these authors covers $\sim700$~K: while \citet{Vaneck2003} derived $T_{\rm eff} = 5320$~K, $\log g = 2.40$, and [Fe/H]~$ = -2.30$, \citet{Masseron2010} obtained $T_{\rm eff} = 4600 $~K, $\log g = 1.70$, and [Fe/H]~$ = -2.36$. Here we derive an intermediate temperature ($T_{\rm eff} = 5000$~K, $\log g = 1.50$, and [Fe/H]~$ = -2.60$). This object shows very large overabundances of both s- and r-process elements. We confirm its CEMP-rs nature, both from the derived [La/Eu]   and the new classification criteria. Its binarity is unknown.
\\
{\bf HD 196944.} This object was one of the lead-rich stars studied by \citet{Vaneck2003}, and also analysed by \citet{aoki2002}, \citet{Masseron2010}, and \citet{Placco2015}. Our atmospheric parameters are in close agreement with those from these  studies, except for $\log g$;  we found $\log g= 1.28$, while previous analyses found $\log g$ in the range from 1.60 to 1.80.
Based on their  criteria, \citet{Masseron2010} and \cite{Bisterzo2011}  classified this object as a CEMP-s star.
Using near-ultraviolet (NUV) spectra from STIS/HST, \citet{Placco2015}  derived atmospheric parameters ($T_{\rm eff}$ = 5170 K, $\log g$ = 1.60, [Fe/H] = $-$2.41) and heavy-element abundances, which they compared with a low-mass low-metallicity model in their Fig. 12.  Their Eu abundance ([Eu/Fe] = $-$0.11),   lower by at least 0.5 dex compared to all other studies, did not match   the model predictions.
We obtain a higher [Eu/Fe] ratio (after NLTE correction) than that of  \citet{aoki2002} (0.78 versus 0.17~dex), and
a similar Er abundance (1.08 versus 0.81~dex). 
Even though heavy elements are enriched in HD~196944, the level of enrichment is not high and is almost the same for s- and r-elements ([La/Fe]~=~0.77, [Ce/Fe]~=~0.89, [Eu/Fe]~=~0.78, [Dy/Fe]~=~0.80, and [Er/Fe]~=~1.08).
This is fully consistent with the present classification of this star as a CEMP-rs star.
This star is a binary of  orbital period 1294~d (Jorissen et al. in prep.).
\\
{\bf HD 198269.} This object was analysed by \citet{Goswami2016} and their atmospheric parameters ($T_{\rm eff}$ = 4500 K, $\log g$ = 1.50, [Fe/H] = $-$2.03) are similar to our values, except that we derive $\log g=0.83$.  With [La/Eu] = 0.55, this object is on the borderline between the CEMP-s and CEMP-rs families. Although \citet{Goswami2016} do not list an Eu abundance, they derive [Er/Fe] = 0.43 and [Os/Fe] = 1.09. In the present analysis Eu, Gd,  and Os show a mild enrichment ([Eu/Fe] = 0.67, [Gd/Fe] = 0.73, and [Os/Fe] = 0.55), while Dy and Er are highly enriched with [X/Fe] values above 1.0, as are Ba, La, Ce, and Pb. This star falls in the CEMP-s category for both the original and new classification schemes. The star has an orbital period of 1295~d.
\\
{\bf HD~201626.} This object was first analysed by \citet{Vanture1992} and subsequently by \citet{Karinkuzhi2014}. These studies derived the abundances of s-process elements only. Here we derive again the  s-process abundances, and extend the analysis to r-process elements. Our atmospheric parameters and abundances are consistent with previous values.  With a [La/Eu] ratio of 0.93, this object naturally falls in the CEMP-s category.  We confirm this assignment with our new classification scheme. The star has an orbital period of 1465~d.
\\
{\bf HD 206983.}
\citet{Masseron2010} analysed this object and derived the parameters ($T_{\rm eff}$ = 4200 K, $\log g$ = 0.60, [Fe/H] = $-$1.00), which are very similar to ours.
This object is only midly enriched in carbon ([C/Fe] = 0.42). The s-process elements similarly  show a moderate enrichment of $\approx$ 0.75~dex. The \citet{Masseron2010} abundances are similar to ours except for Pb. We derive a [Pb/Fe] ratio of 0.88 dex considering the NLTE correction, while \citet{Masseron2010} found a high Pb enrichment ([Pb/Fe] = 1.49~dex).
This star is assigned to the CEMP-s class. It may be a binary.
\\
{\bf HD 209621.} This object has been studied by \citet{Goswamiaoki2010}. Their atmospheric parameters ($T_{\rm eff} = 4500$~K, $\log g = 2.0$, and [Fe/H]~$= -1.93$) are slightly different from ours ($T_{\rm eff} = 4740$~K, $\log g = 1.75$, and [Fe/H]~$= -2.00$). We find high levels of  enrichment for both s- and r-process elements. This is thus another  CEMP-rs star confirmed with both classification schemes. Its orbital period is 407.4~d.  \\
{\bf HD 221170.} This well-studied r-process star was selected as a comparison object (carbon is not enriched in this apparently non-binary object). Comparing our results with the most recent detailed abundance study by \citet{Ivans2006} reveals that our atmospheric parameters ($T_{\rm eff} = 4577$~K, $\log g = 0.77$, and [Fe/H]~$= -2.40$) are in close  agreement with their values ($T_{\rm eff} = 4510$~K, $\log g = 1.00$, and [Fe/H]~$= -2.19$). 
\\
{\bf HD 224959.} This is another of the lead-rich stars of \citet{Vaneck2003}.  \citet{Masseron2010} also analysed this object and obtained $T_{\rm eff} = 4900$~K, $\log g = 2.0$, and [Fe/H]~$= -2.06$. Our estimates are similar for $T_{\rm eff}$ (4969 K) and [Fe/H] ($-$2.36), but not for $\log g$ ($1.26\pm0.3$). With a [La/Eu] ratio of 0.17, HD 224959 belongs to the CEMP-rs group, in agreement with our new classification scheme.\\
{\bf HE 0111$-$1346.} This object was analysed by \citet{Kennedy2011} using low-resolution infrared spectra. These authors derived  $T_{\rm eff} = 4651$~K, $\log g = 1.08$, and [Fe/H]~$= -1.91$ as well as C, N, and O abundances. Using these values, \citet{Hansen2016a} derived the Ba abundance. We present the first detailed chemical abundance pattern.  
HE 0111$-$1346 falls in the CEMP-s group using both  the original and new classifications. It has an orbital period of 402.7~d. 
\\
{\bf HE 0151$-$0341.}  This star was also analysed by \citet{Kennedy2011}, who derived $T_{\rm eff} = 4849$~K, $\log g = 1.42$, and [Fe/H]~$= -2.46$, and \citet{Hansen2016a}, who adopted the same parameters. Although our $T_{\rm eff}$ and  $\log g$ agree with their analyses, we find a lower metallicity of -2.89.  Both s-process and r-process elements are highly enriched: this is another CEMP-rs star in both  classifications. It has an orbital period of 359~d. \\
{\bf HE 0319$-$0215.} This is another object analysed by \citet{Kennedy2011} and \citet{Hansen2016a}. \citet{Kennedy2011} derived $T_{\rm eff} = 4416$~K, $\log g = 0.64$, and [Fe/H]~$= -2.42$. Our effective temperatures and metallicities are slightly different: $T_{\rm eff} = 4738$~K, $\log g = 0.66$, and [Fe/H]~$= -2.90$.  \citet{Hansen2016a} used $T_{\rm eff}$ and $\log g$ determined by \citet{Kennedy2011} to derive C, Fe, and Ba abundances. They found a moderate [Ba/Fe] abundance ratio of 0.52~dex.
Our  abundances indicate high enrichment levels for   s-process and r-process elements, reflecting that HE~0319$-$0215 falls in the CEMP-rs group for both classification schemes. It has an orbital period of 3078~d. \\
{\bf HE 0507$-$1653.} \citet{aoki2007} derived the atmospheric parameters ($T_{\rm eff}= 5000$~K, $\log g = 2.4$, and [Fe/H]~$= -1.50$) well in line with ours.
While we find s-process elements to be highly enriched (with [s/Fe]~$\approx 1.75$), r-process elements are also enhanced with an average [r/Fe] ratio of 1.0~dex.  However, with [La/Eu]~=~0.65, this object satisfies the condition to be a CEMP-s star, as confirmed by the new classification. It has an orbital period of 404~d. 
\\
{\bf HE 1120$-$2122.} 
Based on a series of low-resolution spectral characteristics (the strength of the CH G-band, the strength of the \ion{Ca}{I} feature at 4226~\AA\ and the weakness of the CN band at 4215~\AA) and in comparison with the spectrum of the well-known CH star HD 209621, \citet{Goswami2010} classified this object as a CH star. The two objects  show similar spectral features. There are no former published atmospheric parameters or abundances for this star,
which turns out to be a CEMP-rs object for both  classification schemes. It has an orbital period of 2103~d. 
\\
{\bf HE 1429$-$0551.} \citet{aoki2007} 
derived $T_{\rm eff} = 4700$~K, $\log g = 1.5$, and [Fe/H]~$= -2.50$, close to our values, and presented the abundances of some light elements along with the Ba abundance. We derived abundances for Ba, La, Ce, Pr, Nd, Sm, Eu, and  Gd. This is a borderline  object, previously classified as CEMP-rs (according to its [La/Eu]), that we re-assign to the CEMP-s class based on our multi-element classification scheme.
\\
{\bf HE 2144$-$1832.} \citet{HansenCJ2016} derived the atmospheric parameters $T_{\rm eff} = 4200$~K, $\log g = 0.6$, and [Fe/H]~$= -1.7$, along with the abundances of  Sr and Ba, and suggest it might be an AGB star. Our parameters   match closely.
Even though the derived [La/Eu] ratio points towards the CEMP-rs category, our new classification scheme assigns it to the CEMP-s class.
The carbon abundance from both our and previous analyses indicates that HE~2144$-$1832 is not much enriched in carbon ([C/Fe] = 0.77).\\
{\bf HE 2255$-$1724.} From the inspection of a low-resolution spectrum of this object, \citet{Goswami2010} identified this object as a CH star, for which we present the first  detailed abundance pattern. Although r-process elements (like Sm, Gd, Dy, Er, Os) are enriched in this object, the [La/Eu] ratio of 0.54 makes this object a borderline case. We assign it to the CEMP-s category with our  multi-element classification.

\section{Comparison with nucleosynthesis predictions}
\label{Sect:nucleosynthesis}

\subsection{Models of s- and i-processes in low-mass AGB stars}
Asymptotic giant branch nucleosynthesis predictions have been computed coupling
the STAREVOL code \citep{Siess2008} with an
extended s-process reaction network of 1091 species and the same input physics as in \citet{Goriely18c}.
The solar abundances are   from \citet{Asplund2009}, and correspond to a metallicity $Z = 0.0134$. To describe
the mass-loss rate on the RGB, we use the \citet{Reimers1975} prescription
with $\eta_R=0.4$ from the main sequence up to the beginning
of the AGB and then switch to the \citet{Vassiliadis1993}
rate.  Dedicated models with an initial mass of 1 and 2~\Msun\ have been computed for the different metallicities measured in the present study, namely [Fe/H] = $-$1, $-$2.0, $-$2.5, and $-$3.0.

In the present calculations a diffusion equation is used to compute the partial mixing of protons in the C-rich layers at the time of the third dredge-up (TDU). We follow  Eq.~(9) of \citet{Goriely18c} and use the same diffusive mixing parameters in our simulations as in \citet{Shetye19}, {i.e.} $f_{\rm over}=0.14$, $D_{\rm min} = 10^7\,{\rm cm^2\, s^{-1}}$, and $p = 1/2$, where $f_{\rm over}$ controls the extent of the mixing, $D_{\rm min}$ is the value of the diffusion coefficient at the innermost boundary of the diffusive region, and $p$ is an additional free parameter describing the shape of the diffusion profile. As shown in \citet{Shetye19}, this adopted set of diffusion parameters gives rise to early TDU episodes and s-process enrichments in stars with masses as low as 1~\Msun\ and compatible with observations.

\begin{figure*}
\includegraphics[width=18cm]{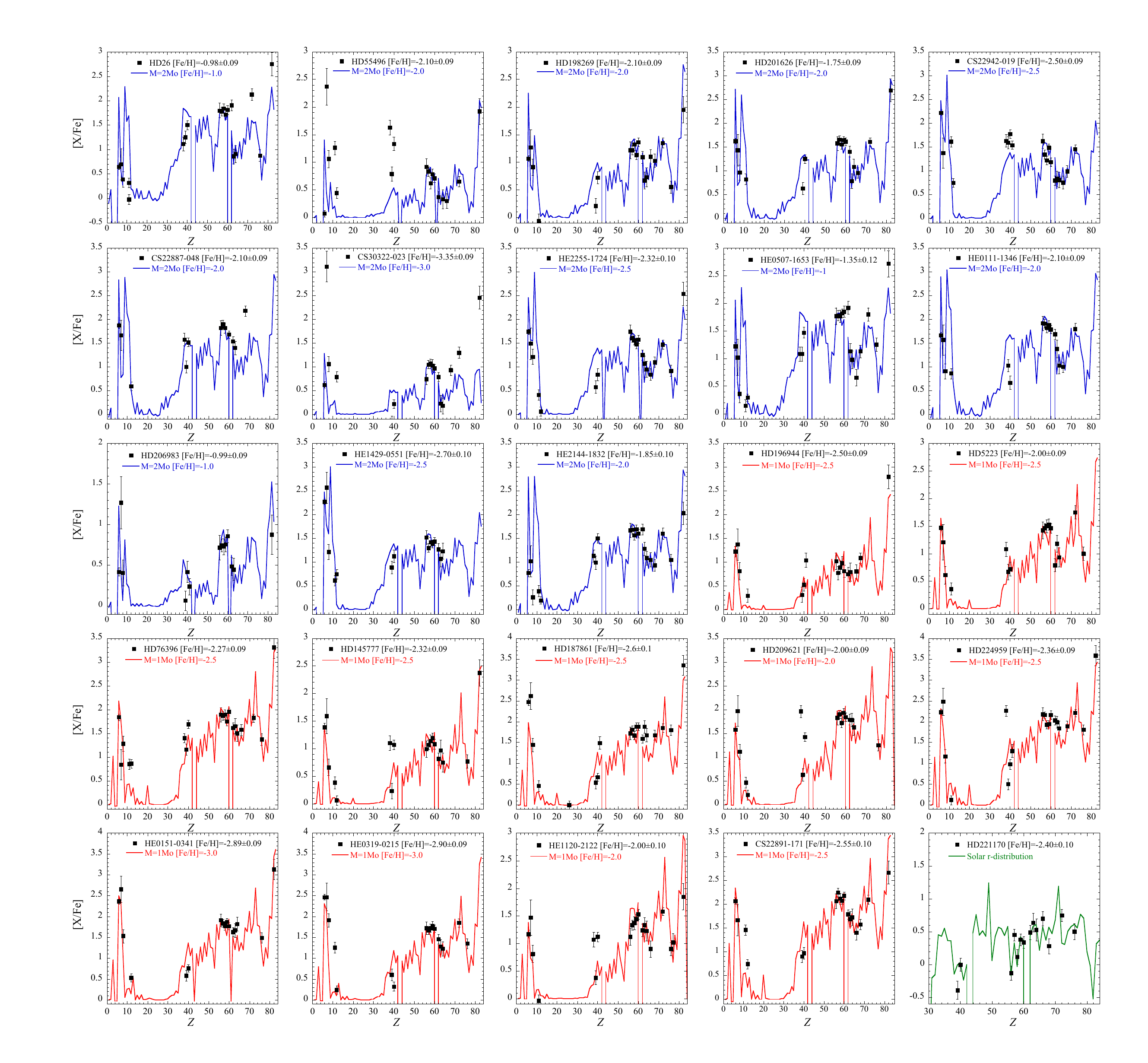}
\caption{Comparison of the abundance pattern in CEMP stars with nucleosynthesis predictions from the STAREVOL code. The 2~\Msun\ model predictions for the 13 CEMP-s stars are shown in blue, the 1~\Msun\ predictions for the 11 CEMP-rs stars in red, and the solar r-distribution for the sole r-star in green. Models are described in Sect.~\ref{Sect:nucleosynthesis}.}
\label{Fig:pattern}
\end{figure*}

The diffusion algorithm adopted in STAREVOL triggers a rather strong overshoot mixing below the convective envelope, especially at the time of the TDU \citep{Goriely18c}. In the case of low-metallicity  2~\Msun\ AGB stars, a strong s-process takes place during the various interpulse phases, leading to a surface enrichment compatible with observations, as seen for the 13 CEMP-s stars in Fig.~\ref{Fig:pattern}.

Interestingly, this mixing is at the origin of the ingestion of protons inside the first thermal pulse of all our low-metallicity 1~\Msun\ AGB stars, as initially found by \citet{Iwamoto04}. 
In this event protons are moved downwards by the convective flow to regions of higher temperature ($T > 10^8$~K) where it is depleted via the reaction $^{12}$C(p,$\gamma$)$^{13}$N. The subsequent decay of $^{13}$N to $^{13}$C is followed by the reaction $^{13}$C($\alpha$,n)$^{16}$O, which produces high neutron densities of up to $N_n \approx 10^{14}-10^{15}$~cm$^{-3}$ \citep{Choplin2021}. After mixing and ensuing rich nucleosynthesis, the convective pulse is found to merge with the stellar envelope, leading to a strong metallicity enrichment of the surface. This surface pollution in carbon and heavy metals tends to accelerate the mass loss, so that no more  thermal pulses occur  in the subsequent evolution along the AGB. More details will be given in a forthcoming paper \citep{Choplin2021}.

\begin{figure}
\includegraphics[clip,trim=2cm 0cm 0cm 0cm, width=9.5cm]{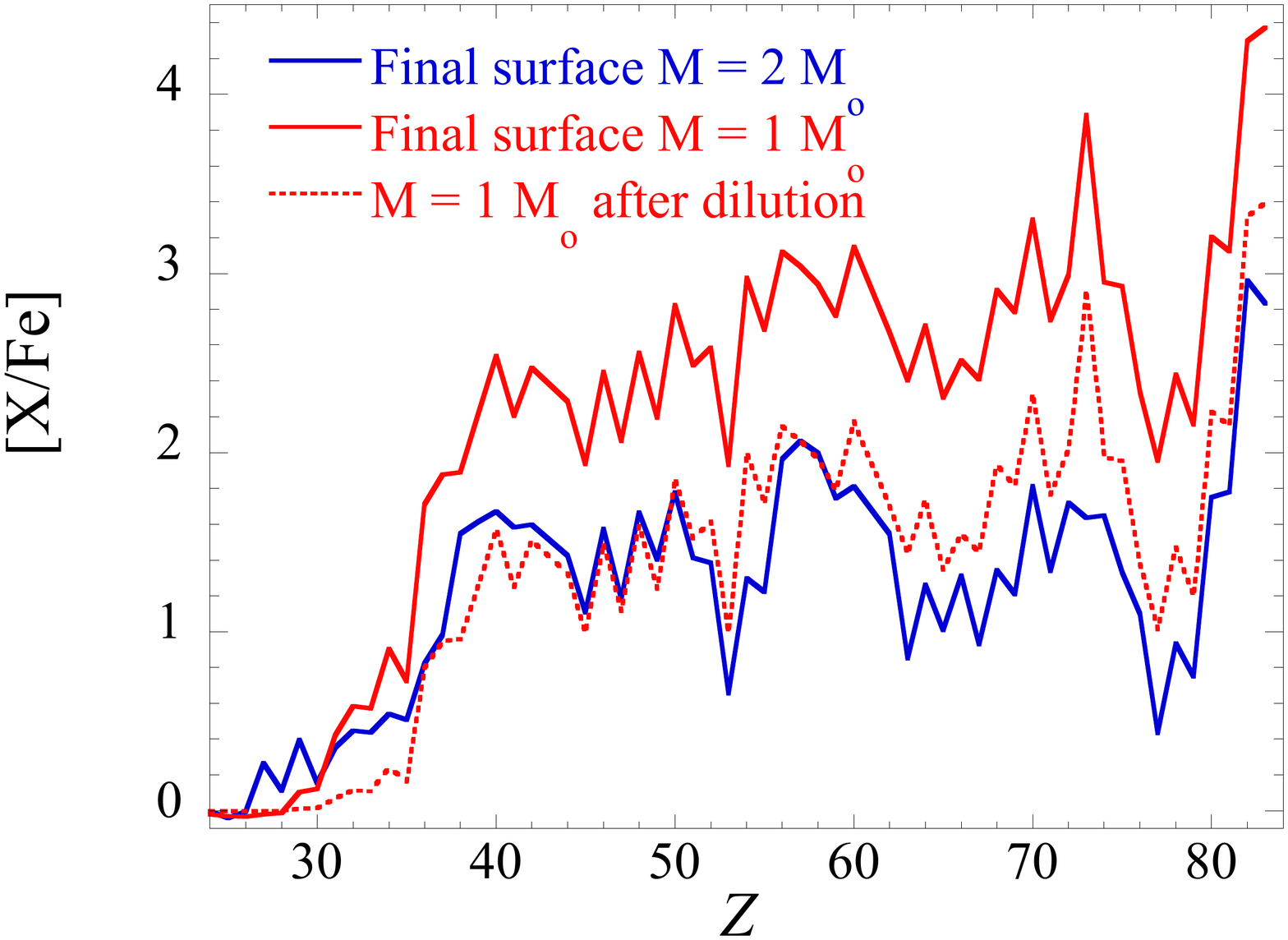}
\caption{Comparison of the final elemental surface distributions [X/Fe] (solid lines) obtained after the development of the neutron-capture processes in 2~\Msun\ (blue line) and 1~\Msun\ (red line) AGB stars of metallicity ${\rm [Fe/H]}=-2.5$. The dashed red line is obtained after dilution of the final surface abundances of the 1~\Msun\ star with a solar-like abundance distribution in order to reproduce the 2~dex La ($Z = 57$) overabundance of the 2~\Msun\ star. This model is called for to explain the CEMP-s stars, while the 1~\Msun\ model explains the CEMP-rs stars, as shown in Fig.~\ref{Fig:pattern}.
\label{Fig:i_vs_s} }
\end{figure}

In Fig.~\ref{Fig:i_vs_s} we compare the final surface enrichments obtained for elements heavier than iron in  1 and 2~\Msun\ AGB stars with metallicity ${\rm [Fe/H]}=-2.5$. Figure~\ref{Fig:i_vs_s} illustrates the high surface enrichment encountered after one single nucleosynthesis episode in the 1~\Msun\ AGB star, while 14 sequences of TDUs are needed in the 2~\Msun\ AGB star to reach a La overabundance of 2~dex. After dilution and renormalization of the final surface abundance of the 1~\Msun\ AGB star on the 2~dex La overabundance in the 2~\Msun\ star, Fig.~\ref{Fig:i_vs_s} also shows the major difference between the two processes, with higher values by about 0.3--0.5~dex in the $51 \le Z \le 55$ and $Z \ge 62$ (including Pb-Bi) regions in the low-mass star. These overproduction factors in the 1~\Msun\ AGB star result from the higher neutron densities (about $N_n \approx 10^{14}$~cm$^{-3}$), which allow short $\beta$-decay branching points to be bypassed and are, in particular, responsible for a significant production of Eu with respect to Ba or La. After renormalization on the same La overabundance, we find proportionally more Eu than La in the 1~\Msun\ star by about 0.6~dex ([La/Eu] = 1.2 and 0.6 for the 2 and 1~\Msun\ star, respectively). This is similar to the extra Eu overabundance found in CEMP-rs stars with respect to CEMP-s stars (see e.g. Fig~\ref{Fig:LaFeEuFe}). The strongest overproduction is found for Ta ($Z=73$) for which, unfortunately, no clean and strong enough line can be accurately detected in the visible wavelength range. The low-metallicity low-mass AGB stars consequently represent a natural site for CEMP-rs stars, as also shown in greater detail in Fig.~\ref{Fig:pattern} for the 11 CEMP-rs stars  studied in the present paper. In particular, the ratio of the Ba-La-Ce to the Sm-Eu-Gd overabundances is seen to be  well reproduced in both classes of stars in Fig.~\ref{Fig:pattern}. The large overabundance of Pb in CEMP-rs stars is also fairly well explained, except for HE~1120$-$2122 and CS~22891$-$171 stars (the latter has an uncertain Pb abundance), which show a rather low Pb enrichment. The similar Pb-to-Ba ratio found in CEMP-s stars like HE~2144$-$1832 and HD~206983 (the latter has an uncertain Pb abundance) also cannot be explained by models; models can reach ${\rm [Pb/hs]}\simeq 0$, but only for lower metallicity stars: [Fe/H] = $-$3.0, whereas HE~2144$-$1832 and HD~206983 have [Fe/H] =  $-$1.85 and $-$1, respectively. 

\subsection{Quantitative comparison of measured abundances with predicted values}
The overall accuracy of the model predictions is given in Table~\ref{Tab:program_stars} where the deviation between observation and model is quantified for each star through the $\chi^2$ indicator, defined as in \citet{Hampel2016}:
\begin{equation}
\chi^2=\sum_X \frac{\left({[\rm X/Fe]_{obs}}-{[\rm X/Fe]_{mod}}\right)^2}{\sigma^2_{\rm X,obs}}\quad .
\label{eq:chi2}
\end{equation}
Here ${[\rm X/Fe]_{obs}}$ and ${[\rm X/Fe]_{mod}}$ are the measured and STAREVOL overproduction factors, respectively, of a given element $X,$ and $\sigma_{\rm X,obs}$ is the associated uncertainty on the measured abundance. We consider the eight elements available in all stars, namely Y, Zr, Ba, La, Ce, Nd, Sm, and Eu\footnote{
As explained in Sect.~\ref{Sect:assignment}, the Eu abundance could not be determined for HD~55496, which was assigned the CEMP-s average [Eu/Sm], resulting in $\log\epsilon_{\rm Eu} = -1.63$~dex for that star.}.

For CEMP-s stars $\chi^2$ ranges between 2.7 and 10.9 with an average value of 6.4, while for CEMP-rs stars values between 1.3 and 10.6 are obtained with  a similar average of 6.1. Similar accuracies are obtained with the parametric canonical i-process model developed by \citet{Hampel2016} (see in particular their Table~2). We note, however, that in the present stellar evolution models, nucleosynthesis is consistently followed by the realistic stellar evolution model and that the only free parameter 
is a dilution factor corresponding to the mixing of the nucleosynthesis yields from the primary star into its 
companion (assumed to be of the same initial composition). In particular, the nucleosynthesis predictions in the 1~\Msun\ models remain  insensitive to the mixing parameters considered, and ingestion of protons is found even without imposing any overshoot \citep{Iwamoto04, Choplin2021}. This shows that the nucleosynthesis associated with the ingestion of protons in the first thermal pulses of low-mass low-metallicity stars can explain CEMP-rs stars with the same accuracy as that reached by the standard s-process in TP-AGB stars (which explains the CEMP-s stars).

In summary, the neutron-capture process taking place in the conditions found here in low-metallicity 1~\Msun\ AGB stars corresponds to an efficient s-process with a relatively high neutron density. Consequently, the origin of CEMP-rs stars may not need to call for exceptional astrophysical sites, such as rapidly accreting C-O or O-Ne white dwarfs in a close binary system  \citep{Denissenkov-2017} or a complex double s+r enrichment scenario \citep{Abate16}.

\section{Abundance profile analysis}
\label{Sect: abund-analysis}

\subsection{The [hs/ls] ratio}
\label{sect:hsls}
The second-to-first s-process peak abundance ratio ([hs/ls]) and the third-to-second s-process peak abundance ratio ([Pb/hs]) give insight on the s-process efficiency (where ls and hs stand for s-process elements of the first (Sr, Y, Zr) and second (Ba, La, Ce) peaks, respectively).

\citet{Hollek2015} noted that the abundance continuum present in CEMP-s and CEMP-rs stars calls for a new classification scheme, and proposed an alternative based on the [Y/Ba] value. Here we  discuss   [Ba/Y]  to better match the usual [hs/ls] index.
\citet{Hollek2015} separated CEMP stars into three groups of increasing [hs/ls]:  CEMP-sA  ($0.3<$ [Ba/Y] $<0.9$),
CEMP-sB  ($0.9<$ [Ba/Y] $<1.5$), and   CEMP-sC  ([Ba/Y] $>1.5$). 
They concluded that their CEMP-sA group mainly concerns CEMP-s stars, whereas CEMP-rs stars belong mostly to the CEMP-sC group.

\begin{figure}
\includegraphics[width=9.5cm]{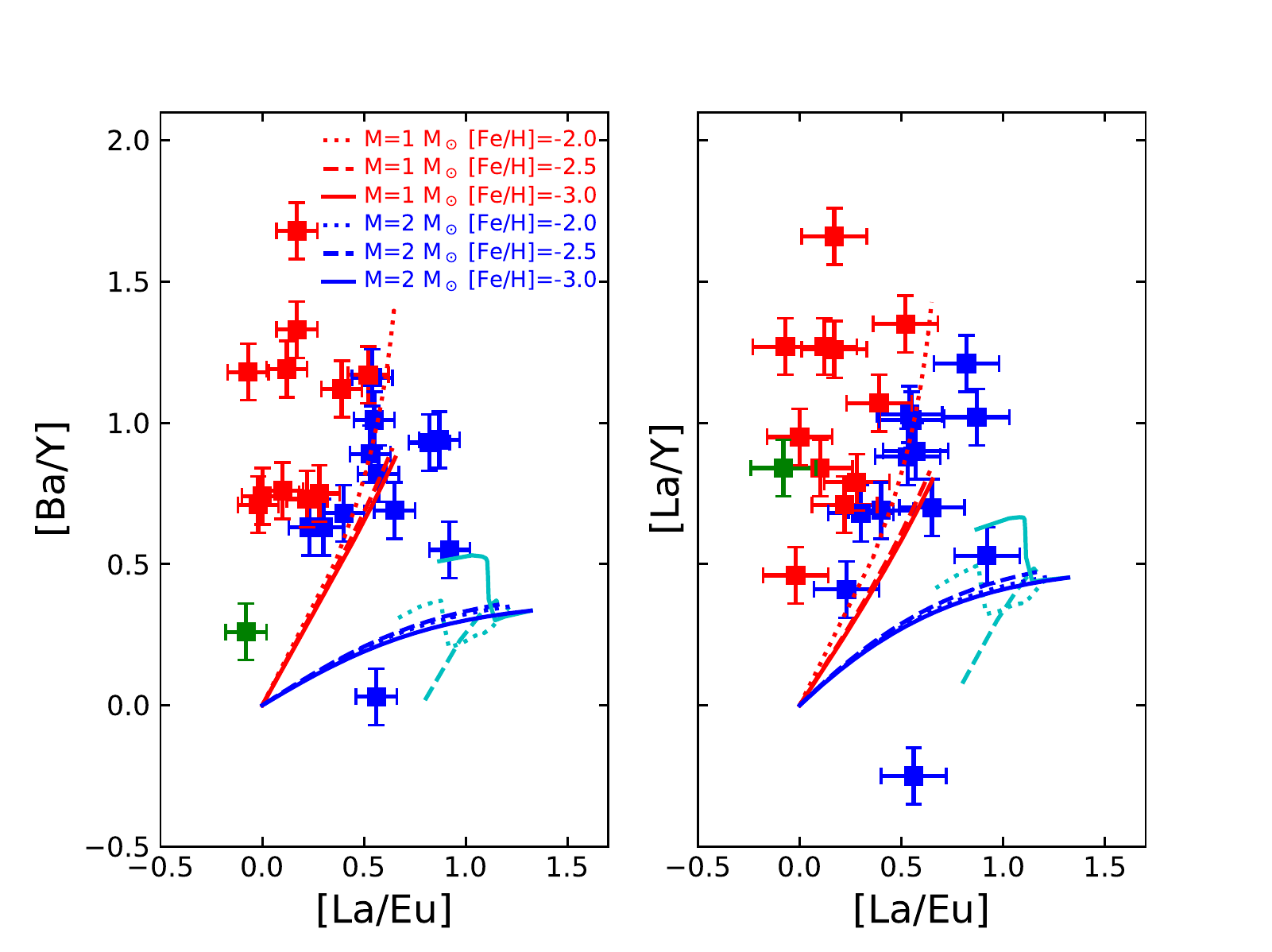}
\includegraphics[width=9.5cm]{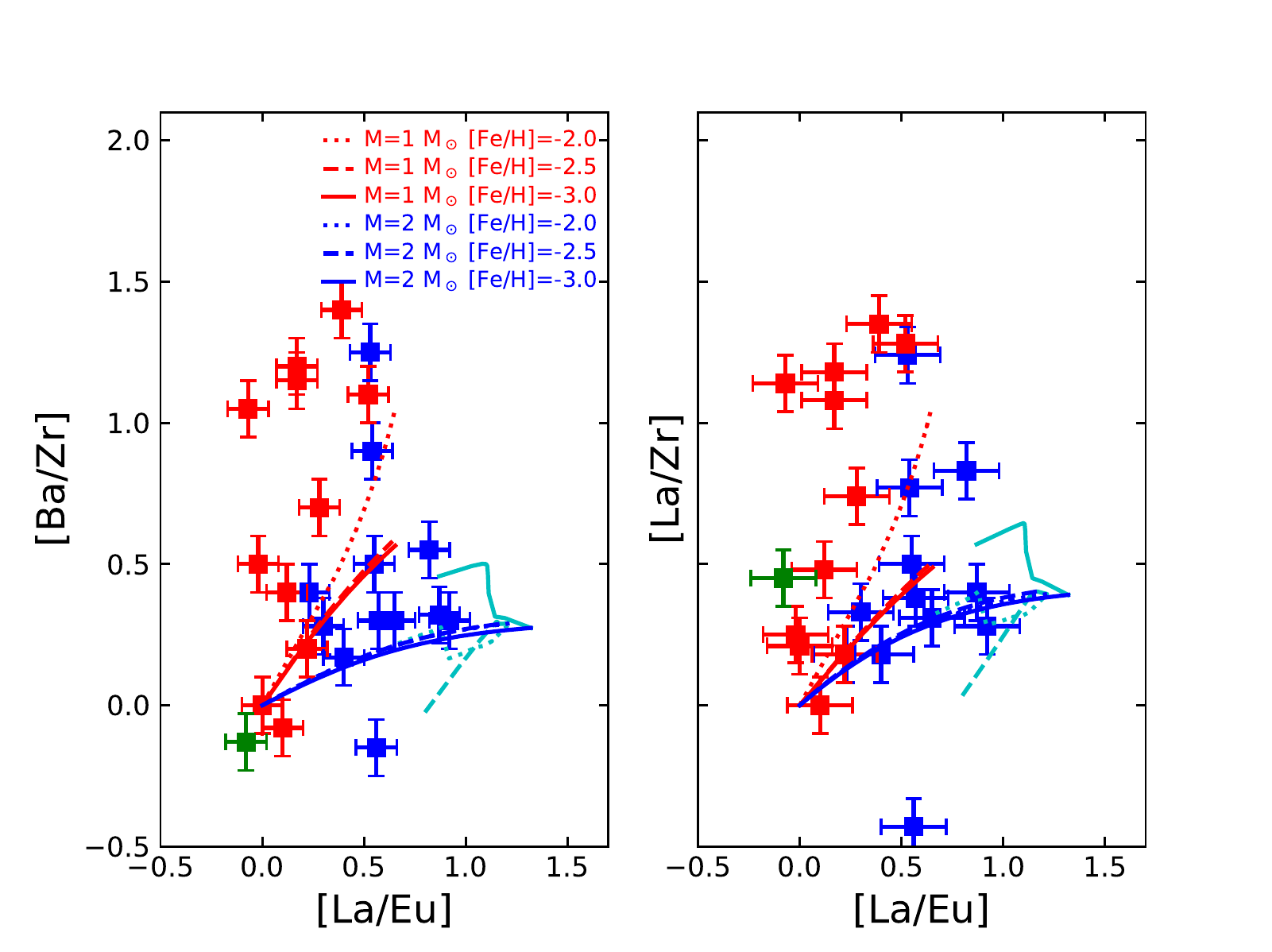}
\caption{Four [hs/ls] indices (where ls = Zr or Y, and hs = Ba or La) as a function of the [s/r] index [La/Eu]. Symbols are as in Fig.~\ref{Fig:CEMP_YEu}.
Theoretical predictions for 1 and 2 \Msun\ AGB stars at different metallicities, as labelled, are overplotted.
For the 2~\Msun\  models 
 the cyan  line represents the evolution of the abundances along the AGB, whereas the blue and red lines represents the dilution of tip-of-the-AGB material in the envelope of the companion, ultimately producing material of solar abundance ([Ba/Zr]=0, [La/Eu]=0, [La/Zr]=0). 
Since the 1~\Msun\ models experience a single thermal pulse, the cyan line reduces to a single point in that case and the red lines represent the mere dilution as above for the blue lines. We note that the 1~\Msun\ model predictions at [Fe/H]~=~$-$2.5 and $-$3 almost exactly superimpose.
\label{Fig:CEMP_YLanew} }
\end{figure}

Figure~\ref{Fig:CEMP_YLanew} presents [Ba/Y] and [La/Y] as a function of [La/Eu]. There is a slight offset of the [hs/ls] distribution of CEMP-rs stars (shifted to higher values) with respect to that of CEMP-s stars. This distribution offset is most clearly seen when using  [Ba/Y].

\begin{figure}[h]
\includegraphics[width=9.5cm]{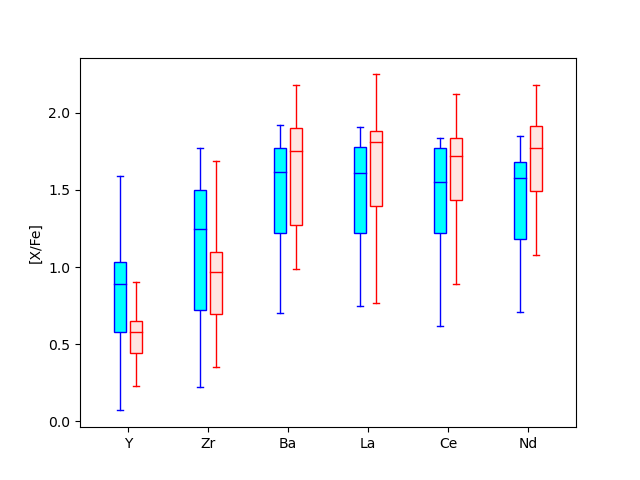}
\caption{Abundance distribution of light-s (X = Y, Zr) and heavy-s (X = Ba, La, Ce, Nd) elements for CEMP-s (blue) and CEMP-rs (red) stars. The box extends from the lower to upper [X/Fe] quartile, with a small horizontal line at the median. The whiskers extend from the box to show the full range of the data. 
\label{Fig:CEMP_boxplot} }
\end{figure}

This offset is tiny, but clearly seen in the boxplots in Fig.~\ref{Fig:CEMP_boxplot}: both the median (indicated as a horizontal line in the blue and red boxes) and the quartiles are shifted, always in the same direction for two light-s elements ([ls/Fe]$_{\rm CEMP-s} > $ [ls/Fe] $_{\rm CEMP-rs} $) and for four heavy-s elements ([hs/Fe]$_{\rm CEMP-s} <$ [hs/Fe]$_{\rm CEMP-rs} $).
This effect can be partly ascribed to metallicity: though the [hs/ls] ratio does not show an extremely marked trend with metallicity (see middle left panel of Fig.~\ref{Fig:CEMP_PbhslsFe}), the seven objects with [hs/ls] $>1$ all have [Fe/H] $<-2$.
However, metallicity alone is not responsible for the offsets in Fig.~\ref{Fig:CEMP_boxplot}, since the  [hs/ls] distributions (where [Fe/H] cancels)  of CEMP-rs and CEMP-s stars are shifted in both Fig.~\ref{Fig:CEMP_YLanew} and Fig.~\ref{Fig:CEMP_PbhslsFe}.
We note that Fig.~\ref{Fig:CEMP_PbhslsFe} presents a more robust estimate of [hs/ls], with  [hs/Fe] computed as $0.5 \times $([La/Fe] + [Ce/Fe])
and [ls/Fe] as $0.5 \times$([Y/Fe] + [Zr/Fe]).
In conclusion, Figs.~\ref{Fig:CEMP_YLanew}, \ref{Fig:CEMP_boxplot}, and \ref{Fig:CEMP_PbhslsFe} show that there is a tendency for CEMP-rs stars to have higher [hs/ls] ratios than CEMP-s stars.

In order to assess whether the difference between the [hs/ls] distributions of CEMP-s and CEMP-rs stars is significant, we performed a Wilcoxon rank-sum test (also called the Mann–Whitney U test), which allows us to determine whether two independent samples are selected from populations having the same distribution.
Let $H_0$ be the null hypothesis that the two groups of measured [hs/ls] ratios for CEMP-rs and CEMP-s stars are drawn from the same distribution.  
Based on the values listed in Table~\ref{Tab:Wilcoxon}, the risk of rejecting $H_0$ even while it is true is 2.6\%. Therefore, the null hypothesis $H_0$ of population identity for   [hs/ls]   can be rejected with a high confidence level. 
For the sake of completeness, the same rank test has also been applied  to individual elemental abundances in order to assess the significance of the differences observed in their distribution displayed in Fig.~\ref{Fig:CEMP_boxplot}. The risk of rejecting $H_0$ even while  true is  always lower than 15\%, except for Sm (Table~\ref{Tab:Wilcoxon}). Considering Pr, Eu, Gd, and Dy (i.e. mostly r-elements), the risk is even lower than 5\%.

Therefore, the null hypothesis $H_0$ of population identity can be rejected with a high confidence level for both light-s and heavy-s elements. It can be safely concluded that CEMP-rs stars have on average higher [hs/Fe] and lower [ls/Fe] than CEMP-s stars, and that this difference is statistically significant.

\begin{table*}
\caption{Results of the Wilcoxon rank-sum test: probability of rejecting, while it would be true, the null hypothesis that the two groups of measured abundances of element $X$ for CEMP-rs and CEMP-s stars are drawn from the same distribution, and the same for the [hs/ls] value, where ls and hs are defined as in Fig.~\ref{Fig:CEMP_boxplot}.}
\label{Tab:Wilcoxon}
\begin{center}
\begin{tabular}{lcccccccccccc}
\hline
Element & Y & Zr & Ba  & La & Ce & Pr & Nd & Sm & Eu & Gd & Dy & [hs/ls]   \\
\hline\\
Probability (\%) & 11.7 &  14.2 & 13.0 & 9.7 &  11.7 &  3.2 &  4.9 &  21.8 &  0.2 &  0.2 &  2.5 & 2.6 \\
\hline
\end{tabular}
\end{center}
\end{table*}

We thus conclude that CEMP-rs stars have a tendency towards higher [hs/ls]   (thereby confirming the \citealt{Hollek2015} results),
 higher [hs/Fe], and lower metallicity when compared to CEMP-s stars, though a large overlap exists between the distributions of the two stellar classes. This continuity between CEMP-s and CEMP-rs abundance patterns does not point to a totally different site for the i-process nucleosynthesis, but rather to the same physical process as the one producing the classical s-process but occurring under slightly different, more extreme conditions (i.e. lower masses, lower metallicities).

Two CEMP-s stars, HD 55496 and CS 22942$-$019, have a negative [hs/ls] index, implying that they have high Y and Zr abundances with respect to La and Ce. It remains difficult to explain theoretically such a trend within the low-mass AGB star models. In contrast, higher mass AGB stars are known to be subject to a non-negligible neutron irradiation within the thermal pulses that give rise to larger production of light s-elements with respect to heavier species \citep[{e.g.}][]{Goriely2005}. Considering an intermediate-mass AGB star would also help to explain the high N abundance found in HD~55496, as discussed in Sect.~\ref{sect:cno}.

In Fig.~\ref{Fig:CEMP_YLanew} the model predictions are also overplotted, for the same mass and metallicity range as in Fig.~\ref{Fig:i_vs_s}.
These predictions start at high [La/Eu] and [hs/ls] values (representing the first TDU), and join the (0,0) point for infinite
dilution of TP-AGB enriched material into the (solar-scaled composition) companion envelope composition.
In the case of 2~\Msun\ models, a specific colour-coding in Figs.~\ref{Fig:CEMP_YLanew}, \ref{Fig:CEMP_LaEuCN}, \ref{Fig:sFe_vs_CO}, \ref{Fig:NNa}, and \ref{Fig:Mghs}  allows us to distinguish  TP-AGB enrichment (in cyan) from dilution (in blue). For the 1~\Msun\ models this distinction was not possible since they experience a single thermal pulse.

Figure~\ref{Fig:CEMP_YLanew} nicely shows that the [hs/ls] predicted abundance ratios of the low-mass AGB stars (in red)
are higher than those of stars experiencing a standard s-process (in blue).
However, because of dilution, all models converge to the origin, which explains the large overlap in [hs/ls] and in [La/Eu] of both modelled and measured abundances. We note here that the measured [hs/ls] ratios, when represented by [Ba/Y], are well above the predictions compared to [Ba/Zr]. We found no explanation for this offset, but it may also be due to the absence of NLTE correction for the Y lines used for the analysis. 

\subsection{The Pb abundance}

\begin{figure}
\includegraphics[width=9.5cm]{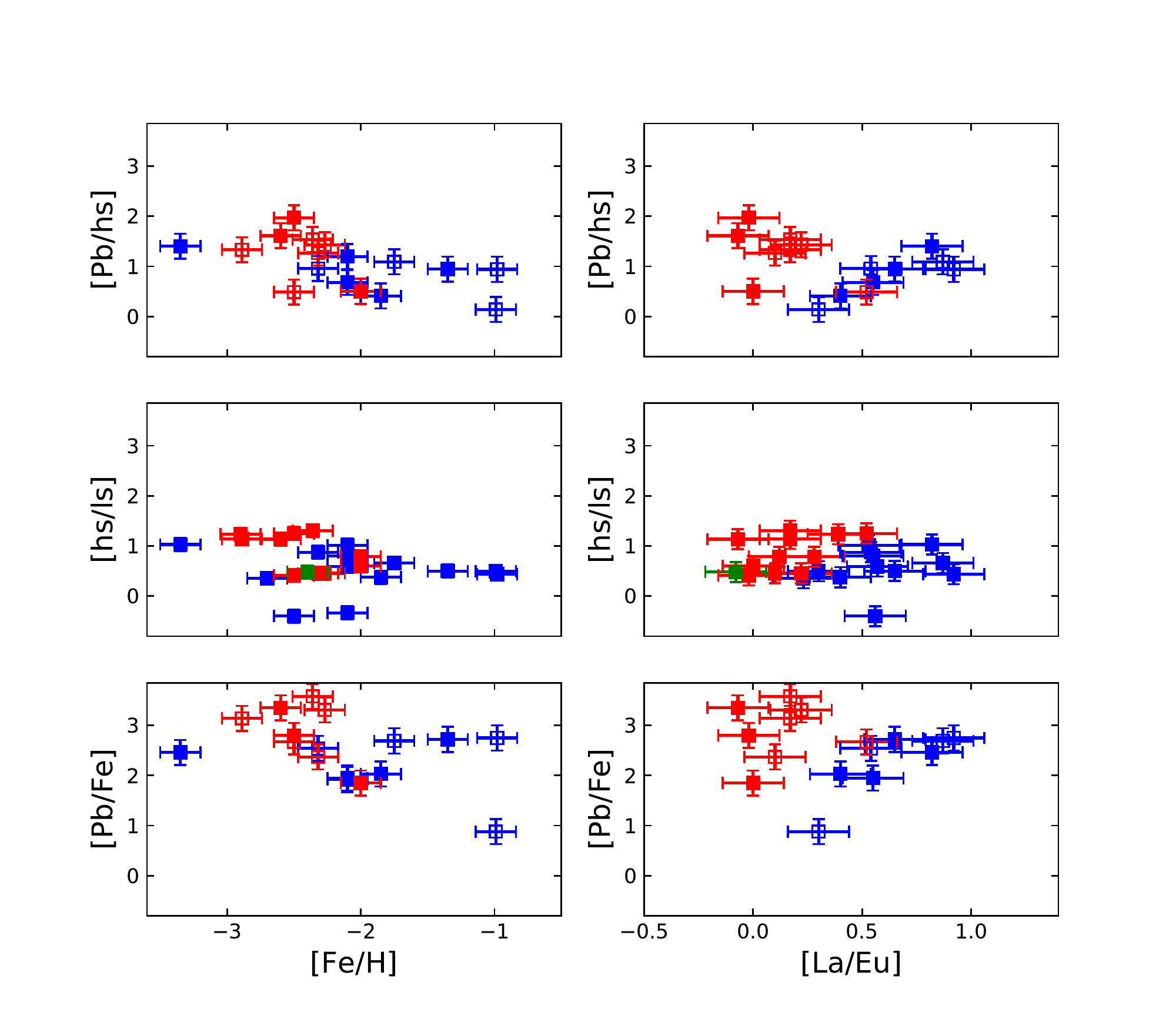}
\caption{[Pb/hs] and [hs/ls] as a function of the metallicity [Fe/H] (left column) and of [La/Eu] (right column). 
The [hs/Fe] ratio is computed as ([La/Fe]+[Ce/Fe])/2 and [ls/Fe] as 
([Y/Fe]+[Zr/Fe])/2.
As usual, blue and red symbols represent CEMP-s and -rs stars, respectively. Empty symbols represent objects with uncertain Pb abundances, whereas filled symbols represent objects with reliable Pb abundances. The non-LTE Pb abundances presented in Table~\ref{Tab:abundances} are used for both cases. 
}
\label{Fig:CEMP_PbhslsFe}
\end{figure}

We now turn to the lead-to-second s-process peak abundance ratio.  
The [Pb/hs] dependence on [Fe/H] is displayed in Fig.~\ref{Fig:CEMP_PbhslsFe}. Non-LTE Pb abundances are plotted with full squares, while uncertain abundances (flagged with `:' in Table~\ref{Tab:abundances}) are represented as empty squares.
For convenience, the [s/Fe], [ls/Fe], [hs/Fe], and [hs/ls] of the programme stars are listed in Table~\ref{Tab:hsls_ratios}. A trend of increasing [Pb/hs] with decreasing metallicity is observed in this figure. The usual explanation is  that the efficiency of the s-process increases when metallicity decreases, more neutrons being available per iron seed nuclei.

The ([Pb/hs], [La/Eu]) panel of Fig.~\ref{Fig:CEMP_PbhslsFe} illustrates the clearest separation between CEMP-s and CEMP-rs stars that we could achieve in this paper: CEMP-s stars have high [La/Eu] and low [Pb/hs], while CEMP-rs stars are characterized by low [La/Eu] and high [Pb/hs].
The only exception is the CEMP-rs star CS~22891$-$171, which is however one of the objects with uncertain Pb abundances.
This tendency (high [Pb/hs] for low [La/Eu]) is somewhat expected because when the La abundance increases, all other things being equal, [Pb/hs] will decrease (because hs = (La + Ce)/2).
However, this trend is not observed among CEMP-s stars alone where [Pb/hs] increases with increasing [La/Eu]. It probably reflects the fact that among CEMP-s stars Eu is not produced and has its solar-scaled abundance, so that stars with high second-peak abundances (La), indicating an efficient s-process, also tend to have high third-peak abundances (Pb). On the contrary, in CEMP-rs objects, a
measurable intrinsic Eu production would disturb this trend by decreasing [La/Eu].
\subsection{CNO and heavy elements}
\label{sect:cno}

\begin{figure}
\includegraphics[width=9.5cm]{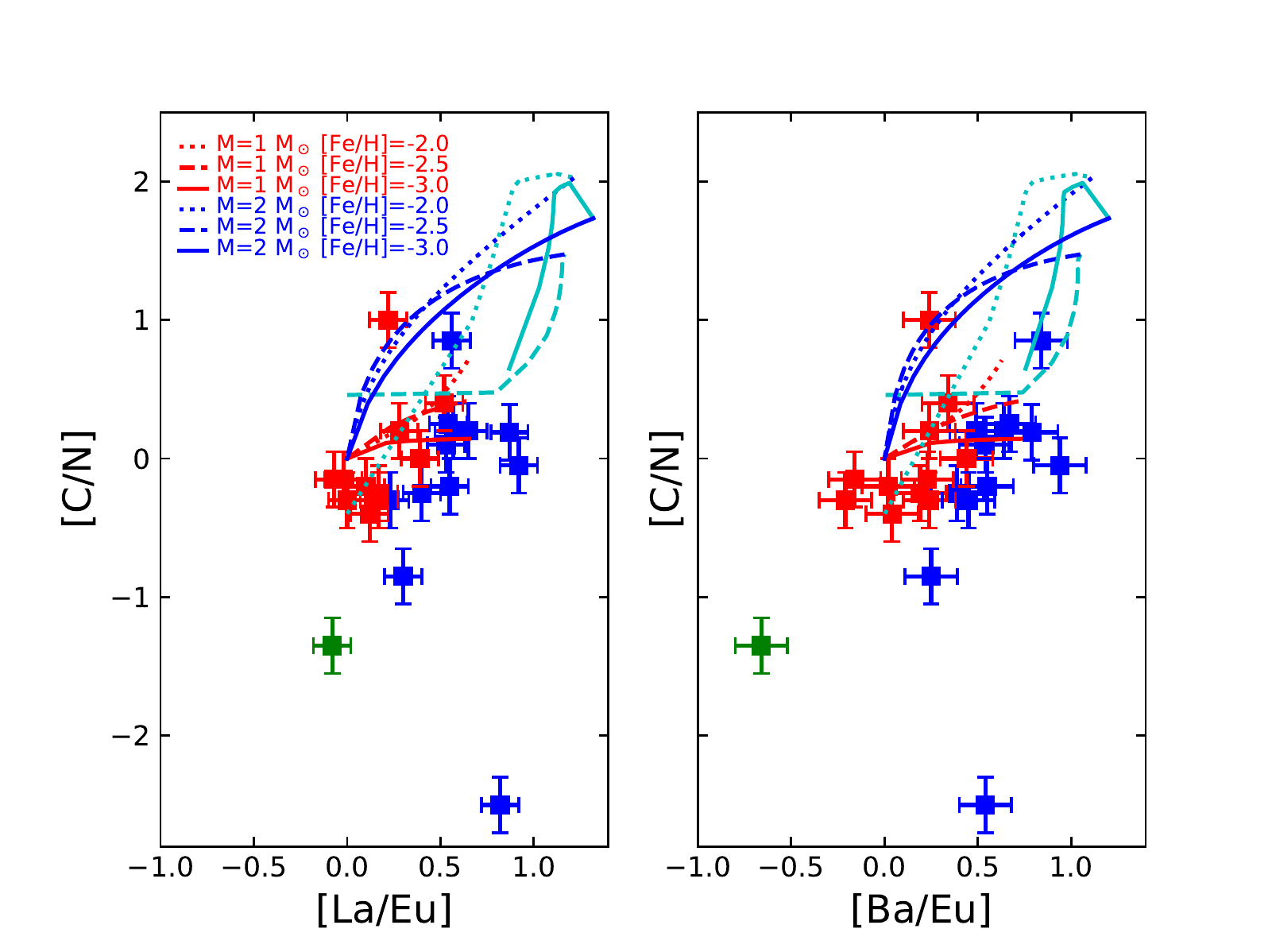}
\includegraphics[width=9.5cm]{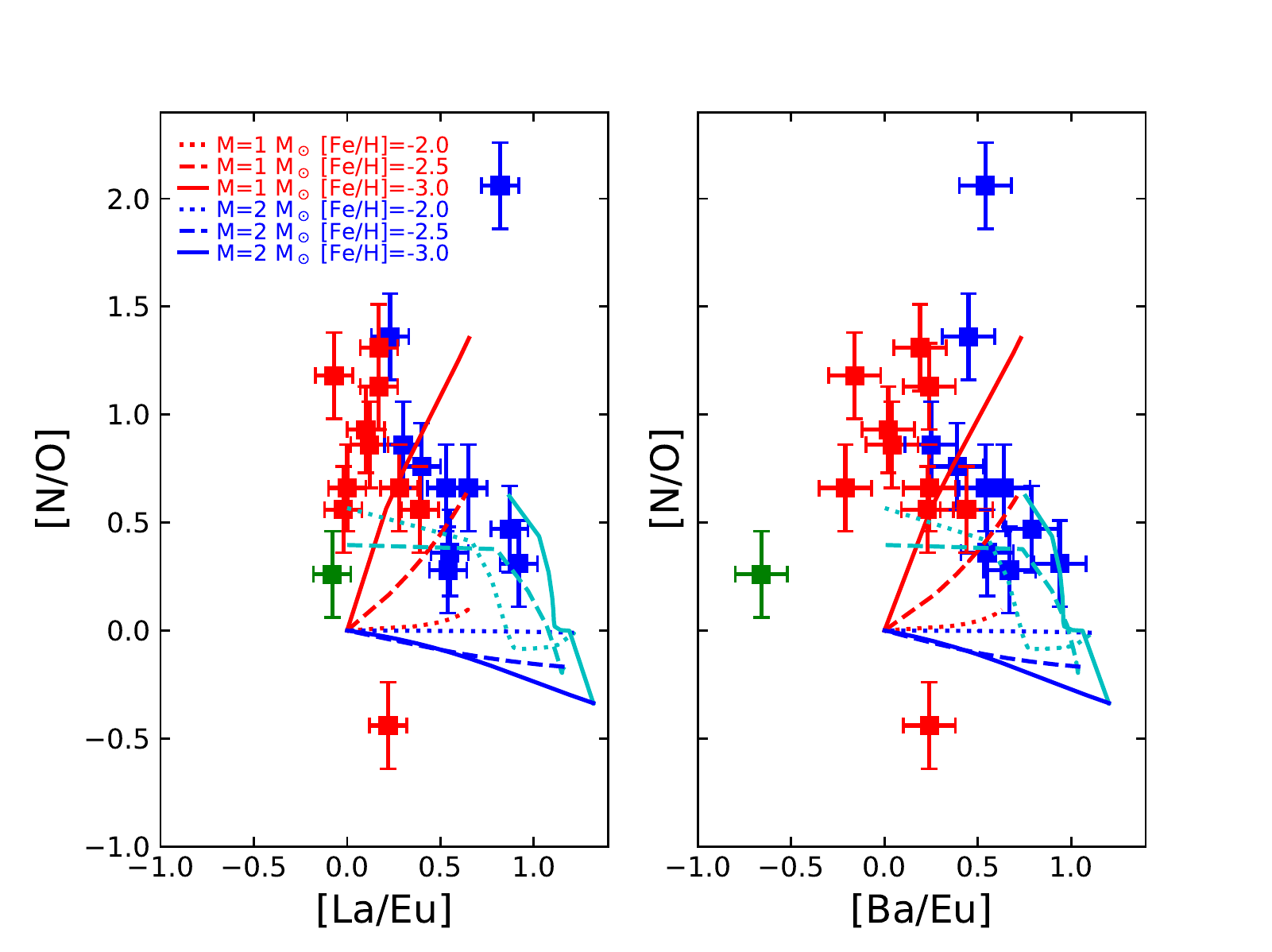}
\caption{Abundance ratios of  [C/N] and [N/O]   as a function of [La/Eu] and [Ba/Eu]. Symbols are as in Fig.~\ref{Fig:CEMP_YEu},
and cyan curves as in Fig.~\ref{Fig:CEMP_YLanew}.
\label{Fig:CEMP_LaEuCN} }
\end{figure}

\begin{figure}
\includegraphics[clip,trim=0.5cm 0cm 0.5cm 0cm, width=9.5cm]{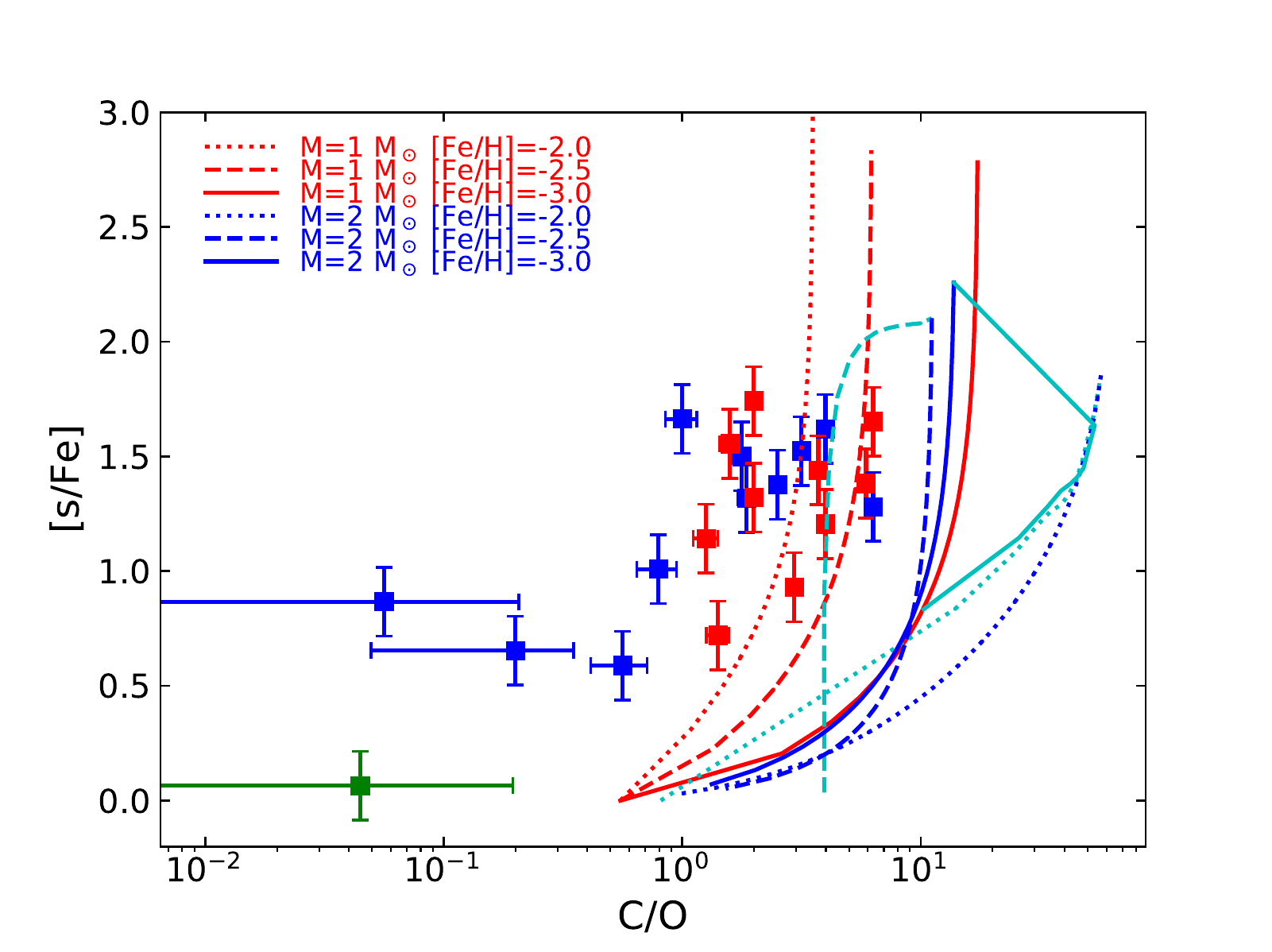}
\caption{Surface s-enrichment [s/Fe] as a function of the surface C/O, with [s/Fe] defined as ([Y/Fe]+[Zr/Fe]+[Ba/Fe]+[La/Fe]+[Ce/Fe]+[Nd/Fe])/6. The measured abundances are colour-coded as in Fig.~\ref{Fig:CEMP_YEu}, and the model predictions as in Fig.~\ref{Fig:CEMP_YLanew}.
\label{Fig:sFe_vs_CO} }
\end{figure}

Figure~\ref{Fig:CEMP_LaEuCN} shows the CNO abundances with [La/Eu] and [Ba/Eu]. The good separation between the CEMP-s and -rs subgroups in the ([C/N],[La/Eu]) or ([C/N],[Ba/Eu]) planes simply reflects the separating power of [La/Eu] or [Ba/Eu].

Nitrogen-enhanced metal-poor  stars have been defined by \citet{Johnson-07} as CEMP stars with [C/N]~$<-0.5$ and [N/Fe]~$>0.5$.
According to this definition, two NEMP stars are present in our sample, both also classified as CEMP-s stars: CS~30322$-$023 and HD~55496.
Unfortunately, the latter has no Eu measurement available, but its Sm abundance can be used as a proxy. If, as we do in Sect.~\ref{Sect:assignment}, we assign to HD~55496 the average [Eu/Sm] of CEMP-s stars ([Eu/Sm]$_{\rm av,s}=-0.42$), we infer [La/Eu] = 0.92 and [C/N]~$= -2.3$ for HD 55496, which makes it a twin of CS~30322$-$023   ([La/Eu]~=~0.82, [C/N]~$=-2.5$). 
This high nitrogen abundance points towards  HBB \cite[]{Boothroyd1995} in massive AGB stars (M~$\gtrsim 4$M$_\odot$), but the positive [hs/ls] ratio of CS~30322$-$023 does not comply with an s-process nucleosynthesis powered by $^{22}$Ne($\alpha$,n)$^{25}$Mg in the convective pulse. The situation is more understandable with HD~55496 where [hs/ls] is negative, as already discussed in Sect.~\ref{sect:hsls}.

We note that \citet{Masseron2006} interpreted CS~30322$-$023 as a possible low-metallicity TP-AGB star since the distance estimated from the spectroscopic parameters led to a luminosity compatible with the TP-AGB in the HR diagram. This interpretation is difficult to reconcile with its high nitrogen abundance implying an intermediate-mass star, if the high nitrogen abundance is indeed due to HBB. With the Gaia DR2 parallax, the position of CS 30322$-$023 in the HR diagram or in a ($\log g$, $T_{\rm eff}$) diagram is compatible with that of a tip-of-the-RGB star (see Figs.~\ref{Fig:HRnormal}, \ref{Fig:HRanormal}, \ref{Fig:loggTeffnormal_1}, and \ref{Fig:loggTeffanormal_2}, where CS 30322$-$023 is located at $\log T_{\rm eff}=3.65$ and $\log g= 1$).

Considering now the group of objects with [C/N]~$>-1$ in  Fig.~\ref{Fig:CEMP_LaEuCN}, there might be a loose trend (among CEMP-s and CEMP-rs stars separately) of increasing  [C/N] with increasing [La/Eu] (and [Ba/Eu]). This trend is expected in low-mass stars as third dredge-ups in the absence of HBB contribute to increase both s-process elements and C. 

Finally, we note that CEMP-s and CEMP-rs stars do not seem to belong to separate classes as far as their C, N and O abundances are concerned. Instead, they  show a continuity in their [C/N] and [N/O] abundance ratios.

The comparison between model predictions and abundance determinations
is shown in Fig.~\ref{Fig:CEMP_LaEuCN}.
A significant number of CEMP-rs stars have ${\rm [C/N]} \la 0$, a feature that is not reproduced by the low-mass (1~\Msun) models.
Furthemore, the [C/N] and [N/O] ratios are hardly explained by the more traditional nucleosynthesis in the 2~\Msun\ AGB star.

Figure~\ref{Fig:sFe_vs_CO} illustrates the correlation between the surface s-process enrichment and the C/O ratio. We note that the 1~\Msun\ predictions are in better agreement that the 2~\Msun\ values.
The large C overabundances predicted in AGB stars, but not confirmed by derived abundances, is a long-standing problem \citep[see  e.g.][]{Shetye19}.

\begin{figure}
\includegraphics[width=9.5cm]{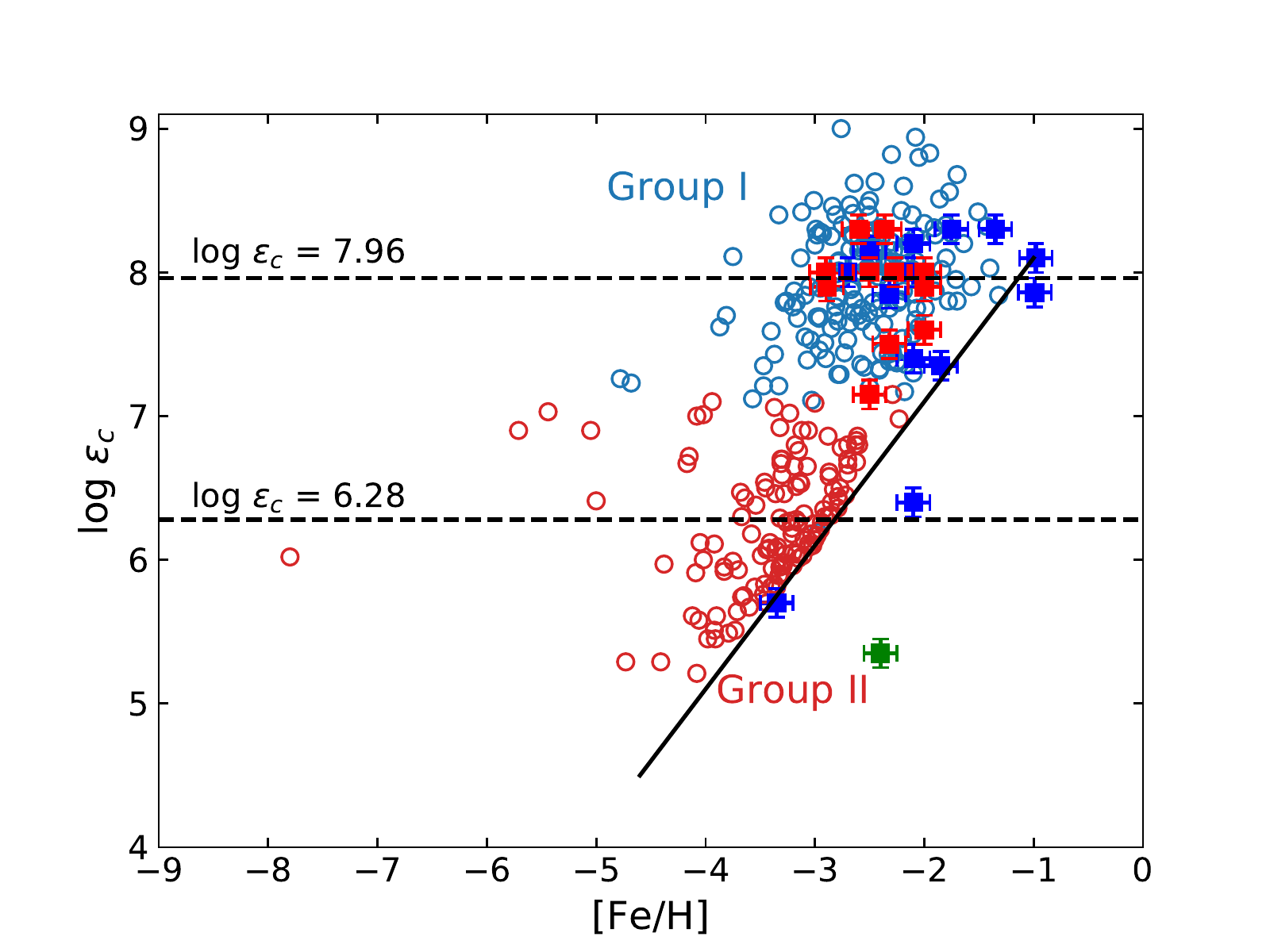}
\caption{Scatter plot of the carbon abundance ($\log \epsilon_C$) as a function of metallicity. The diagonal line traces the Galactic evolution of carbon with metallicity, in the absence of any in situ enhancement.
The stars in this sample are indicated by blue (CEMP-s) and red (CEMP-rs) squares.
CEMP stars from the literature as compiled by \citet{Yoon2016} are superimposed: CEMP-s stars are represented as blue circles, they constitute Group I, with an average C abundance as indicated by the upper dashed horizontal line.
CEMP-no stars are plotted as red circles, and  define Group II.
\label{Fig:epsilonC}}
\end{figure}

A figure displaying $\log \epsilon_{\rm C}$ as a function of [Fe/H] was first presented by \citet{Spite2013} and then further discussed by \citet{Hansen2016b} and  \citet{Yoon2016}.
Here Fig.~\ref{Fig:epsilonC} represents our stellar sample, together with the compilation of CEMP-s, -rs, and -no stars of \citet{Yoon2016}.
It clearly shows the bimodal carbon abundance distribution among CEMP stars, with a high-C group (centred on  $\log \epsilon_{\rm C}=7.96$)
containing mostly CEMP-s stars and a low-C group (centred on  $\log \epsilon_{\rm C}=6.28$)  composed of CEMP-no stars.
As noted by \citet{Yoon2016}, the $\log \epsilon_{\rm C}$ distribution of the CEMP-rs stars exhibits no clear difference from that of the CEMP-s stars. However, two CEMP-s stars  in our sample, not present in the \citet{Yoon2016} sample,
surprisingly fall in the low-C (group II) zone, among CEMP-no objects; they are CS 30322$-$023 and HD 55496, the two NEMP stars where
some carbon might have been turned into nitrogen by HBB, as mentioned above.
Actually the carbon abundance of HD~55496 falls {below} the diagonal line tracing the Galactic evolution of carbon with metallicity, in the absence of any in situ enhancement. This sub-Galactic carbon abundance   also  points to the operation of HBB. 
To be reconciled with the average carbon abundance of CEMP-s stars ($\log \epsilon_{\rm C}= 7.96$, see Fig.~\ref{Fig:epsilonC}), CS~30322$-$023 and HD~55496 lack respectively $\Delta \log \epsilon_{\rm C}=$ 2.26 and 1.56 dex of carbon. 
Given that this amount of carbon is entirely transformed to nitrogen by the CNO cycle 
($\Delta \log\epsilon_N = - \Delta \log\epsilon_C  $), we have
[C/N]$_{\rm measured} = $
[C/N]$_{\rm without~HBB} - 2 \Delta \log\epsilon_C$. 
These two stars have a low [C/N] ratio ([C/N] = $-$2.5 and $-$2.3 for CS~30322$-$023 and HD~55496, respectively; see Table~\ref{Tab:abundances}).
Without HBB, 
the [C/N] ratio of the AGB progenitors of these two stars would thus be around [C/N]~=~2.
This is exactly the value predicted for low-mass stars not experiencing HBB \citep{Johnson-07}, as can also be seen in the upper panel of Fig.~\ref{Fig:CEMP_LaEuCN}, top of the cyan curve, prior to dilution (in blue) in the companion envelope.
After dilution, the  [C/N] ratio for these two objects, had they not   experienced HBB, would then be fully compatible with that of the bulk of CEMP-s stars.
The measured low [C/N] ratio of these two stars, and the fact that they do not fall above the Galactic $\log \epsilon_C$ trend in Fig.~\ref{Fig:epsilonC},  can thus probably be explained by HBB operation.

\subsection{Na and heavy elements}

\begin{figure}
\includegraphics[width=9cm]{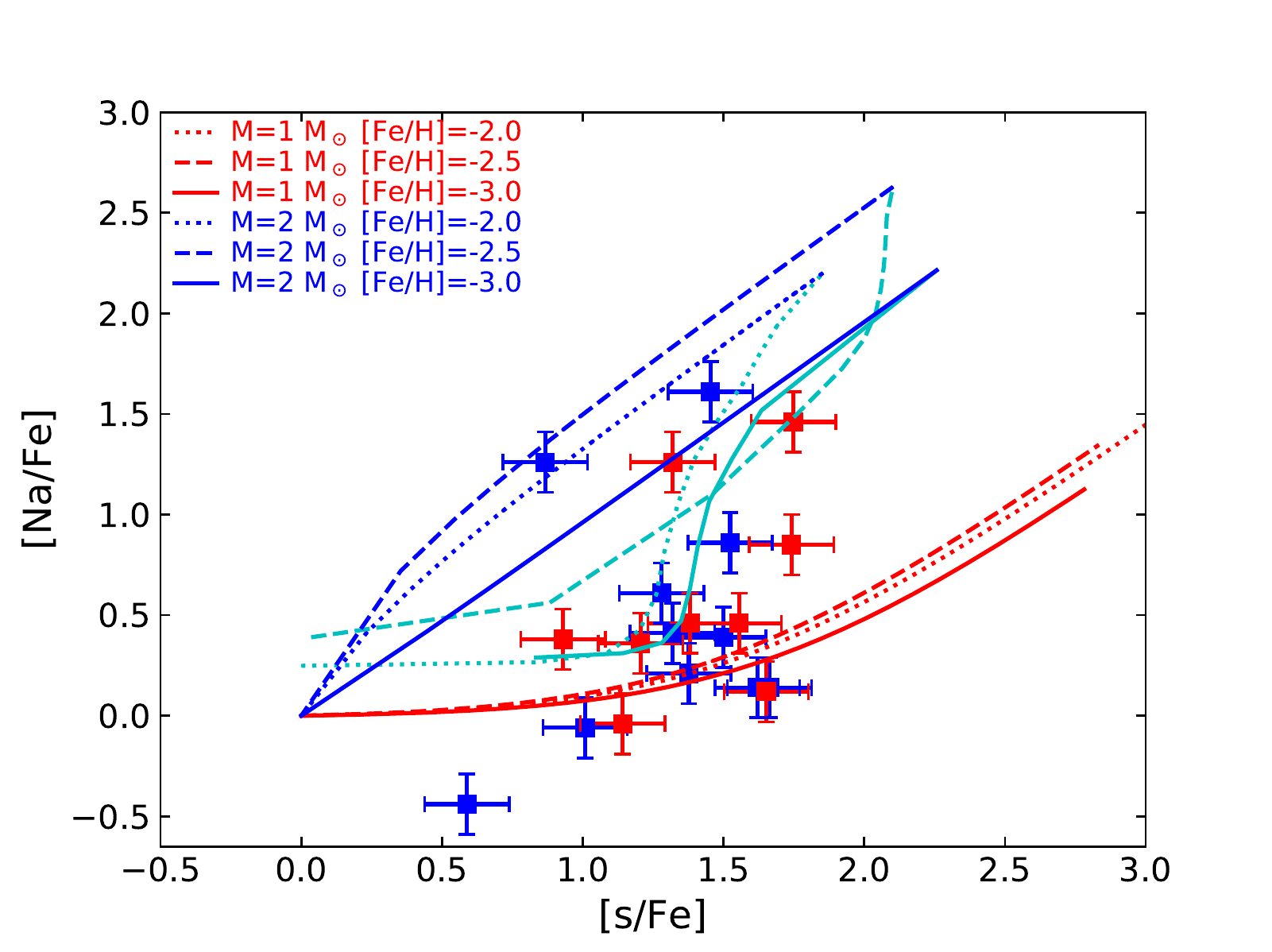}
\caption{Abundances of [Na/Fe]   shown as a function of [s/Fe], where [s/Fe] is computed as ([Y/Fe]+[Zr/Fe]+[Ba/Fe]+[La/Fe]+[Ce/Fe]+[Nd/Fe])/6. 
The measured abundances are colour-coded as in Fig.~\ref{Fig:CEMP_YEu}, and the model predictions as in Fig.~\ref{Fig:CEMP_YLanew}.
\label{Fig:NNa} }
\end{figure}

Figure~\ref{Fig:NNa} compares the [Na/Fe] overproduction with the surface s-enrichment. Stars of 1~\Msun\ are expected to be less enriched in Na than 2~\Msun\ stars, where the partial mixing of protons is known to be a   favourable site for Na production \citep{Goriely2000}.
As seen in Fig.~\ref{Fig:NNa},
the predictions  broadly  cover the region occupied by the measured abundances. However, the most Na-enriched CEMP-rs stars can hardly be explained, similarly to the less Na-rich CEMP-s stars.
Within the model of the partial mixing of protons at the time of the TDU, the Na production takes place in a tiny layer above the one responsible for the s-process \citep{Goriely2000}. For this reason the [Na/Fe] to [s/Fe] correlation depends on the profile of protons mixed into the C-rich region. The overprediction of the Na enrichment (Fig.~\ref{Fig:NNa}) might be an indication that the parametrization used for the mixing of protons may not be totally adequate. More studies, involving multidimensional simulations, are needed to solve this issue.

\subsection{[ls/hs] versus [Mg/hs]}

\begin{figure}
\includegraphics[width=9.5cm]{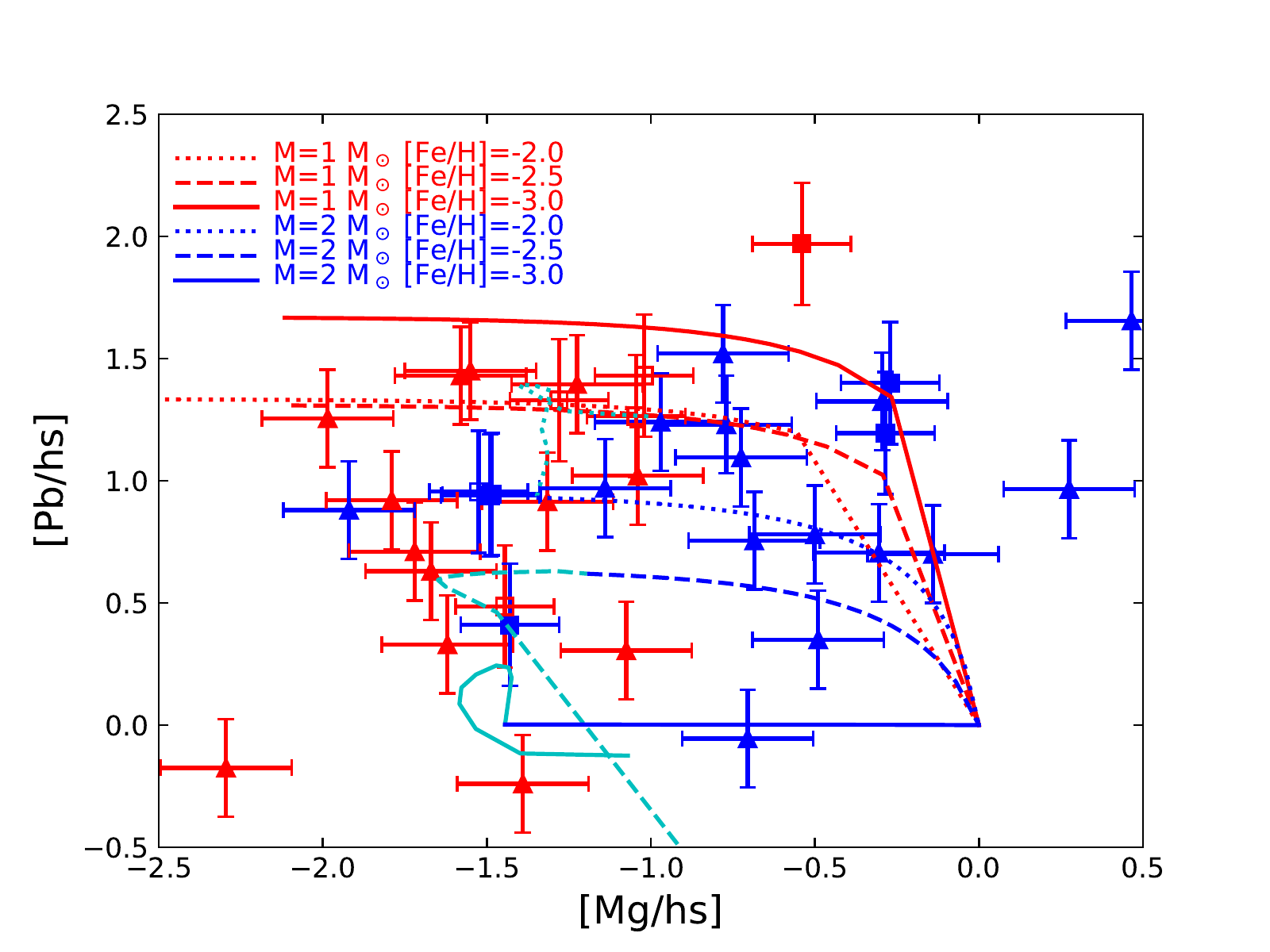}
\includegraphics[width=9.5cm]{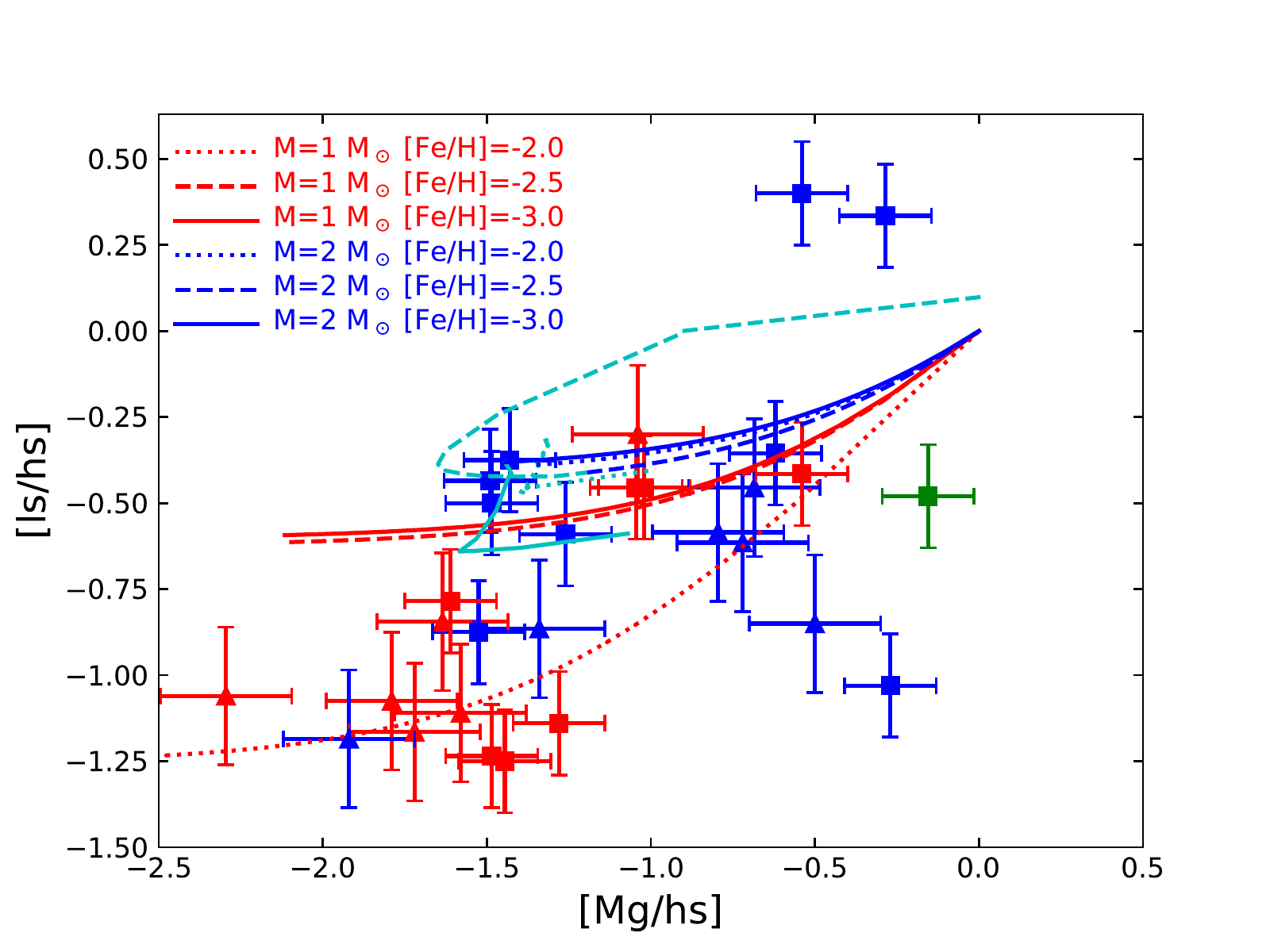}
\caption{Abundances [Pb/hs] and [ls/hs] as a function of [Mg/hs], where [ls] = 1/2([Y/Fe]+[Zr/Fe]) and [hs/Fe] = 1/2([La/Fe]+[Ce/Fe]). The measured abundances are colour-coded as in Fig.~\ref{Fig:CEMP_YEu}, empty squares as in Fig.~\ref{Fig:CEMP_PbhslsFe}, and the model predictions as in Fig.~\ref{Fig:CEMP_YLanew}. Blue and red triangles represent CEMP-s and CEMP-rs stars from the literature. 
\label{Fig:Mghs} }
\end{figure}

\begin{figure}
\includegraphics[width=9.5cm]{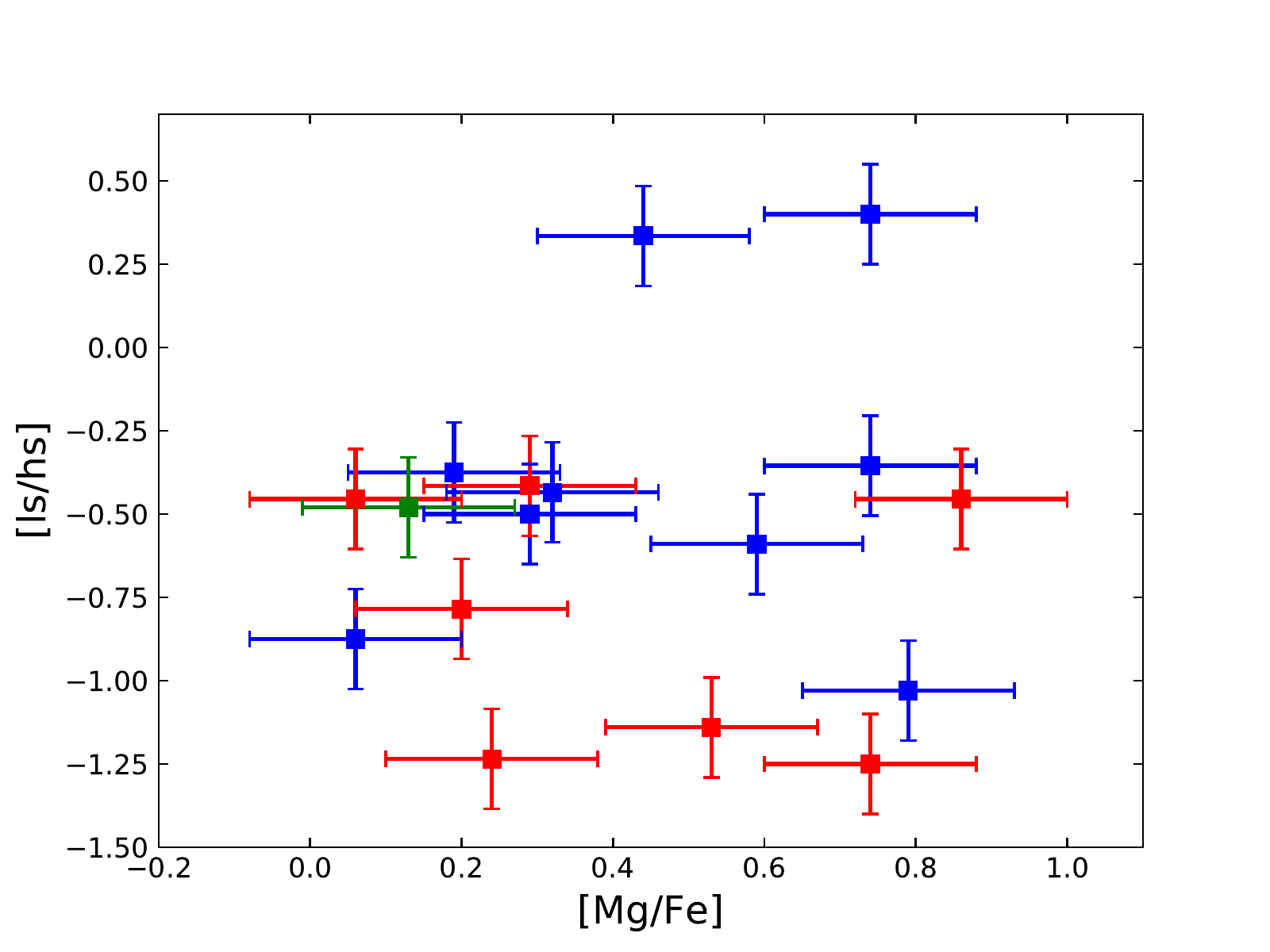}
\caption{[Mg/Fe] scatter plot of CEMP-s and CEMP-rs stars. Symbols are as in Fig.~\ref{Fig:CEMP_YEu}.
\label{Fig:MgFe} }
\end{figure}

\cite{Lugaro-2012} illustrated how, in a plane built with two intrinsic axes,
namely the ([ls/hs], [Mg/hs]) plane (i.e. where both axes are dominated by in situ nucleosynthetic processes rather than by the chemical evolution of the Galaxy), CEMP-s and CEMP-rs stars occupy different locations. 

We note, however, that in such a plane, a linear correlation of slope 1 is expected 
since
\begin{equation}
[{\rm ls/hs}] = [{\rm Mg/hs}] + \log({\rm ls}/{\rm Mg}) - \log({\rm ls}/{\rm Mg})_\odot ,
\end{equation}
except if there is a star-to-star variation of [ls/Mg].

Nevertheless, in both Fig.~7 of \citet{Lugaro-2012} and in the bottom panel of  Fig.~\ref{Fig:Mghs}, CEMP-rs stars seem to cluster at lower [ls/hs] and [Mg/hs] ratios than CEMP-s stars. One might wonder whether this effect is merely due to the higher [hs/Fe] of CEMP-rs stars or to a difference in [Mg/Fe].

From a theoretical point of view, CEMP-s and CEMP-rs stars can be expected to behave as shown in Fig.~\ref{Fig:Mghs}: (i)   Mg-low   CEMP-s stars, where s-process proceeds with the $^{13}$C($\alpha$,n)$^{16}$O neutron source activated in radiative conditions during the interpulse in low-mass stars;
(ii)  Mg-high  CEMP-s stars, where the $^{22}$Ne($\alpha$,n)$^{25}$Mg neutron source also contributes to the s-process nucleosynthesis during the convective thermal pulses; and
(iii) Mg-low CEMP-rs stars that activate $^{13}$C($\alpha$,n)$^{16}$O in convective conditions during proton ingestion episodes in the pulses.

However, in our sample we could not find any difference in the [Mg/Fe] ratios of CEMP-s and CEMP-rs stars (Fig.~\ref{Fig:MgFe}). We conclude that Fig.~7 of \citet{Lugaro-2012} and our Fig.~\ref{Fig:MgFe} illustrate the difference in [hs/Fe] of CEMP-s and CEMP-rs stars, as already demonstrated in Sect.~\ref{sect:hsls}.

Globally, model predictions are in fair agreement with observations, as shown in Fig.~\ref{Fig:Mghs}.
Models predict  higher [Pb/hs] and lower [ls/hs] overabundances for the 1~\Msun\ AGB stars relative to the 2~\Msun\ ones, though in both cases [Mg/hs] ratios span a large range  typically between $-$2 and 0~dex.  We note that the predictions regarding the light s-elements [ls] vary quite significantly if we adopt Y or Zr as the proxy, as seen in Fig.~\ref{Fig:CEMP_YLanew}.

\subsection{The Nb-Zr thermometer}
\label{Sect:Nb_Zr_thermometer}

\begin{figure}[h]
\includegraphics[width=9.5cm]{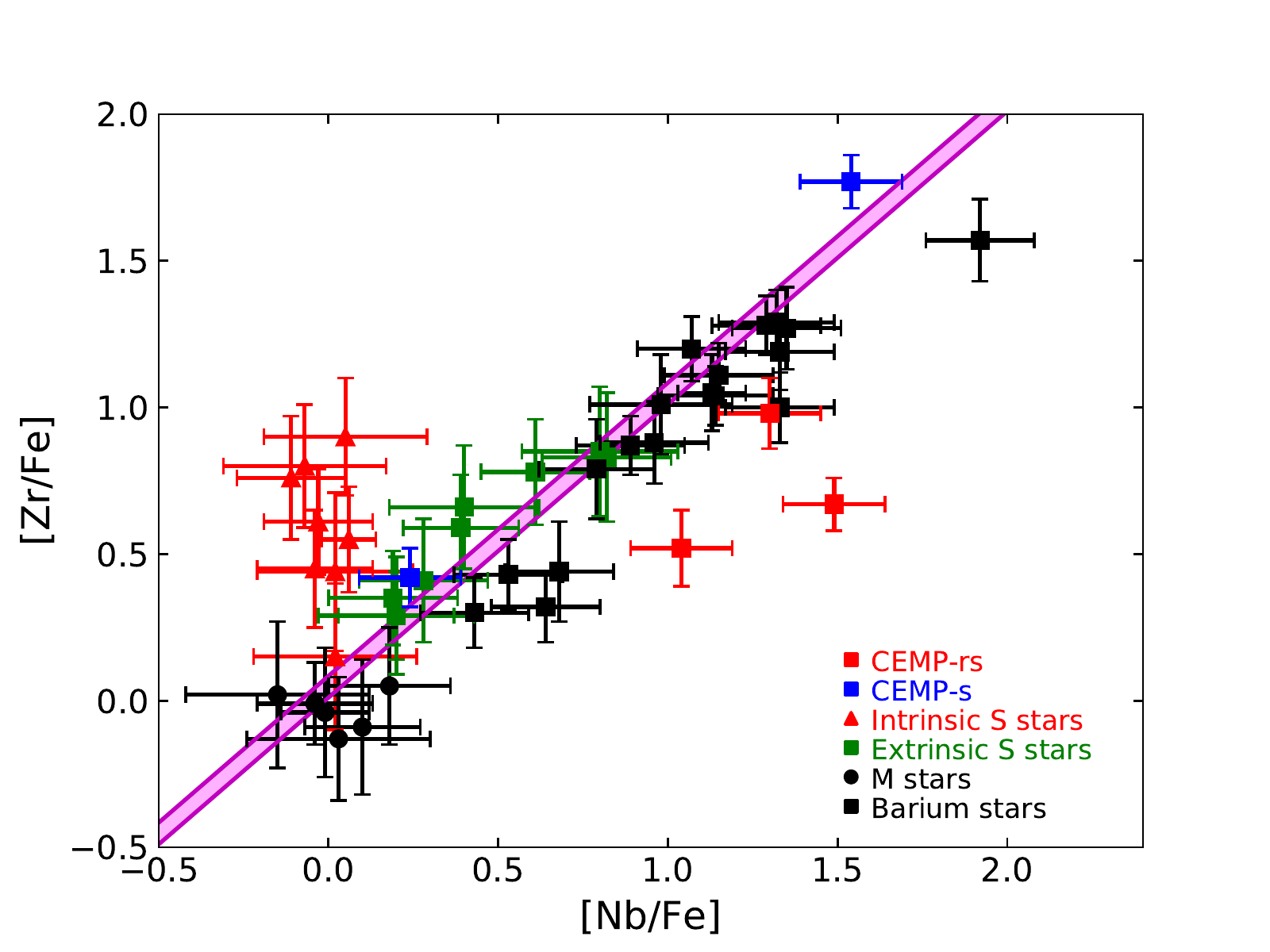}
\caption{Nb - Zr plane for CEMP-s and -rs stars with measurable Nb abundances, along with barium stars and intrinsic or extrinsic S-type stars from \citet{Karinkuzhi2018} and \citet{neyskens2015}, respectively. The magenta shaded band indicates the expected location of stars polluted by material resulting from the $s$-process  operating at temperatures between 1.0 (upper line) and 3.0$\times 10^8$~K (lower line). These simple predictions assume that  the chain of neutron captures along the Zr isotopes is in local equilibrium, as explained in \citet{Karinkuzhi2018}.
\label{Fig:NbZr} }
\end{figure}

The Nb/Zr thermometer offers an interesting diagnostic, allowing us to constrain the nucleosynthesis operation temperature and thus the stellar mass. 
In extrinsic stars, $^{93}$Zr produced by the s-process has had time to decay (since it was produced in the companion star long ago) into mono-isotopic $^{93}$Nb. Therefore, Nb/Zr ratio in extrinsic stars is equivalent to the $^{93}$Zr/Zr isotopic ratio in intrinsic AGB stars, which in turn   probes the s-process operation temperature. This diagnostic has been investigated in extrinsic S stars of near solar metallicity \citep{neyskens2015,Shetye2018} and barium stars of sub-solar metallicity \citep{Karinkuzhi2018}. These studies have provided a new confirmation that at such moderate metallicities the $^{13}$C neutron source is operating in low-mass stars. We  hoped to apply this thermometer to CEMP-rs stars to constrain the putative i-process, its neutron source ($^{22}$Ne or $^{13}$C), and its site.
Unfortunately, we could not measure Nb abundances in most objects since at the metallicity and temperature ranges of our programme stars Nb is mostly ionized, and all the sensitive Nb~II lines are located below 4000~\AA\ where HERMES spectra generally have a poor signal-to-noise ratio. This prevented us from drawing a robust conclusion about the behaviour of CEMP stars in the Nb - Zr plane. Nevertheless, Fig.~\ref{Fig:NbZr} shows that the two CEMP-s stars for which a niobium abundance could be safely derived follow the trend of extrinsic S stars and barium stars, as expected. The three CEMP-rs stars with measured Nb abundances also lie within the extrinsic-star region, and have a slightly higher [Nb/Fe] ratio than CEMP-s stars for a given [Zr/Fe]. However, this trend should   be validated by a larger sample of niobium measurements in CEMP-rs objects.

\section{HR diagram}
\label{Sect: HR}
In this section the programme stars are located in the HR diagram using Gaia DR2 parallaxes \citep{Gaia2018}.
The Bayesian estimates of the distances is taken from \cite{Bailer-Jones-2018}. 
STAREVOL \citep{Siess2008} evolutionary tracks using a metallicity scaled from  the \cite{Asplund2009} solar mixture are overplotted from the Hertzsprung gap to the RGB tip. When the evolutionary time exceeds the Hubble time, the tracks are truncated.

\begin{figure}
\includegraphics[width=10cm]{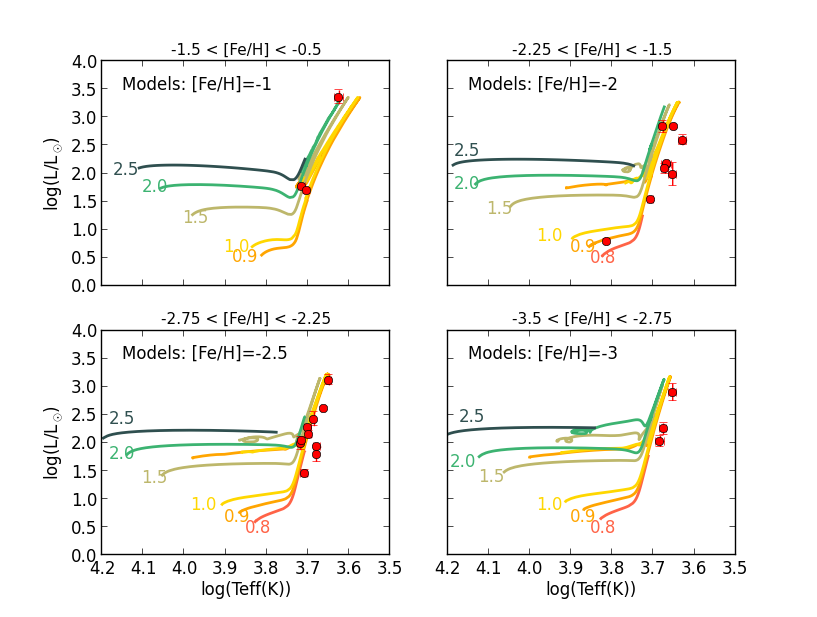}
\caption{HR  diagrams of the programme stars (red dots) split into four metallicity bins, as labelled at the top of each subpanel. Stars with a lower limit on their luminosity are marked with an arrow.  STAREVOL (Hertzsprung gap and RGB) evolutionary tracks are overplotted; their metallicity is indicated in each subpanel. The (initial) stellar masses are indicated in colour beside the corresponding track.
\label{Fig:HRnormal} }
\end{figure}

Figure~\ref{Fig:HRnormal} displays the resulting HR diagrams. While the luminosities of CEMP-rs stars 
are compatible with the RGB phase of evolution (except for CS 22887$-$048, which is in the Hertzsprung gap), 
the derived temperatures (respectively luminosities) seem too cool (respectively too low)  with respect to track temperatures (respectively luminosities) for metallicities  [Fe/H]~$< -1.5$.

\begin{figure}
\includegraphics[width=10cm]{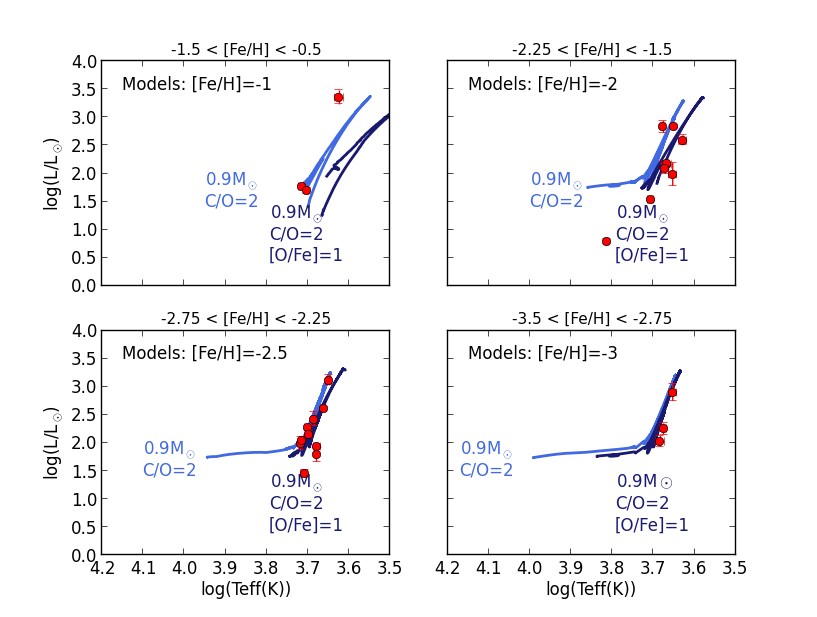}
\caption{Same as Fig.~\ref{Fig:HRnormal}, but for 0.9~\Msun\ STAREVOL evolutionary tracks of modified composition, as indicated.
\label{Fig:HRanormal} }
\end{figure}

\begin{figure}
\includegraphics[width=10cm]{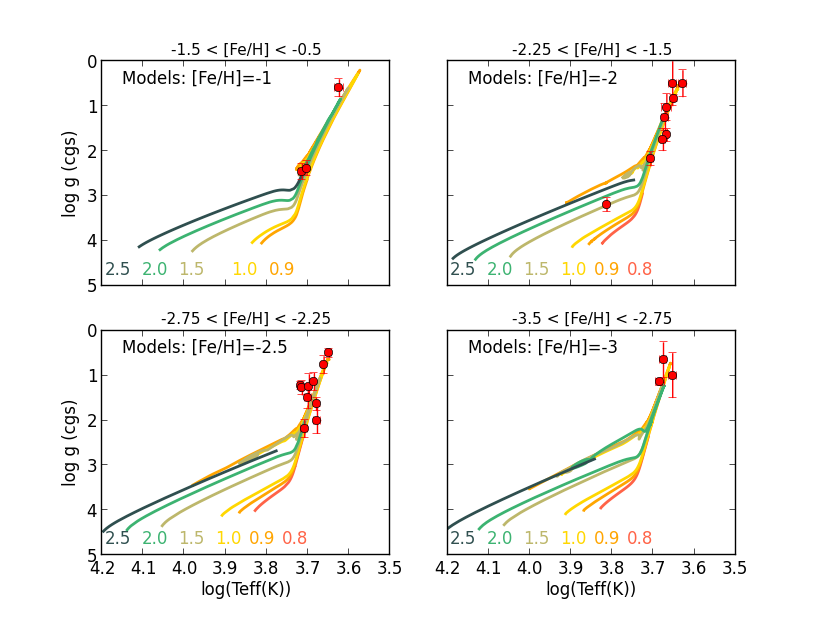}
\caption{Same as   Fig.~\ref{Fig:HRnormal}, but for the ($\log g, \log T_{\rm eff}$) diagram with the same  STAREVOL evolutionary tracks.
\label{Fig:loggTeffnormal_1} }
\end{figure}

\begin{figure}
\includegraphics[width=10cm]{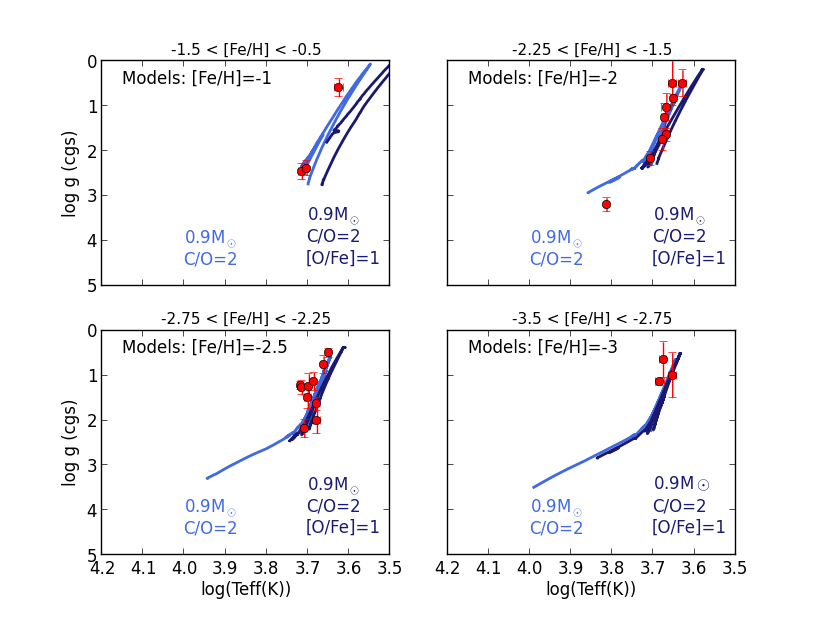}
\caption{Same as Fig.~\ref{Fig:loggTeffnormal_1}, but with the STAREVOL evolutionary tracks of modified composition  used in Fig.~\ref{Fig:HRanormal}.
\label{Fig:loggTeffanormal_2} }
\end{figure}

To reach a better agreement, 0.9~\Msun\ stellar-evolution models with a different initial C and O composition were computed.
Indeed CEMP stars are enriched in carbon and sometime have excess oxygen ([O/Fe]~$> 0$, as a result of Galactic chemical evolution), which impacts the photospheric opacities and thus the model effective temperature. When we only change the C/O ratio, we keep the initial O abundance and increase that of carbon to reach the desired C/O value. The mass fractions are then renormalized and the excess material removed from the most abundant species. When both [O/Fe] and C/O are imposed, we first set the O mass fraction to the desired value, then the carbon to get the required C/O ratio and renormalized as before. 
Our CEMP-s and -rs stars have on average  C/O~=~2.4 and [O/Fe]~=~0.96.
The comparison between Figs.~\ref{Fig:HRnormal} and \ref{Fig:HRanormal}
illustrates how the RGB is shifted towards lower temperatures when C/O~=~2, and even further  when an oxygen abundance of [O/Fe]~=~1 is adopted. In this way the agreement with the location of CEMP-rs stars in the HR diagram is improved. 
For completeness, the ($\log g, \log T_{\rm eff}$) diagrams, which are independent of distance and extinction, are also presented in Figs.~\ref{Fig:loggTeffnormal_1} and \ref{Fig:loggTeffanormal_2}.
These examples illustrate the importance of computing stellar evolution models with the proper chemical composition, especially when considering extrinsic stars that have been substantially  polluted by material with a non-solar composition.
Given the degeneracy of the evolutionary tracks along the RGB, no firm conclusion can be drawn on the stellar masses.

\section{Conclusions}
\label{Sect: conclusion}
The homogeneous analysis of a sample of 13 CEMP-s stars, 11 CEMP-rs stars, and one r-process-enriched star has shown that:
\begin{enumerate}
\item Using several elemental s- and r-process abundances is important to properly classify stars as either CEMP-s or CEMP-rs, given the abundance uncertainties that can affect individual determinations. We propose here a classification  based on an `abundance distance' to the r-process abundance profile (assumed to be universal), based on eight s- and r-process elements.
\item Whereas it is relatively easy to isolate r-process-enriched objects, it is far less obvious to draw a clear separation between CEMP-rs and CEMP-s stars. This difficulty is strengthened by the recent  abundance  studies (including ours) measuring r-process abundances in stars presenting a low s-process enrichment. These studies have uncovered a growing number of CEMP-rs stars characterized by modest overabundance levels\footnote{CEMP-s stars  with low enrichment levels also exist.}.
Distinguishing, at low metallicities, the origin of a r-contribution explained either as the star's pristine composition or as a consequence of 
mass transfer from an AGB having experienced proton-ingestion, will represent a challenge for future abundance studies.

\item The measured abundances of CEMP-rs stars can be reproduced by models of low-metallicity, low-mass stars experiencing a proton ingestion during a thermal pulse, yielding the intermediate neutron densities required for the i-process. We show that the quality of the fit is as good as that  obtained when s-process predictions are compared to CEMP-s abundance profiles. Since there is no doubt that the s-process nucleosynthesis takes place in low- and intermediate-mass TP-AGB stars (e.g. because of technetium detection),  similarly there should be no doubt  that the i-process responsible for the abundance peculiarities of CEMP-rs stars could also take place in low-mass, low-metalliticy AGB stars. We therefore propose to call them CEMP-sr stars instead, since they represent a particular manifestation of the s-process at low-metallicities.

\item It is important to use evolutionary tracks with a chemical composition matching that of extrinsic stars, in particular with the correct [C/Fe] and [O/Fe] ratios, to correctly reproduce the temperatures of CEMP stars in the HR diagram. Most CEMP-rs stars are found to lie on the RGB; since they have a large binary fraction (at least 82\% in our sample), they stand at the low-metallicity tail of the extrinsic-star family, with CEMP-s, CH, Ba, and extrinsic S stars as the other members of that family (along a sequence of increasing metallicity).
\end{enumerate}

\begin{acknowledgements}
We thank the anonymous referee for useful suggestions. D.K. acknowledges the financial support from the Science and Engineering Research Board (SERB), DST, India through  file number PDF/2017/002338, and CSIR-India through file No.13(9086-A)2019-Pool. A visiting research fellowship from F.R.S.-FNRS (Belgium) is also gratefully acknowledged. SVE thanks the Fondation ULB for its support. LS and SG are senior FNRS research associates. T.Me. is supported by a grant from the Fondation ULB. T.Ma. acknowledges support from MINECO under grant AYA-2017-88254-P.
 The {\it Mercator} telescope is operated thanks to grant number G.0C31.13 of the FWO under the “Big Science” initiative of the Flemish governement. Based on observations obtained with the HERMES spectrograph, supported by the Fund for Scientific Research of Flanders (FWO), the Research Council of K.U.Leuven, the Fonds National de la Recherche Scientifique (F.R.S.- FNRS), Belgium, the Royal Observatory of Belgium, the Observatoire de Genève, Switzerland and the Thüringer Landessternwarte Tautenburg, Germany. This research has made use of the SIMBAD database, operated at CDS, Strasbourg, France and NASA ADS, USA.

\end{acknowledgements}

\bibliographystyle{aa}
\bibliography{CEMP-ref}

\begin{appendix}
\section{Line list}

Table~\ref{Tab:linelist} presents the lines used in the present abundance analysis.\\
\topcaption{Lines used in the abundance analysis.\label{Tab:linelist}}
\begin{supertabular}{llr}
\hline
$\lambda$ & $\chi_{\rm low}$ & $\log gf$ \\
(\AA) & (eV) & \\
\hline\\
\tablehead{
\hline\\
$\lambda$ & $\chi_{\rm low}$ & $\log gf$ \\
(\AA) & (eV) & \\
\hline\\}

O I&      &      \\
6300.304& 0.000 &-9.715\\
7771.941&9.146&0.369\\
7774.161&9.146&0.223\\
7775.338&9.146&0.001\\
\end{supertabular}

\section{Elemental abundances for the program stars}
\begin{table}[h!]
\caption{Light element abundances.
\label{Tab:light_abundances}}
\begin{tabular}{lrrrrrrrrr}
\hline
\\
Star Name   &[Ca/Fe] &[Sc/Fe]  &[Ti/Fe] & [Cr/Fe]    &[Mn/Fe]  & [Co/Fe] &[Ni/Fe]&[Cu/Fe] & [Zn/Fe]\\
\hline
\\
CS 22887$-$048  &  0.66 &   -- &   0.45  &   0.16   &   $-$0.33   & 0.11  &0.18   &   --  & 0.39   \\
CS 22891$-$171  & 0.81  &   0.10  &   0.69  &    $-$0.09  &   $-$0.68 &   0.06  & 0.43  &     --    & 0.29 \\
CS 22942$-$019  & 0.31  &   0.35  &   0.40  &    $-$0.14  &   $-$0.23 &   0.11  & 0.18  &      --   & 0.24 \\
CS 30322$-$023  & 0.36  &      -  &   0.43  &    $-$0.39  &   $-$0.33 &  $-$0.04  & 0.13  &      --   & 0.34 \\
HD 26        & 0.19  &   0.38  &   0.63  &      0.07   &          0.05 &--     &0.11          & 0.09       &  0.42     \\
\hline
\end{tabular}
\end{table}

{\footnotesize
\begin{table*}[h!]
\caption{Elemental abundances
\label{Tab:abundances}}
\begin{tabular}{lclrrrrrrrr}
\hline
\\
\multicolumn{2}{c}{}& \multicolumn{4}{c}{CS 22891$-$171} && \multicolumn{3}{c}{CS 22887$-$048} \\
\cline{4-6}\cline{8-10}\\
 &    Z  &    log$_{\odot}{\epsilon}^a$ & log${\epsilon}$&$\sigma_{l}$ (N)& [X/Fe] $\pm~\sigma_{t}$ & & log${\epsilon}$&$\sigma_{l}$(N)& [X/Fe]~$\pm~ \sigma_{t}$  \\
 &       &                         &                  &        &        \\
\hline
C  & 6  & 8.43 &   8.00  &   0.08(3)  &   2.07 $\pm$ 0.07 & &8.20  &  0.08(3)   &   1.87 $\pm$ 0.06  &  \\
$^{12}$C/$^{13}$C& --& -- & -- &  --      &  6      & &    --      &    --       &   12         &  \\
N  & 7  & 7.83 &   7.00  & 	 0.08(5)  &   1.67 $\pm$ 0.33 & &     7.40  &  0.05(5)   &   1.67 $\pm$ 0.32  &  \\
\hline
\end{tabular}

$^{a}$ Asplund et al. (2009) \\
$:$ Uncertain abundances due to noisy/blended region\\
\end{table*}
}
\setcounter{table}{1}
{\footnotesize
\begin{table*}
\caption{Continued.}
\begin{tabular}{lclrrrrrrrr}
\hline
\\
\multicolumn{2}{c}{}& \multicolumn{4}{c}{HD~26} && \multicolumn{3}{c}{HD~5223} \\
\cline{4-6}\cline{8-10}\\
 &    Z  &    log$_{\odot}{\epsilon}^a$ & log${\epsilon}$&$\sigma_{l}$ (N)& [X/Fe]~$\pm~ \sigma_{t}$ & &log${\epsilon}$&$\sigma_{l}$(N)  & [X/Fe]~$\pm~ \sigma_{t}$  \\
 &       &                         &                  &        &        \\
\hline

C  &  6   & 8.43 &  8.10  &  0.06(6)  &   0.65 $\pm$ 0.06   &  &   7.90   &  0.03(4)  &  1.47 $\pm$ 0.05  &  \\
$^{12}$C/$^{13}$C& --&--  & -- &--        &  16         & &   --        &      --      &   9         &   \\
N  &  7   & 7.83 &  7.55  &  0.03(10) &   0.70 $\pm$ 0.32   &  &   7.10   &  0.05(5)  &  1.27 $\pm$ 0.32  &  \\
\hline
\end{tabular}

$^{a}$ Asplund et al. (2009) \\
$:$ Uncertain abundances due to noisy/blended region\\
\end{table*}
}
\setcounter{table}{1}
{\footnotesize
\begin{table*}
\caption{Continued.}
\begin{tabular}{lclrrrrrrrr}
\hline\\
\multicolumn{2}{c}{}& \multicolumn{4}{c}{HD~145777} && \multicolumn{3}{c}{HD~187861} \\
\cline{4-6}\cline{8-10}\\
&    Z  &    log$_{\odot}{\epsilon}^a$ & log${\epsilon}$&$\sigma_{l}$ (N)&  [X/Fe]~$\pm~ \sigma_{t}$ & &log${\epsilon}$&$\sigma_{l}$(N)&  [X/Fe]~$\pm~ \sigma_{t}$  \\
 &       &                         &                  &        &        \\
\hline
C  &  6  & 8.43 & 7.50  &  0.10(4)  &  1.39 $\pm$ 0.07   & & 8.30  &  0.09(3)  &  2.47 $\pm$ 0.08  &  \\
$^{12}$C/$^{13}$C& -- & -- & --  &   --      &  6        & &      --     &     --       &   16         &  \\
N  &  7  & 7.83 & 7.20  &  0.05(8)  &  1.59 $\pm$ 0.32   & & 7.85  &  0.03(4)  &  2.62 $\pm$ 0.32  & \\
\hline
\end{tabular}

$^{a}$ Asplund et al. (2009) \\
$:$ Uncertain abundances due to noisy/blended region\\
\end{table*}
}
\setcounter{table}{1}
{\footnotesize
\begin{table*}
\caption{Continued.}
\begin{tabular}{lrlrrrrrrrr}
\hline
\\
\multicolumn{2}{c}{}& \multicolumn{4}{c}{HD~201626} && \multicolumn{3}{c}{HD~206983} \\
\cline{4-6}\cline{8-10}\\
 &    Z  &    log$_{\odot}{\epsilon}^a$ & log${\epsilon}$&$\sigma_{l}$ (N)& [X/Fe]~$\pm~ \sigma_{t}$ & &log${\epsilon}$&$\sigma_{l}$(N) & [X/Fe]~$\pm~ \sigma_{t}$  \\
 &       &                         &                  &        &        \\
\hline

C  &   6  & 8.43 & 8.30  &   0.05(3)   &  1.62 $\pm$ 0.06 & &    7.85  &    0.04(5)  &   0.42 $\pm$ 0.05  &    \\
$^{12}$C/$^{13}$C& -- & -- & -- &     --       &  9         & &   --        &      --      &   24        &   \\
N  &   7  & 7.83 & 7.30  &   0.04(5)   &  1.43 $\pm$ 0.33 & &    8.10  &    0.18(10) &   1.27 $\pm$ 0.32  & \\
\hline
\end{tabular}

$^{a}$ Asplund et al. (2009) \\
$:$ Uncertain abundances due to noisy/blended region\\
\end{table*}
}
\setcounter{table}{1}
{\footnotesize
\begin{table*}
\caption{Continued.}
\begin{tabular}{lclrrrrrrrr}
\hline
\\
\multicolumn{2}{c}{}& \multicolumn{4}{c}{HD 224959} && \multicolumn{3}{c}{HE 0111$-$1346} \\
\cline{4-6}\cline{8-10}\\
 &    Z  &    log$_{\odot}{\epsilon}^a$ & log${\epsilon}$&$\sigma_{l}$ (N)&  [X/Fe]~$\pm~ \sigma_{t}$ & &log${\epsilon}$&$\sigma_{l}$(N)& [X/Fe]~$\pm~ \sigma_{t}$  \\
 &       &                         &                  &        &        \\
\hline

C  &  6  & 8.43 &  8.30  &   0.05(3)  &   2.23 $\pm$ 0.07 & &   8.00  &  0.04(3)  &   1.67 $\pm$ 0.06  &\\
$^{12}$C/$^{13}$C&-- & -- & -- &      --      &       12    & &   --      & --      &   16         & \\
N  &  7  & 7.83 &  7.95  &   0.06(10) &   2.48 $\pm$ 0.33 & &   7.30  &  0.05(4)  &   1.57 $\pm$ 0.32  &  \\
\hline
\end{tabular}

$^{a}$ Asplund et al. (2009) \\
$:$ Uncertain abundances due to noisy/blended region \\
\end{table*}
}
\setcounter{table}{1}
{\footnotesize
\begin{table*}
\caption{Continued.}
\begin{tabular}{lclrrrrrrrr}
\hline
\\
\multicolumn{2}{c}{}& \multicolumn{4}{c}{HE 0507$-$1653} && \multicolumn{3}{c}{HE 1120$-$2122} \\
\cline{4-6}\cline{8-10}\\
 &    Z  &    log$_{\odot}{\epsilon}^a$ & log${\epsilon}$&$\sigma_{l}$ (N) & [X/Fe]~$\pm~\sigma_{t}$ & &log${\epsilon}$&$\sigma_{l}$(N) & [X/Fe]~$\pm~\sigma_{t}$  \\
 &       &                         &                  &        &        \\
\hline

C  &  6  & 8.43 &    8.30  &  0.10(5)  &   1.22 $\pm$ 0.07  &&    7.60  &  0.08(4)     &  1.17 $\pm$ 0.07  &  \\
$^{12}$C/$^{13}$C&-- & -- & -- &   --  &  32          &&   --       &    --       &   9         &   \\
N  &  7  & 7.83 &    7.50  &  0.09(4)  &   1.02 $\pm$ 0.33  &&    7.30  &  0.04(6)	 &  1.47 $\pm$ 0.32  &  \\
\end{tabular}

$^{a}$ Asplund et al. (2009) \\
$:$ Uncertain abundances due to noisy/blended region \\
\end{table*}
}
\setcounter{table}{1}
{\footnotesize
\begin{table}
\caption{Continued.}
\begin{tabular}{lclrrr}
\hline
\\
\multicolumn{2}{c}{}& \multicolumn{4}{c}{HE 2255$-$1724}  \\
\cline{4-6}\\
 &    Z  &    log$_{\odot}{\epsilon}^a$ & log${\epsilon}$&$\sigma_{l}$ (N) & [X/Fe]~$\pm~\sigma_{t}$  \\
 &       &                         &                  &        &        \\
\hline

C  &  6  & 8.43 &    7.85  &  0.08(5)  &   1.74 $\pm$ 0.06    \\
$^{12}$C/$^{13}$C& -- &  --  & --         &       --      &  7 \\
N  &  7  & 7.83 &    7.00  &  0.12(4)  &   1.49 $\pm$ 0.33    \\
\end{tabular}

$^{a}$ Asplund et al. (2009) \\
$:$ Uncertain abundances due to noisy/blended region\\
\end{table}

{\footnotesize
\begin{table}
\caption{Abundance ratios for the programme stars}
\label{Tab:hsls_ratios}
\begin{tabular}{lrrrr}
\hline
\\
Star name &[ls/Fe] & [hs/Fe]&  [hs/ls] & [s/Fe] \\
\hline\\

CS 22887$-$048 & 1.26  & 1.85  & 0.59 & 1.62\\
CS 22891$-$171 & 0.94  & 2.19  & 1.25 & 1.75\\
CS 22942$-$019 & 1.68  & 1.28  &$-$0.40 & 1.45\\
CS 30322$-$023 & 0.03  & 1.06  & 1.03 & 0.65\\
HD 26        & 1.38  & 1.81  & 0.43 & 1.66\\
HD 5223      & 0.70  & 1.48  & 0.78 & 1.21\\
HD 55496     & 1.06  & 0.73  &$-$0.33 & 0.87\\
HD 76396     & 1.43  & 1.88  & 0.45 & 1.74\\
HD 145777    & 0.65  & 1.11  & 0.46 & 0.93\\
HD 187861    & 0.61  & 1.74  & 1.13 & 1.38\\
HD 196944    & 0.42  & 0.83  & 0.41 & 0.72\\
HD 198269    & 0.47  & 1.27  & 0.80 & 1.01\\
HD 201626    & 0.94  & 1.60  & 0.66 & 1.38\\
HD 206983    & 0.25  & 0.74  & 0.49 & 0.59\\
HD 209621    & 1.03  & 1.81  & 0.78 & 1.56\\
HD 221170    &$-$0.20  & 0.29  & 0.49 & 0.07\\
HD 224959    & 0.74  & 2.05  & 1.31 & 1.65\\
HE 0111$-$1346 & 0.85  & 1.87  & 1.02 & 1.52\\
HE 0151$-$0341 & 0.67  & 1.81  & 1.14 & 1.44\\
HE 0319$-$0215 & 0.49  & 1.73  & 1.24 & 1.32\\
HE 0507$-$1653 & 1.28  & 1.78  & 0.50 & 1.62\\
HE 1120$-$2122 & 0.75  & 1.35  & 0.60 & 1.14\\
HE 1429$-$0551 & 1.01  & 1.36  & 0.35 & 1.28\\
HE 2144$-$1832 & 1.25  & 1.62  & 0.37 & 1.50\\
HE 2255$-$1724 & 0.71  & 1.59  & 0.88 & 1.32\\
\hline\\
\end{tabular}

[ls/Fe] = ([Y/Fe]+[Zr/Fe])/2

[hs/Fe] = ([La/Fe]+[Ce/Fe])/2

[s/Fe]  = ([Y/Fe]+[Zr/Fe]+[Ba/Fe]+[La/Fe]+[Ce/Fe]+[Nd/Fe])/6

\end{table}
}

\begin{table}\small
\caption{Standard, minimum and maximum s- and r-contributions to the Solar System abundances, adapted from Table~3 of \cite{Goriely1999}, here presented per element (El.) and  in the scale  $\log \epsilon_{x_i} = \log_{10} (n_{x_i}/n_{\rm H}) + 12$. A dash means that the corresponding abundance limit is totally negligible}
\label{Tab:r-process}
\begin{tabular}{llrrrrrr}
\hline
\\
Z & El. &   Stand s & Min s & Max s & Stand r & Min r & Max r \\
\hline\\
30 & Zn & 2.44 & 2.40 & 2.48  & 2.44 & 2.38 & 2.48  \\
31 & Ga & 3.12 & 3.10 & 3.14  & 2.46 &  --  & 2.81  \\
32 & Ge & 3.52 & 3.47 & 3.56  & 3.09 & 2.80 & 3.30  \\
33 & As & 2.35 & 2.33 & 2.37  & 2.13 & 2.06 & 2.22  \\
34 & Se & 3.35 & 3.33 & 3.37  & 3.13 & 3.09 & 3.28  \\
\end{tabular}
\end{table}

}

\end{appendix}

\end{document}